%% file: main.tex
\documentclass[12pt,twoside,doublespace,leftblank,vi]{mitthesis}

\usepackage{cmap}
\usepackage[T1]{fontenc}
\usepackage{longtable}

\usepackage{booktabs}
\usepackage{array}
\usepackage[table]{xcolor}
\usepackage[nottoc]{tocbibind} %to include list of figures and list of tables in the TOC
\usepackage{multicol}
\usepackage{hyperref} %to include hyperlinks
\usepackage{paralist} %to create inline lists
\usepackage{caption} 
\usepackage{subcaption} %to create subfigures
\usepackage[final]{microtype} %for better typography - specially overfull boxes in bibliography

\usepackage{tabularx}

\usepackage[
    backend=biber,
    style=numeric,
    sorting=none,
    sortcites=false,
    block=space
]{biblatex}
\AtEveryBibitem{%
  \clearfield{urlyear}%
  \clearfield{urlmonth}%
  \clearfield{urlday}%
  \clearfield{urlendyear}%
  \clearfield{urlendmonth}%
  \clearfield{urlendday}%
}
\usepackage{pgfplots, pgfplotstable} %to make plots
\usepgfplotslibrary{dateplot}
\usepgfplotslibrary{fillbetween}
\usepackage{amsmath}
\usepackage{amssymb} % for \lesssim and other math symbols
\pgfplotsset{compat=1.3}
\usepackage{tikz}
\usetikzlibrary{chains,shapes.symbols}

\begin{document}

\include{cover}
% Some departments (e.g. 5) require an additional signature page.  See
% signature.tex for more information and uncomment the following line if
% applicable.
% \include{signature}
\pagestyle{plain}

\include{contents}
\include{chap1}
\include{chap2}
\include{chap3}
\include{chap4}

\include{chap5}
\include{chap6}
\include{chap7}

\appendix
\include{appa}

\include{biblio}
\end{document}

%% file: cover.tex
% -*-latex-*-
%
% Drop-in replacement for cover.tex.
% This version follows the newer MIT example style in dual-degree-committee.pdf:
%   - no dotted signature lines on the title page
%   - "Authored by / Certified by / Approved by" blocks
%   - optional thesis committee page
% It deliberately does NOT load any font package, so it preserves the font
% already used by your mitthesis document.

\title{Measuring the electric dipole moment of the neutron using neutron star spin-down}
\author{Abriana Lyda}

% Leave this blank unless you want previous degrees printed below your name.
% Example:
% \prevdegrees{B.S. Physics, Massachusetts Institute of Technology, 2026}
\prevdegrees{}

\department{Department of Physics}
\degree{BACHELOR OF SCIENCE IN PHYSICS}
\degreemonth{May}
\degreeyear{2026}
\thesisdate{May 8, 2026}

\makeatletter

% Patch only the supervisor metadata behavior locally.  The uploaded mitthesis.cls
% had a hard-coded supervisor title in its abstract-page macro; this version uses
% the title supplied in the second argument below and includes the co-supervisor
% on the abstract page.
\def\supervisor#1#2{\setbox\@titlesupervisor\vbox{\unvbox\@titlesupervisor
  \vskip 10pt \def\baselinestretch{1}\large
  \fontsize{12}{12}\selectfont\signature{Certified by}{#1\\ Thesis Supervisor \\ #2}}%
  \setbox\@abstractsupervisor\vbox{\unvbox\@abstractsupervisor
  \vskip\baselineskip \def\baselinestretch{1}\@normalsize
  \par\noindent Thesis Supervisor: #1 \\ #2}}

\def\cosupervisor#1#2{\setbox\@titlecosupervisor\vbox{\unvbox\@titlecosupervisor
  \vskip 10pt \def\baselinestretch{1}\large
  \fontsize{12}{12}\selectfont\signature{Certified by}{#1\\ Co-Thesis Supervisor \\ #2}}%
  \setbox\@abstractcosupervisor\vbox{\unvbox\@abstractcosupervisor
  \vskip\baselineskip \def\baselinestretch{1}\@normalsize
  \par\noindent Co-Thesis Supervisor: #1 \\ #2}}

\def\endabstractpage{\end{abstract}\noindent
  \unvcopy\@abstractsupervisor
  \unvcopy\@abstractcosupervisor
  \newpage}

\makeatother

% These supervisor commands are retained so that the abstract page still knows
% who the thesis supervisors are.  The custom title page below prints the title
% page explicitly, rather than relying on the older \maketitle layout.
\supervisor{Dr.~Prajwal Mohan Murthy}{Senior Research Fellow}
\cosupervisor{Professor Robert Redwine}{Professor of Physics, Emeritus}
\chairman{Professor Scott Hughes}{ Associate Head, Department of Physics}

\makeatletter

% Current MIT permission language, matching the structure of the attached sample.
\newcommand{\MITModernPermission}{%
The author hereby grants to MIT a nonexclusive, worldwide, irrevocable,
royalty-free license to exercise any and all rights under copyright,
including to reproduce, preserve, distribute and publicly display copies of
this thesis, or release the thesis under an open-access license.%
}

% One aligned role block on the title page.
\newcommand{\MITCoverRole}[3]{%
  \par\noindent
  \begin{minipage}[t]{1.20in}\raggedright #1:\end{minipage}%
  \begin{minipage}[t]{4.25in}\raggedright #2\\#3\end{minipage}%
  \par\vspace{0.55\baselineskip}%
}

% A committee-list entry.  Used on the optional committee page.
\newcommand{\MITCommitteeEntry}[2]{%
  {#1\par}%
  {#2\par}%
  \vspace{1.25\baselineskip}%
}

% Optional committee page.  The second argument is for extra/non-supervisor
% committee members.  If it is empty, the "Other Committee Members" section is
% suppressed.
\newcommand{\MITThesisCommitteePage}[2]{%
  \clearpage
  \thispagestyle{empty}%
  \begingroup
  \begin{singlespace}
  \vspace*{0.75in}%
  \begin{flushright}
  \begin{minipage}{0.60\textwidth}
  \centering
  {\Large THESIS COMMITTEE\par}
  \vspace{1.65\baselineskip}

  {\Large Thesis Supervisors\par}
  \vspace{1.25\baselineskip}
  #1

  \if\relax\detokenize{#2}\relax
  \else
    \vspace{1.7\baselineskip}
    {\Large Other Committee Members\par}
    \vspace{1.25\baselineskip}
    #2
  \fi
  \end{minipage}
  \end{flushright}
  \end{singlespace}
  \endgroup
  \clearpage
}

% Title page matched to the attached PDF layout, but set in the thesis font.
\newcommand{\MITModernTitlePage}{%
  \clearpage
  \thispagestyle{empty}%
  \begingroup
  \begin{singlespace}
  \centering
  \vspace*{0.25in}

  {\Large \@title\par}
  \vspace{0.55\baselineskip}
  {By\par}
  \vspace{0.45\baselineskip}
  {\Large \@author\par}
  \ifx\@prevdegrees\@empty
  \else
    \vspace{0.35\baselineskip}
    {\@prevdegrees\par}
  \fi

  \vspace{2.0\baselineskip}
  {\MakeUppercase{Submitted to the \@department}\par}
  \vspace{0.25\baselineskip}
  {\MakeUppercase{in partial fulfillment of the requirements for the \@degreeword\ of}\par}
  \vspace{0.25\baselineskip}
  {\MakeUppercase{\@degree}\par}
  \vspace{0.25\baselineskip}
  {\MakeUppercase{at the}\par}
  {\MIT\par}
  \vspace{0.25\baselineskip}
  {\MakeUppercase{\@degreemonth\ \@degreeyear}\par}

  \vspace{1.35\baselineskip}
  {\copyright\@degreeyear\ \@author. All rights reserved.\par}
  \vspace{0.45\baselineskip}
  \begin{minipage}{0.92\textwidth}
    \centering
    \MITModernPermission
  \end{minipage}

  \vspace{1.25\baselineskip}
  \begin{minipage}{0.96\textwidth}
  \MITCoverRole{Authored by}{\@author}{\@department, \@thesisdate}
  \MITCoverRole{Certified by}{Dr. Prajwal Mohan Murthy}{Senior Research Fellow, Thesis Supervisor}
  \MITCoverRole{Certified by}{Professor Robert Redwine}{Professor of Physics, Emeritus, Co-Thesis Supervisor}
  \MITCoverRole{Approved by}{Professor Scott Hughes}{Associate Professor of Physics \\Associate Head, Department of Physics}
  \end{minipage}

  \end{singlespace}
  \endgroup
  \clearpage
}

\makeatother

% -----------------------------------------------------------------------------
% FRONT MATTER
% -----------------------------------------------------------------------------

\MITModernTitlePage

% Keep this page if your department/thesis office wants a committee page.  If you
% only want the title page, comment out the whole \MITThesisCommitteePage block.
\MITThesisCommitteePage{% thesis supervisors
  \MITCommitteeEntry{Dr.~Prajwal Mohan Murthy}{Senior Research Fellow}
  \MITCommitteeEntry{Professor Robert Redwine}{Professor of Physics, Emeritus}
}{% other committee members; leave this argument empty if not needed
  % \MITCommitteeEntry{Committee Member Name}{Title or Department}
}

% The abstractpage environment below is still the one from mitthesis.cls.
\cleardoublepage
% Uncomment the next line if you do NOT want a page number on your
% abstract and acknowledgments pages.
% \pagestyle{empty}
\setcounter{savepage}{\thepage}
\begin{abstractpage}
\input{abstract}
\end{abstractpage}

\cleardoublepage

\section*{Acknowledgments}

I would first and foremost like to thank Dr.~Prajwal Mohan Murthy for his mentorship over the past three years. I have been extraordinarily fortunate to receive such dedicated one-on-one mentorship as an undergraduate, and I will carry the lessons from this experience with me throughout my career as a physicist. I also gratefully acknowledge that Dr.~Mohan Murthy's contributions to this work were supported by a Phi Kappa Phi Fellowship and BNL Award No.~460913.

I would also like to thank Professor Robert Redwine for his steadfast support during my time in his research group, as well as for the funding provided through the MIT Laboratory for Nuclear Science. I am additionally grateful for support from NASA, MIT UROP, and MIT Student Aid.

Finally, I would like to thank my family---Mom, Dad, Josh, and Linda---whose support I have come to understand and appreciate more deeply throughout my time at MIT. Their unwavering pride and encouragement have been a pillar throughout these years. To James, Fiona, Diego, Izzy, Tyra, and Maggie: I could not have asked for a better group of people to go through MIT with.

%% file: abstract.tex
% $Log: abstract.tex,v $
% Revision 1.1  93/05/14  14:56:25  starflt
% Initial revision
% 
% Revision 1.1  90/05/04  10:41:01  lwvanels
% Initial revision
% 
%
%% The text of your abstract and nothing else (other than comments) goes here.
%% It will be single-spaced and the rest of the text that is supposed to go on
%% the abstract page will be generated by the abstractpage environment.  This
%% file should be \input (not \include 'd) from cover.tex.

\noindent\textit{Context.}
The neutron electric dipole moment (nEDM) is a sensitive probe of physics beyond the Standard Model and of new sources of CP violation. While laboratory searches currently provide the strongest direct limits, neutron stars offer a complementary environment in which large neutron populations, strong magnetic fields, and long-term spin evolution can be used to test additional energy-loss channels.

\smallskip
\noindent\textit{Aims.}
This work develops a source-specific framework for constraining the nEDM using the nearby millisecond pulsar PSR J0437-4715. The goal is to determine how much of its intrinsic spin-down power could be attributed to electric dipole radiation from polarized crustal neutrons.

\smallskip
\noindent\textit{Methods.}
The analysis combines neutron-star structure modeling, radio-polarization constraints, surface-geometry information, and present-day spin-down energetics. Mass and radius constraints are used to construct an individualized stellar structure profile and estimate the inner-crust free-neutron reservoir. Folded and single-pulse radio polarization data provide a surface magnetic-field estimate independent of the observed spin-down, while hot-spot constraints are used to compare candidate magnetic geometries. These inputs are incorporated into a present-day torque budget, and the residual spin-down power not accounted for by standard loss channels is interpreted as the maximum power available to a possible nEDM-driven radiation channel. The direct EDR interpretation is then examined for crustal and magnetospheric screening, motivating a screening-aware alternative in which the residual power is assigned to an effective CP-odd magnetic-quadrupole channel.

\smallskip
\noindent\textit{Results.}
Using the screening-aware magnetic quadrupolar interpretation, the residual-power analysis gives a 90\% upper limit on the neutron-scale CP-odd magnetic quadrupole moment of $|M_0^n| < 7.47\times10^{-38}\ e\,{\rm cm}^2$. Under the adopted QCD conversion coefficients, this corresponds to $|\bar{\theta}| < 2.99\times10^{-9}$ and to an equivalent neutron electric dipole moment limit of $|d_n| < 4.42\times10^{-25}\ e\ {\rm cm}\ (90\%~{\rm C.L.})$.

\smallskip
\noindent\textit{Conclusions.}
The work first obtains an unscreened direct-EDR benchmark and then derives a screening-aware bound on an effective neutron MQM. The associated nEDM value from the latter is conditional on a pure-$\bar\theta$ interpretation. Although weaker than current laboratory limits, this bound demonstrates that source-specific neutron-star modeling can provide a complementary astrophysical constraint on the nEDM. 

%% file: contents.tex
  % -*- Mode:TeX -*-
%% This file simply contains the commands that actually generate the table of
%% contents and lists of figures and tables.  You can omit any or all of
%% these files by simply taking out the appropriate command.  For more
%% information on these files, see appendix C.3.3 of the LaTeX manual. 
\tableofcontents
\newpage
\listoffigures
\newpage
\listoftables

%% file: chap1.tex
%% This is an example first chapter.  You should put chapter/appendix that you
%% write into a separate file, and add a line \include{yourfilename} to
%% main.tex, where `yourfilename.tex' is the name of the chapter/appendix file.
%% You can process specific files by typing their names in at the 
%% \files=
%% prompt when you run the file main.tex through LaTeX.
\chapter{Introduction}

One of the goals of modern particle physics is to understand why the observed universe is made predominantly of matter rather than antimatter. Within the Standard Model, this question is tied to the violation of discrete symmetries, namely charge conjugation $C$, parity $P$, and time reversal $T$. These symmetry operations probe how the laws of physics behave under fundamental transformations of particles, space, and time. The CPT theorem states that in any local, Lorentz-invariant quantum field theory with a Hermitian Hamiltonian, the combined transformation $CPT$ must be an exact symmetry. In other words, although nature need not be invariant under charge conjugation, parity, or time reversal separately, it is expected to remain invariant when all three are applied. Sakharov's classic conditions for generating a baryon asymmetry require baryon number violation, $C$ and $CP$ violations, as well as a departure from thermal equilibrium \cite{sakharov_violation_1967}. However, the Standard Model fails to account for matter-antimatter asymmetry. Because of this, CP violation is one of the most important probes of physics beyond the Standard Model. Electric dipole moments are among the most sensitive known tests of CP-violating physics. A neutron electric dipole moment would provide direct evidence for time-reversal and parity-violating interactions, and—assuming $CPT$ invariance—thus for CP violation.

The neutron is electrically neutral overall, yet it is composed of charged quarks. A nonzero EDM corresponds to a separation of positive and negative charge inside the neutron, aligned with its spin. For a non-degenerate spin-1/2 system, such a permanent dipole moment is forbidden by both parity and time reversal symmetry. 

CP violation in the weak interaction has been established experimentally. The first observation came in 1964 in the neutral kaon system, where the decay $K_L \rightarrow \pi\pi$ revealed that CP symmetry is not exact in weak processes \cite{christenson_evidence_1964}. In the Standard Model these effects are described by the Cabibbo–Kobayashi–Maskawa (CKM) matrix appearing in the charged-current weak interaction \cite{kobayashi_cp_1973}. 
\begin{figure*}
\centering
\includegraphics[width=\textwidth,clip]{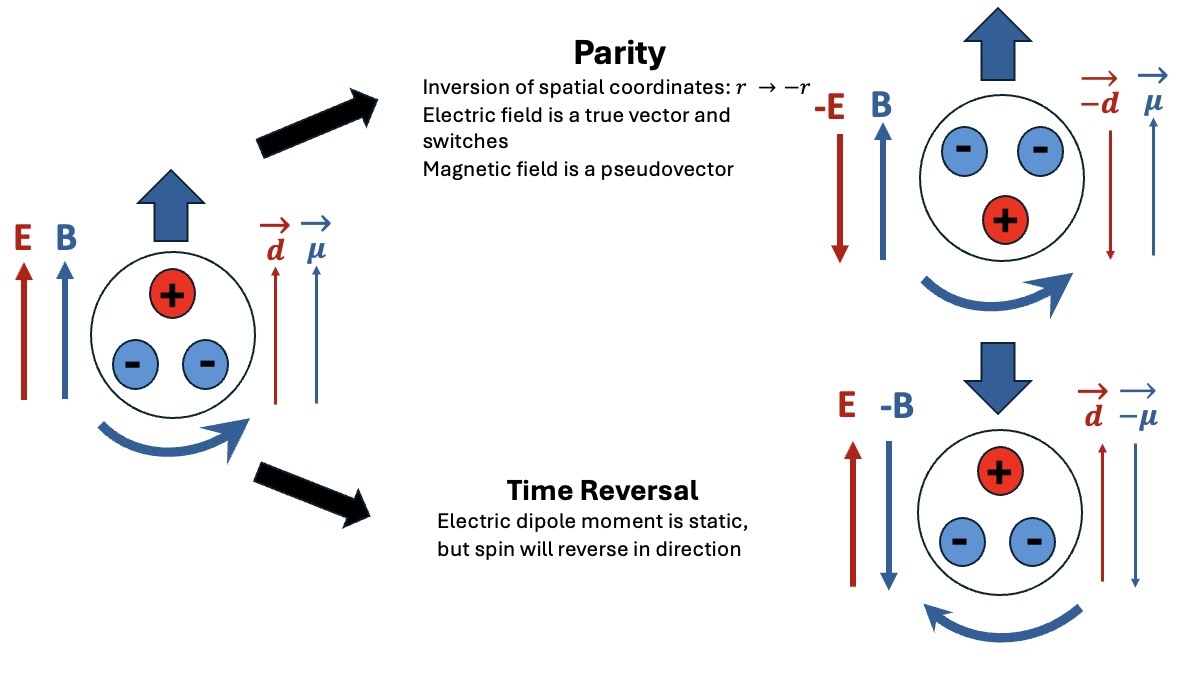}
\caption{Illustration of the discrete-symmetry argument for a permanent neutron electric dipole moment (nEDM). In external electric and magnetic fields, the electric dipole moment $\mathbf{d}$ behaves as a true vector, while the magnetic moment $\mathbf{\mu}$ and spin $\mathbf{J}$ behave as pseudo-vectors. Under time reversal $T$, the spin and magnetic moment reverse direction while a permanent electric dipole moment remains unchanged, so a nonzero nEDM is not invariant under $T$. Under the combined charge-conjugation and parity transformation $CP$, the electric dipole moment changes sign while the spin remains unchanged, again showing that a permanent nEDM violates $CP$. Assuming CPT symmetry is conserved, observation of a permanent nEDM would therefore imply $T$ and $CP$ violation.}
\label{fig-1} 
\end{figure*}
Quantum chromodynamics (QCD) allows a CP-violating term in its Lagrangian; however, unlike the weak CP violation, no comparable CP-violating signal has been observed in the strong sector. This implies that $\bar\theta$, the physical CP-violating angle of QCD, must be extremely small. Currently the experimental upper bound is $\bar \theta < 6.2 \times 10^{-11}$. However, nothing in the Standard Model requires it \cite{liu_lattice_2024, benabou_clearing_2025}. This is referred to as the strong CP problem, and the neutron electric dipole moment is thus a distinctly important observable, as it is a clean probe for CP violation in the strong sector. The current best experimental limit is $|d_n|< 1.8\times 10^{-26} e ~\rm cm$ \cite{abel_measurement_2020}. The CKM contribution to the neutron EDM is expected to be between $10^{-32}-10^{-31} e ~\rm cm$. Next-generation ultracold-neutron experiments aim to improve the sensitivity by roughly an order of magnitude; however, the current experimental methods are limited by the number of ultracold neutrons that can be produced and stored. These experiments predict measurements around $3\times 10^{-27}e\rm~cm$. However, the statistical uncertainty limits measurements to be $>1\times 10^{-28} e\rm~cm$ \cite{ayres_design_2021}. This ultimately calls for new methods for measuring the neutron EDM, and one promising approach is to use neutron stars. 

\section{Motivations for introducing an astrophysical method for measuring the nEDM}
Neutron stars immediately combat the issue of statistical uncertainty, as they contain nearly \(10^{57}\) neutrons \cite{piekarewicz_nuclear_2022}. More broadly, astrophysical and cosmological systems have repeatedly provided some of the strongest constraints on fundamental particle properties. A particularly important example is the sum of neutrino masses: observations of the cosmic microwave background and baryon acoustic oscillations constrain the total neutrino mass to \(\sum m_\nu < 0.12~{\rm eV}\), demonstrating that astrophysical data can probe particle-physics parameters beyond the reach of many direct laboratory measurements. Further, the sum of neutrino masses and the nEDM are both quantities that have lower and upper bounds set by either experimental or theoretical constraints, making the nEDM a natural target for astrophysical methods. Stellar systems provide similar examples. The neutrino magnetic moment has been constrained to \(\mu_{\nu}<1.5\times 10^{-12}\mu_B\) from stellar evolution near the tip of the red-giant branch \cite{capozzi_axion_2020}, while stellar cooling gives some of the strongest axion bounds, with globular-cluster red-giant data implying \(g_{ae}\lesssim 2.5\times 10^{-13}\) and horizontal-branch star counts implying \(g_{a\gamma\gamma}<0.65\times 10^{-10}\) \cite{caputo_astrophysical_2024, ayala_revisiting_2014}. These examples motivate the use of neutron stars as complementary laboratories for constraining fundamental quantities such as the nEDM.
\begin{figure*}
\centering
\includegraphics[width=\textwidth,clip]{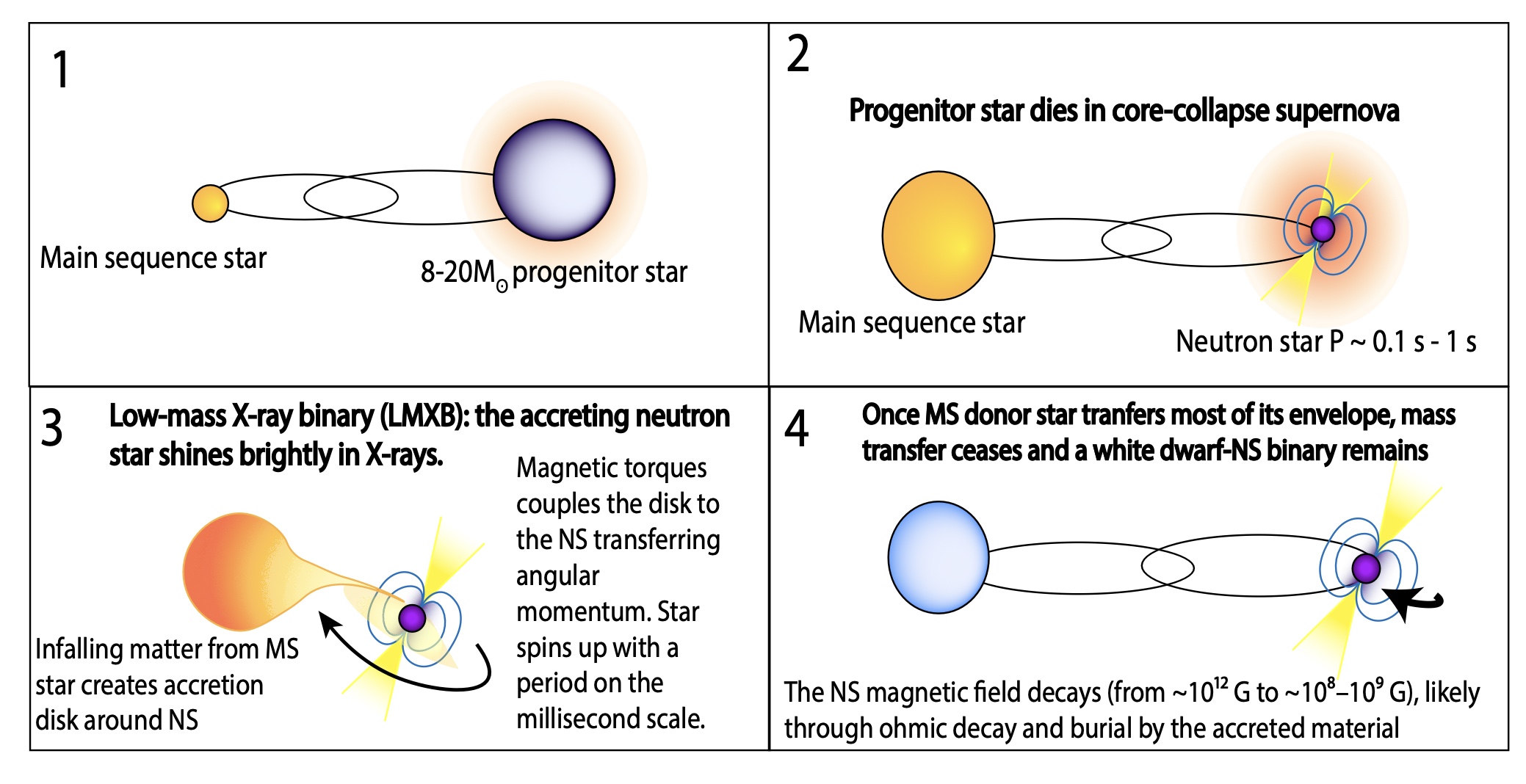}
\caption{Illustration depicting the recycling process of a millisecond pulsar, specifically for an MSP similar to PSR J0437-4715 that has a white dwarf companion. As elaborated later, we will use the age of the companion to model energy loss evolution post-recycling \cite{lorimer_binary_2008}.}
\label{recyled} 
\end{figure*}
\section{Background on PSR J0437-4715}
Neutron stars are the compact remnants left behind when massive stars end their lives in core collapse events. Because they compress more than a solar mass into a stellar remnant only $\approx 10 \rm~km$ in radius, they provide a unique laboratory for studying matter, gravity, magnetism, and radiation under such extreme conditions. A particularly important subset of neutron stars is millisecond pulsars (MSPs), which are old neutron stars that have been spun up through accretion from a binary companion, as shown in Fig.~\ref{recyled}. Their short spin period, low spin-down rates, and extreme rotational stability make them valuable for studying dense-matter physics, binary evolution, precision timing, and relativistic gravity.

Among known MSPs, PSR J0437-4715 is especially important. Discovered in the Parkes 70-cm survey, it was immediately recognized as an unusually bright and nearby system \cite{johnston_discovery_1993}. Subsequent long-baseline timing has made it one of the best characterized neutron stars, as seen in Table \ref{tab:j0437_observables}. Recent Parkes Pulsar Timing Array analysis gives a distance of $156.96\pm 0.11~\rm pc$ and an orbital inclination of $137.506\pm 0.016^{\circ}$ \cite{reardon_neutron_2024}. PSR J0437-4715 is a recycled pulsar in a binary with a low-mass white dwarf companion. Multiwavelength observations show that the companion is well described by a cool hydrogen-atmosphere white dwarf with mass $0.25\pm 0.02~M_{\odot}$ and an age of $6.0\pm 0.5~\rm Gyr$ \cite{durant_spectrum_2012}. Ultraviolet and X-ray observations reveal thermal emission associated with the neutron star itself. In the ultraviolet, the inferred bulk surface temperature lies in the range $(1.25-3.5)\times 10^5$ K \cite{durant_spectrum_2012}, which is surprisingly high for a star of this age and requires that some internal heating mechanisms must be present.

% ---------- Preamble ----------

\definecolor{sectiongray}{gray}{0.92}

\newcolumntype{L}[1]{>{\raggedright\arraybackslash}p{#1}}

% ---------- Table ----------
\begin{table}[p]
\centering
\scriptsize
\setlength{\tabcolsep}{3.5pt}
\renewcommand{\arraystretch}{1.08}

\caption{\textbf{Measured and inferred observables for PSR J0437$-$4715.}
Parentheses denote $1\sigma$ uncertainties in the last quoted digits.}
\label{tab:j0437_observables}

\vspace{0.3em}

\begin{tabular}{@{}r L{0.33\textwidth} L{0.27\textwidth} L{0.32\textwidth}@{}}
\toprule
\multicolumn{2}{c}{\textbf{Quantity}} & \textbf{Value} & \textbf{Source} \\
\midrule

\rowcolor{sectiongray}
\multicolumn{4}{@{}l}{\textbf{Discovery and radio timing}} \\

1. & Discovery / binary MSP identification
& Bright nearby binary MSP; $P \simeq 5.75~\mathrm{ms}$
& Johnston et al. \cite{johnston_discovery_1993} \\

2. & Spin frequency, $f$
& $173.6879456649439(4)~\mathrm{s^{-1}}$
& Reardon et al. \cite{reardon_neutron_2024} \\

3. & Spin period, $P=1/f$
& $5.75745194159340~\mathrm{ms}$
& Derived from Reardon et al. \cite{reardon_neutron_2024} \\

4. & Spin-frequency derivative, $\dot f$
& $-1.728367(3)\times10^{-15}~\mathrm{s^{-2}}$
& Reardon et al. \cite{reardon_neutron_2024} \\

5. & Spin-period derivative, $\dot P=-\dot f/f^2$
& $5.72923\times10^{-20}$
& Derived from Reardon et al. \cite{reardon_neutron_2024} \\

6. & Spin-frequency second derivative, $\ddot f$
& $-61(15)\times10^{-30}~\mathrm{s^{-3}}$
& Reardon et al. \cite{reardon_neutron_2024} \\

7. & Right ascension, $\alpha_{\rm J2000}$
& $04^{\rm h}37^{\rm m}15.9284042(4)^{\rm s}$
& Reardon et al. \cite{reardon_neutron_2024} \\

8. & Declination, $\delta_{\rm J2000}$
& $-47^\circ15'09.303700(4)''$
& Reardon et al. \cite{reardon_neutron_2024} \\

9. & Dispersion measure, DM
& $2.64539(3)~\mathrm{pc\,cm^{-3}}$
& Reardon et al. \cite{reardon_neutron_2024} \\

10. & Proper motion in right ascension, $\mu_\alpha\cos\delta$
& $121.4420(6)~\mathrm{mas\,yr^{-1}}$
& Reardon et al. \cite{reardon_neutron_2024} \\

11. & Proper motion in declination, $\mu_\delta$
& $-71.4717(7)~\mathrm{mas\,yr^{-1}}$
& Reardon et al. \cite{reardon_neutron_2024} \\

12. & Total proper motion, $\mu$
& $140.913~\mathrm{mas\,yr^{-1}}$
& Derived from Reardon et al. \cite{reardon_neutron_2024} \\

13. & Timing parallax, $\pi$
& $6.43(5)~\mathrm{mas}$
& Reardon et al. \cite{reardon_neutron_2024} \\

\addlinespace[2pt]
\rowcolor{sectiongray}
\multicolumn{4}{@{}l}{\textbf{Binary-orbit parameters}} \\

14. & Binary model
& T2
& Reardon et al. \cite{reardon_neutron_2024} \\

15. & Orbital period, $P_b$
& $5.74104635(19)~\mathrm{d}$
& Reardon et al. \cite{reardon_neutron_2024} \\

16. & Epoch of periastron, $T_0$
& $\mathrm{MJD}~54530.1722(2)$
& Reardon et al. \cite{reardon_neutron_2024} \\

17. & Projected semimajor axis, $x$
& $3.36671466(3)~\mathrm{lt\mbox{-}s}$
& Reardon et al. \cite{reardon_neutron_2024} \\

18. & Orbital eccentricity, $e$
& $1.91805(10)\times10^{-5}$
& Reardon et al. \cite{reardon_neutron_2024} \\

19. & Longitude of periastron, $\omega$
& $1.359(13)^\circ$
& Reardon et al. \cite{reardon_neutron_2024} \\

20. & Observed orbital-period derivative, $\dot P_b$
& $3.733(3)\times10^{-12}$
& Reardon et al. \cite{reardon_neutron_2024} \\

21. & Orbital inclination, $i$
& $137.506(16)^\circ$
& Reardon et al. \cite{reardon_neutron_2024}; original geometry: van Straten et al. \cite{straten_test_2001} \\

22. & Companion mass, $M_c$
& $0.221(4)~M_\odot$
& Reardon et al. \cite{reardon_neutron_2024} \\

23. & Pulsar mass from timing/Shapiro delay, $M_p$
& $1.418(44)~M_\odot$
& Reardon et al. \cite{reardon_neutron_2024} \\

24. & Distance from Shklovskii/orbital-period derivative, $D_{\rm shk}$
& $156.96(11)~\mathrm{pc}$
& Reardon et al. \cite{reardon_neutron_2024} \\

25. & Distance from timing parallax, $D_\pi$
& $155.4(12)~\mathrm{pc}$
& Reardon et al. \cite{reardon_neutron_2024} \\

26. & Radial velocity, $v_r$
& $-75(18)~\mathrm{km\,s^{-1}}$
& Reardon et al. \cite{reardon_neutron_2024} \\

27. & Transverse velocity, $v_\perp$
& $\approx104.9~\mathrm{km\,s^{-1}}$
& Derived from Reardon et al. \cite{reardon_neutron_2024} \\

\addlinespace[2pt]
\rowcolor{sectiongray}
\multicolumn{4}{@{}l}{\textbf{White-dwarf companion}} \\

28. & Optical detection of companion
& Low-mass helium white dwarf companion
& Bell et al. \cite{bell_optical_1993}; Danziger et al. \cite{danziger_optical_1993-1} \\

29. & White-dwarf effective temperature, $T_{\rm eff,WD}$
& $3950\pm150~\mathrm{K}$
& Durant et al. \cite{durant_spectrum_2012} \\

30. & White-dwarf radius, $R_{\rm WD}$
& $(1.9\pm0.2)\times10^9~\mathrm{cm}$
& Durant et al. \cite{durant_spectrum_2012} \\

31. & White-dwarf mass, $M_{\rm WD}$
& $0.25\pm0.02~M_\odot$
& Durant et al. \cite{durant_spectrum_2012} \\

32. & White-dwarf cooling age, $\tau_{\rm WD}$
& $6.0\pm0.5~\mathrm{Gyr}$
& Durant et al. \cite{durant_spectrum_2012} \\

\addlinespace[2pt]
\rowcolor{sectiongray}
\multicolumn{4}{@{}l}{\textbf{X-ray, UV, and gamma-ray constraints}} \\

33. & Bulk neutron-star surface temperature from UV component
& $T_{\rm NS}\sim(1.25$--$3.5)\times10^5~\mathrm{K}$
& Durant et al. \cite{durant_spectrum_2012} \\

34. & NICER-inferred equatorial radius, $R_{\rm eq}$
& $11.36\pm 0.90~\mathrm{km}$
& Choudhury et al. \cite{choudhury_nicer_2024} \\

35. & NICER/PPTA-informed mass, $M$
& $1.418\pm0.037~M_\odot$
& Choudhury et al. \cite{choudhury_nicer_2024} \\

\bottomrule
\multicolumn{4}{@{}l}{\footnotesize\textit{Note.} Derived quantities are computed from values reported in the cited source.}

\end{tabular}
\end{table}

Using NICER X-ray pulse-profile modeling informed by the precise radio timing priors, Choudhury et al. \cite{choudhury_nicer_2024} inferred a mass of $1.418\pm 0.037~M_{\odot}$ and an equatorial radius of $11.36\pm 0.90~\rm km $. Their analysis also found that the hot emitting regions are inconsistent with a purely centered dipolar magnetic field. These results make PSR J0437-4715 a very compelling star to study: it is close enough, bright enough, and sufficiently well constrained externally that one can investigate the thermal state of an older neutron star, the geometry of surface emission, the structure of the magnetic field, and the equation of state of dense matter. For these reasons, PSR J0437-4715 is the ideal star to investigate the use of neutron stars to place an upper bound on the neutron EDM.
\section{Setting an upper bound on the nEDM using PSR J0437-4715}

The goal of this work is to determine how strongly the observed spin-down of PSR~J0437-4715 can constrain the neutron electric dipole moment (nEDM). Conceptually, the method is straightforward: if the standard channels of energy loss from the pulsar can be modeled well enough, then any remaining positive residual power can be interpreted as the maximum contribution of an additional channel. In this work, that additional channel is taken to be electric dipole radiation produced by a bulk neutron electric dipole moment aligned with the magnetic field axis, and later another CP-odd channel: the neutron magnetic quadrupole moment. The challenge is therefore not simply to measure the spin-down of the pulsar, but to construct a source-specific physical model thorough enough to identify which fraction of the observed energy loss is already accounted for by known mechanisms.

This method requires two major ingredients. The first is a model of the neutron star itself: its mass, radius, density structure, crust thickness, and neutron inventory. These determine where the relevant neutron population resides and how many neutrons are available to participate in a bulk polarized signal. The second is a model of the star's energy loss channels, especially the electromagnetic torque. Since the nEDM constraint is obtained only after subtracting the modeled standard losses from the observed intrinsic spin-down power, the analysis must avoid circularly inferring the magnetic field from spin-down and then using that same inferred field to explain the spin-down. For this reason, a central theme of this work is the construction of a
magnetic-field estimate that is as independent as possible from the timing data alone.

Chapter 2 develops the neutron-star structure model used throughout this work. Using observational constraints on the mass and radius of PSR J0437-4715, we construct an individualized density profile that resolves the core, inner crust, and outer crust. This allows the crust thickness, moment of inertia, and neutron inventory to be estimated in a way that is specific to the source rather than assumed from a canonical neutron-star model. Since our nEDM interpretation ultimately depends on how many neutrons may plausibly contribute to a coherent bulk electric dipole signal, the structure chapter establishes the geometric and microphysical foundation for the later polarization and spin-down analysis.

Chapter~3 uses wideband radio polarimetry to construct a
propagation-informed surface-field posterior that is not obtained by
equating the observed spin-down power to an electromagnetic torque.
Folded, frequency-resolved polarization profiles identify the
orthogonal-polarization-mode transition and define the adopted
$\nu_{\rm OPM}$ interval, while SEARCH-mode single-pulse products are
used to validate the selected OPM-core branch. The measured transition
interval is then combined with emission-height constraints, a
polarization-limiting-radius prior, and explicit pair-plasma and
multiplicity assumptions to infer the surface-field scale $B_{\rm surf}$.
The resulting constraint is model dependent and retains limited timing
dependence through the adopted pair-luminosity prior, but it avoids
reusing the conventional $P\dot P$-derived magnetic field in the
subsequent torque subtraction.

Chapter 4 then examines the large-scale surface-field geometry of PSR J0437-4715. Motivated by the hotspot structure inferred from X-ray observations, we compare several candidate field families, including a centered dipole, an offset dipole, and higher-order multipolar extensions. This comparison is used to determine which magnetic geometry best supports the observed hotspot asymmetry and therefore which field morphology is most physically plausible for the source. The outcome of this chapter is important for two reasons: it constrains the surface magnetic configuration relevant for the torque calculation, and it clarifies how the local crustal field may differ from the idealized large-scale dipole moment often assumed in simple pulsar models.

With those components in place, Chapter 5 assembles the spin-down interpretation. First, the standard torque channels are modeled, with force-free magnetic dipole radiation taken as the dominant present-day contribution and additional channels,
such as gravitational-wave losses, included as subdominant possibilities. Next, the polarized neutron inventory is refined by moving from a purely global neutron-counting
picture to an inner-crust free-neutron reservoir weighted by the local polarization response. This yields an effective participating neutron count, $N_{\rm{pol}}$, appropriate for mapping a bulk electric dipole moment to an individual-neutron nEDM. The modeled standard losses are compared against the observed intrinsic spin-down power. Any remaining positive residual is treated as the maximum power available to electric dipole radiation, which then gives an upper bound on $d_n$.

Chapter 6 tests the unscreened assumption underlying the Chapter 5 result. After examining crustal and magnetospheric screening, it develops an effective magnetic-quadrupole channel as a screening-aware alternative that preserves the source-specific spin-down energy-budget framework.

The logic of the final bound can therefore be summarized conceptually as follows. The observed spin-down sets the total available power budget. The structure model
and polarization model determine how many neutrons can effectively participate in a bulk-aligned electric dipole signal. The radio and hotspot analyses constrain the
magnetic field and geometry needed to model the dominant standard torque channels. Once those channels are subtracted, the surviving residual power defines the maximum
allowed CP-odd radiation. In this way, the final nEDM constraint emerges not from any single observable, but from a chain of inference connecting neutron-star
structure, radio propagation, magnetic geometry, torque evolution, and polarized neutron microphysics within one source-specific framework.

This work, therefore, does not treat PSR~J0437-4715 merely as a pulsar with measured timing parameters. Instead, it uses the star as a fully modeled physical system whose structure and magnetosphere can be constrained well enough to translate spin-down into an astrophysical upper bound on the neutron electric dipole moment.

%% file: chap2.tex
\chapter{Neutron star structure}

\begin{figure*}
\centering
\includegraphics[width=12cm,clip]{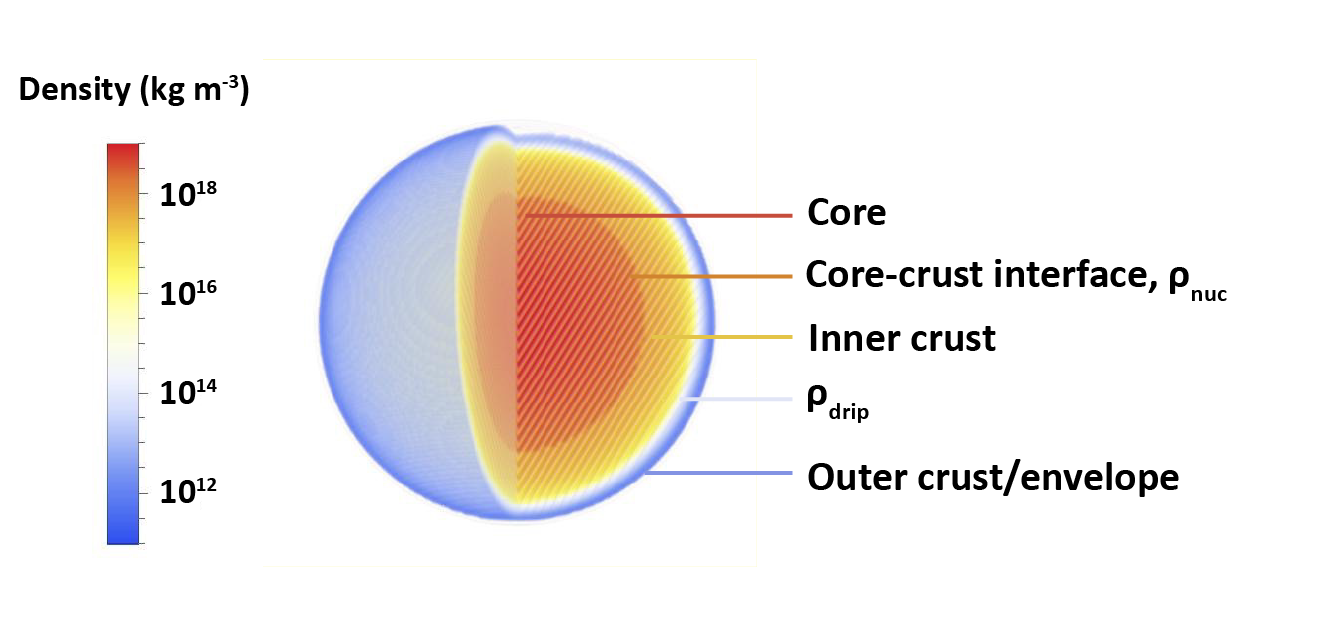}
\caption{Model of PSR J0437-4715 using the density model described below. This model uses a global charge neutrality condition, which can be seen with the mass discontinuity at the core-crust interface. Note the nuclear density $\rho_{nuc}$ marks the end of the core and the drip density $\rho_{drip}$ marks the end of the inner crust.}
\label{fig:ch2_density_model}
\end{figure*}

The neutron star equation of state is typically solved under a local charge-neutrality condition, in which $n_p(r)=n_e(r)$ at every point. Recent models have relaxed this condition to global neutrality, where the only requirement is $\int \rho_{ch}~ dV =0$ \cite{belvedere_neutron_2012}. This allows a thin, ultra-relativistic electron layer at the core-crust interface with a slightly proton-rich core. The electric field at the core-crust interface stores an energy density that consequently creates a jump in the mass-energy density. The discontinuity in the mass density alters mass-radius relations, as well as core radius estimates. Figure~\ref{fig:ch2_density_model} illustrates the density discontinuity at the core–crust interface within the model used below.

The outer crust of neutron stars is the outermost, solid layer of the star with densities less than the drip density. The atoms are stripped of electrons due to high pressure, and a crystalline lattice of neutron-rich nuclei is left. Free electrons then move through the lattice for pressure support. Approaching the inner crust, electron capture drives nuclei to be more neutron-rich \cite{chamel_physics_2008, lorenz_neutron_1993}.

The inner crust of the neutron star lies between the neutron drip density, where free neutrons begin to leak out of nuclei, and the nuclear saturation density, where nuclei dissolve into uniform nuclear matter \cite{negele_neutron_1973}. The inner crust contains neutron-rich nuclei, in which heavy nuclei are packed into a Coulomb lattice, and degenerate electrons provide charge neutrality and contribute to pressure. Free neutrons drip out of the nuclei and form a superfluid in the interstitial space of the lattice \cite{watanabe_thermodynamic_2000}. Near the core-crust interface, there is the nuclear pasta state of matter where nuclei rearrange into exotic shapes due to the balance between nuclear attraction and Coulomb repulsion. The nuclear pasta phase behaves differently from both solid and liquid matter. This modifies the crust's conductivity and magnetic diffusivity. The inner crust ultimately acts as a resistive, anisotropic layer that strongly shapes the magnetic field's long-term evolution \cite{nandi_transport_2018, vigano_unifying_2013}.

The core of the neutron star consists of densities greater than the nuclear saturation density, and its contents, specifically of the inner core, are relatively unknown. The core does contain superconducting protons \cite{douchin_unified_2001}. This raises questions about the magnetic properties of neutron star cores. Many neutron star models, specifically early models, claim that the superfluid neutrons and superconducting protons cannot support bulk polarization \cite{vigano_unifying_2013}. Thus, the core does not significantly contribute to the magnetic dipole of a neutron star \cite{cumming_magnetic_2004, wood_three_2015, gourgouliatos_magnetic_2016, lander_magnetic-field_2019}. 

However, observations of large magnetic field neutron stars such as magnetars suggest that the core may contribute to the field through fluxoids that are pinned to the proton superconductor. The investigation of neutron star magnetic field origins and core contributions is ongoing \cite{ho_dynamical_2017, gourgouliatos_magnetic_2022, pons_pulsar_2012, gusakov_magnetic_2020}. However, the bulk of this research looks into core contributions for high-magnetic field stars. This work returns to earlier models where the magnetic field is crust-confined, specifically to determine if ordinary and millisecond pulsars are capable of containing the entirety of their magnetic field in their crusts, suggesting whether or not one can treat these stars as if the main contribution of their magnetic field comes from the crust.

\section{Mass-radius relation}
The Equation of State (EoS) models the relationship between pressure and gravity in a neutron star. Currently, there is no complete EoS that describes all behaviors of the neutron star, but the Tolman–Oppenheimer–Volkoff (TOV) \cite{tolman_static_1939, oppenheimer_massive_1939} equations are used to provide the relativistic generalization of hydrostatic equilibrium \cite{lattimer_neutron_2001, nambo_static_2021}.
The TOV equations are solved for different central densities and yield different neutron star models, with unique masses and radii. The mass-radius (M-R) relation is very sensitive to the EoS chosen.

Measurements of the mass and radius of neutron stars provide insight into which EoS matches observation, but are limited. Mass measurements are typically determined through pulsar timing of binary systems. Observing effects such as Shapiro delay, periastron advance, and orbital decay from gravitational wave emission allows rather precise mass measurements \cite{demorest_two-solar-mass_2010}. However, pulsar timing often does not provide much information on the radius of a star.

\begin{figure*}
\centering
\includegraphics[width=\textwidth,clip]{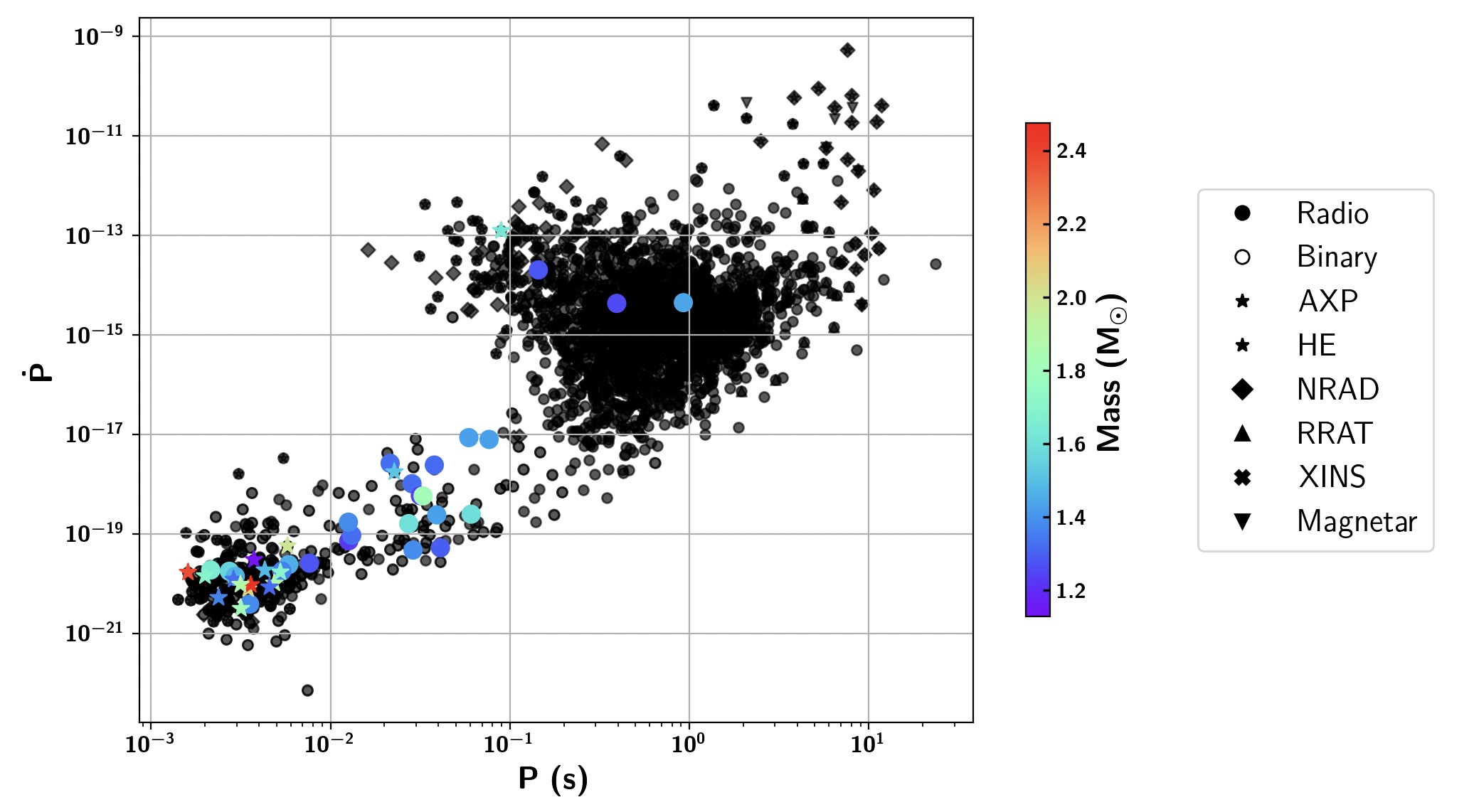}
\caption{Period and period derivative diagram for all stars in the ATNF catalog \cite{manchester_australia_2005} with their respective categories, noting that the majority of neutron stars with documented mass values, denoted by markers with color correlating to the color scale on the right, are considered millisecond pulsars. }
\label{fig-2} 
\end{figure*}

The radius is much more difficult to estimate and relies on X-ray observations. Thermal X-ray spectra of isolated neutron stars or X-ray binaries are fit with atmosphere models to estimate the apparent radius. Newer techniques, such as pulse-profile modeling from the NICER mission, observe the relativistic light bending and how it affects the shape of pulsations in hot spots on the surface of neutron stars. The shape of these hot spots depends on the compactness M/R, and sometimes this can be used to measure both the mass and radius \cite{miller_psr_2019, miller_radius_2021}. This limits investigations to ordinary pulsars, usually millisecond pulsars, which can be seen in the lower left corner of Figure \ref{fig-2}, as they have very stable, short periods that one can precisely measure. 

\section{Neutron star density models}
\label{sec-1}

Many other density models use local-neutrality, single-branch polytropes that smear the core–crust interface and miss (i) the interface energy jump from global neutrality \cite{belvedere_neutron_2012}, (ii) stiffening across nuclear pasta that modifies inner-crust curvature \cite{negele_neutron_1973, chamel_physics_2008}, and (iii) the observational values of $M$ and $R$ to provide accurate limiting parameters \cite{miller_psr_2019, miller_radius_2021}. For testing if the large-scale dipole can be crust-confined in ordinary/MSP pulsars, a piecewise profile with explicit drip and saturation landmarks and a well-defined $R_{\rm core}$ is preferred.

\subsection{Density model}
As seen in Figure~\ref{fig-2}, millisecond pulsars are the majority of stars with documented masses, and thus are used as the basis of this model. The central density and the average crust density were used to parameterize the model. For the central density, 8 different EoS models (BPAL12 \cite{ferrari_neutron_2010}, BGN1H1 \cite{potekhin_physics_2010}, FPS \cite{lorenz_neutron_1993}, BBB2 \cite{datta_equilibrium_1998}, SLy \cite{douchin_unified_2001}, BGN1 \cite{potekhin_physics_2010}, APR \cite{akmal_equation_1998}, BGN2 \cite{potekhin_physics_2010}) central density-mass relations were compared \cite{haensel_neutron_2007}. The SLy (Skyrme Lyon) model contains a central density of approximately $1\times 10^{18}~\mathrm{kg~m^{-3}}$ and is the average central density of the 8 models. Further, this model is often used for M-R relations and supports stars with masses above $2~M\odot$, where many other models do not support a maximum mass that high. The core was modeled using the gravitational binding energy and the central density. 

As mentioned in the introduction, the floor of the core is set at the nuclear saturation $\rho_{nuc} \simeq 2.8\times 10^{17}~\mathrm{kg~m^{-3}}$, where the crystalline crust melts, causing nuclei to overlap and nuclear-pasta structures to disappear. At this point, the sparse protons will form a superconductor, marking the beginning of the outer core \cite{nandi_transport_2018, watanabe_thermodynamic_2000, yakovlev_electron_2015}. Due to the unknown physics of the inner core of neutron stars, our model does not reliably constrain small radius values; however, the mass between 0 and 2 km only accounts for approximately 5$\%$ of the star's entire mass. This makes a minimal contribution to our estimations of the core radii, and as mentioned before, it should not largely affect the magnetic field. 
\begin{equation}
\rho(r)=
\begin{cases}
4\pi^2\!\left(\rho_{0,~\rm core}\!\left(1-\dfrac{G M r^2}{c^2R^4}\right)^{\!a}\right) r^4, & 0<r<R_{\rm core}\\[0.8em]
4\pi \left(b+c\ln{\rm r}\right) r^4, & R_{\rm core}<r<R_{\rm inner~crust}\\[0.8em]
4\pi r^4\rho_{\rm 0,~outer~crust} e^{dr}, & R_{\rm inner~crust}<r<R
\end{cases}
\label{eq:rho}
\end{equation}
The inner crust’s log–log curvature is consistent with the fact that EoS continuously stiffens across nuclear pasta phases \cite{haensel_neutron_2007, negele_neutron_1973}. The average crust density was used because there may be a mass discontinuity at the core-crust interface, and this discontinuity may vary by star to preserve the global neutrality condition. In the case that the inner-core density ceiling was $\rho_{nuc}$, this would fulfill the local neutrality condition. The crust floor is set at the neutron-drip density $\rho_{drip}=4\times 10^{14}~\mathrm{kg~m^{-3}}$ because at this point the nuclei become neutron-rich until free neutrons appear. At densities lower than the drip density, this becomes the star's envelope, whose mass exponentially decays.

\begin{figure*}[th!]
\centering
\includegraphics[width=\textwidth,clip]{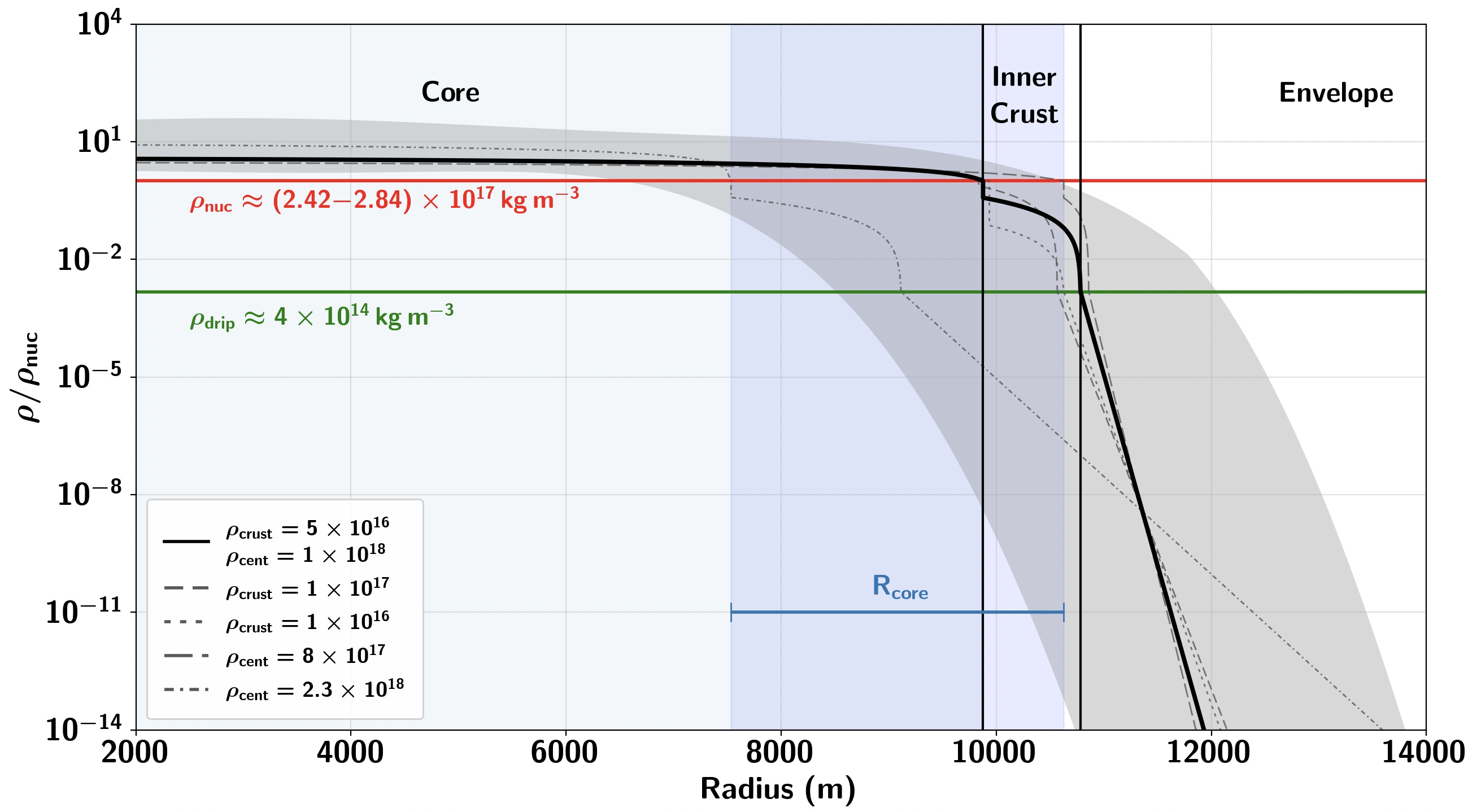}
\caption{Density model of J0437-4715 with documented mass $1.418\pm 0.037~M_{\odot}$ and radius $11.36\pm0.90$ km. The red line shows nuclear saturation, $\rho_{nuc}$ (core floor) \cite{douchin_unified_2001}, and the green line shows neutron drip, $\rho_{drip}$ (crust floor) \cite{negele_neutron_1973}. The dotted lines show the density model with the extremes of the central and average density values from the 8 different EOS models \cite{haensel_neutron_2007}. The grey shaded region includes the propagated uncertainties from $(M,R)$ and density choices. The blue strip marks $R_{\rm core}$ uncertainties within the $2\sigma$ range.}
\label{fig-3}
\end{figure*}

\noindent\textit{Justification and precedent.}
Equation~\eqref{eq:rho} and Figure \ref{fig-3} show a piecewise, reduced-form model that (i) preserves the interface jump permitted by global neutrality \cite{belvedere_neutron_2012}, (ii) encodes the well-known inner–crust curvature across nuclear pasta \cite{negele_neutron_1973, watanabe_thermodynamic_2000, nandi_transport_2018, vigano_unifying_2013}, and (iii) treats the thin outer layers with an exponential scale characteristic of standard envelope/crust models \cite{baym_ground_1971, gudmundsson_structure_1983}. Using reduced-form models to match $(M,R)$ and interior trends is standard practice (e.g., piecewise-polytrope fits) \cite{read_constraints_2009, lattimer_neutron_2001}.

The core and crust branches in Eq.~\eqref{eq:rho} are joined at $R_{\rm core}$ with
(i) continuity of the enclosed mass $m(r)$ and (ii) allowance for a small, upward
density jump on the crust side motivated by global neutrality \cite{belvedere_neutron_2012}.
We enforce the integral mass constraint $\int_0^R 4\pi r^2 \rho(r)\,dr = M$ and
anchor the landmarks at nuclear saturation and neutron-drip.
The free shape parameters $(a,b,c,d)$ and normalizations (e.g., $\rho_{0,\mathrm{core}}$) are
determined by minimizing a residual that combines (i) the mass constraint,
(ii) the core and inner crust density floor constraints, and
(iii) the outer-envelope cutoff near $\rho\!\sim\!10^9~\mathrm{kg\,m^{-3}}$ \cite{haensel_neutron_2007}.
Dimensional consistency is ensured by absorbing scale factors into the fit coefficients; the explicit $r^4$ factors are a bookkeeping choice that anticipates the spherical shell Jacobian in later volume integrals (e.g., $I$), and do not alter the physical
content of the density profile.

Standard parametric models either use an analytic Tolman-type density profile for the core, joined smoothly to polytropic descriptions of the crust, or they approximate a full tabulated equation of state using several polytropic segments fitted over different density ranges \cite{read_constraints_2009,lattimer_neutron_2001}.
Equation~\eqref{eq:rho} differs in three ways that are helpful for the present goal:
(1) it explicitly permits the small density jump at the core--crust boundary expected under
global neutrality \cite{belvedere_neutron_2012}; (2) it encodes the observed logarithmic
curvature across the inner crust with a weak $(b+c\ln r)$ dependence, rather than a single
power law \cite{negele_neutron_1973,watanabe_thermodynamic_2000,nandi_transport_2018,vigano_unifying_2013};
and (3) it treats the thin outer layers with an exponential scale height in line with
outer-crust/envelope physics \cite{baym_ground_1971,gudmundsson_structure_1983}. These choices keep the model compact,
transparent, and directly anchored to $(M,R)$ while retaining the key microphysical landmarks.
\begin{figure*}[th!]
\centering
\includegraphics[width=\textwidth,clip]{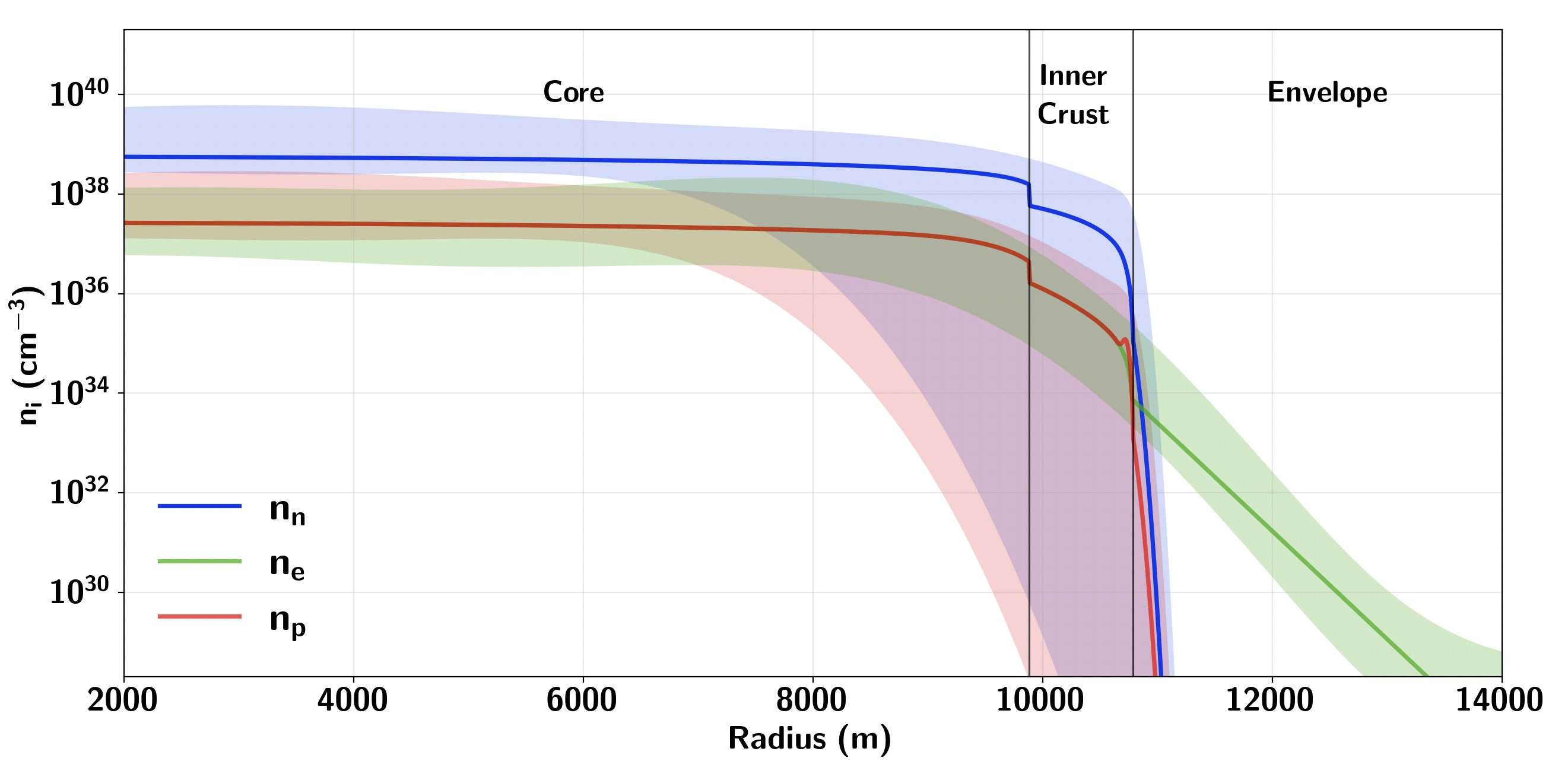}
\caption{Diagram depicting the neutron, proton, and electron density breakdown of the PSR J0437-4715 density model consistent with other global neutrality density models \cite{belvedere_neutron_2012}. }
\label{fig-4}
\end{figure*}
\begin{table}
\small
\centering
\caption{Adopted mass, radius, and inferred core-radius values for the neutron stars considered in this chapter.}
\label{tab-1}
\begin{tabular}{llll}
\hline
NS& Mass ($M_{\odot}$) & R (km) & $R_{core}$ (km)\\\hline
J0437-4715 \cite{choudhury_nicer_2024} & $1.418\pm0.037$ & $11.36 \pm0.90$ & $9.65 \pm0.22$ \\
J0030+0451 \cite{miller_psr_2019}& $1.44 \pm 0.15$& $13.02 \pm 0.125$&$ 11.03 \pm0.04$\\
J0348+0432 \cite{huo_moment_2018}&$2.01\pm0.04$&$12.55 \pm 0.40$&$9.56 \pm 0.05$\\
J0740+6620 \cite{miller_radius_2021}&$2.08 \pm 0.07$&$13.70 \pm 0.26$&$11.04 \pm 0.11$\\\hline
\end{tabular}
\end{table}

\subsection{Interpretation of the model}
With this density model, we can find the radius of the core and thus the thickness of the crust given a measured mass and radius, as seen in Figure~\ref{fig-3} and Table~\ref{tab-1}. The crust thickness is $\Delta R = R_{\rm star} - R_{\rm core}$. More specifically, the radius is defined by when the outer crust density becomes $10^9~\rm kg~m^{-3}$ \cite{haensel_neutron_2007}. The crust thickness, coupled with the density model, allows one to know how many neutrons are within the crust. 

\begin{figure*}[t!]
\centering
\includegraphics[width=\textwidth,clip]{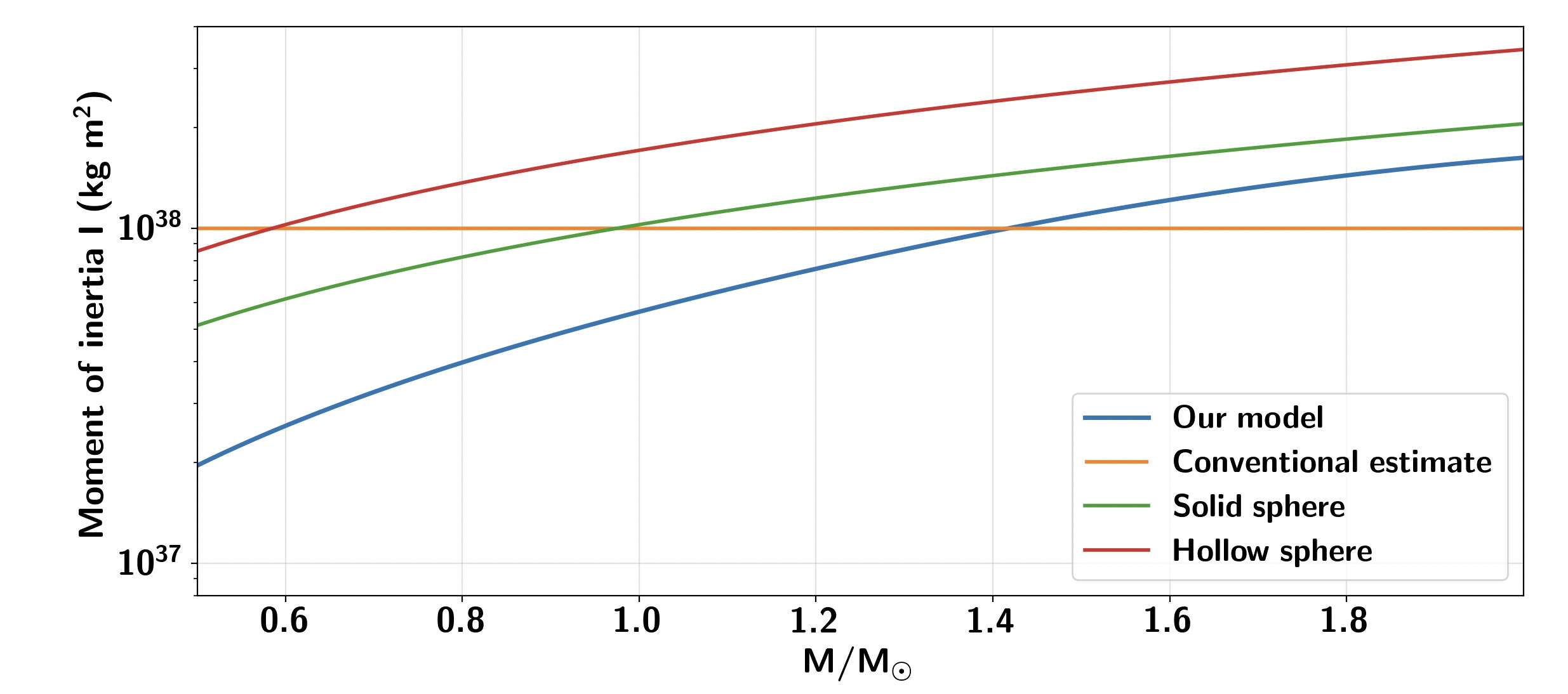}
\caption{Moment of inertia of an 11.36 km star for masses varying from $ 0.5- 2~M_{\odot}$. This interval spans the range relevant for typical observed neutron-star masses and includes the $\sim 2M_\odot$ mass scale of the heaviest well-measured pulsars. For reference, the canonical moment of inertia ($1\times 10^{38}~\rm kg~m^{2}$) is shown in orange, as well as the moment of inertia for a solid and hollow sphere in green and red, respectively.}
\label{fig-5}
\end{figure*}

Further, improvements can be made in the moment of inertia given $I=\int~\rho r^2 dV$. Canonically, a moment of inertia of $1\times 10^{38}~ \rm kg~m^2$ is used \cite{lattimer_physics_2004, potekhin_neutron_2015, haensel_neutron_2007}. However, as seen in Figure~\ref{fig-5}, this estimation is not accurate for stars with extreme masses. Given a more precise moment of inertia, the magnetic field values may also be improved, as well as our investigation into crustal polarization.  

\begin{equation}
I=\int_0^R 4\pi r^4 \rho(r)\,dr
\label{eq:I}
\end{equation}

Equation~\eqref{eq:I} is the standard spherical-shell expression $I=\int r^2\,dM$ with $dM=4\pi r^2\rho\,dr$. Using the source-specific $I$ derived from Eq.~\eqref{eq:rho} rather than a canonical value follows standard guidance on the $I(M,R)$ variability expected across EoS \cite{lattimer_neutron_2001, potekhin_magnetic_2018}.

\section{Crust and magnetic field origins}
With the core radius values and thus the crust thickness, one can find the percentage of mass that originates from the crust as well as the number of neutrons in the crust, as shown in Fig. \ref{fig-4}. Further, given the magnetic dipole of the entire star and the total neutrons in the crust, one can find the percentage of neutrons in the star that are aligned with the star's magnetic dipole.

\begin{equation}
I\Omega\dot{\Omega} \;=\; -\,\frac{B_p^{\,2} R^6 \Omega^{4}}{6c^3}\,\sin^2\alpha,
\qquad
B_p \;=\; \sqrt{\frac{6c^3 I\,|\dot{\Omega}|}{R^6 \Omega^{3}\,\sin^2\alpha}}
\label{eq:Bp}
\end{equation}

Equation~\eqref{eq:Bp} is the textbook vacuum rotating-dipole relation connecting spin-down power to $B_p$ and geometry (obliquity $\alpha$). The other magnetic field model is the force-free magnetospheric model, which alters a numerical prefactor and the $\alpha$-dependence. However, a change in model would only cause shifts, not a relative difference between stars \cite{camilo_observations_2000, spitkovsky_time-dependent_2006}.

\begin{equation}
\mu_{\rm NS} \;=\; \frac{B_p R^3}{2}
\label{eq:mu}
\end{equation}

\noindent\textit{Context.}
Equation~\eqref{eq:mu} gives the dipole moment associated with a polar field $B_p$ and radius $R$ for a dipole configuration (standard in pulsar phenomenology; see also \cite{camilo_observations_2000}).

\begin{equation}
N_{n,\mathrm{crust}} \;=\; \int_{R_{\rm core}}^{R}\!\frac{\,\rho(r)}{m_n}\,4\pi r^2\,dr
\label{eq:Nn}
\end{equation}

\noindent\textit{Justification.}
Equation~\eqref{eq:Nn} uses the same $\rho(r)$ as Eq.~\eqref{eq:rho} and the mass of the neutron to calculate the total number of neutrons in the crust.
\begin{equation}
\mathcal{P}_{\rm req} \;=\; \frac{\mu_{\rm NS}}{N_{n,\mathrm{crust}}\,|\mu_n|}
\label{eq:Pol}
\end{equation}

\noindent\textit{Interpretation.}
Equation~\eqref{eq:Pol} is a bookkeeping ratio: the fraction of crustal neutrons that would need to align with $|\mu_n|$ to supply the global dipole $\mu_{\rm NS}$ inferred from timing. Values $\ll 1$ indicate a feasible crust-confined dipole budget under the adopted assumptions.

\noindent\textit{Magnetosphere systematics.}
Equation~\eqref{eq:Bp} uses the vacuum rotating-dipole normalization. Force-free
magnetospheres modify the numerical prefactor and the $\alpha$-dependence at the tens-of-percent
level \cite{spitkovsky_time-dependent_2006}. Here the vacuum form is retained for a
conservative, consistent comparison across sources; adopting a force-free calibration
would shift $B_p$ coherently without changing the qualitative polarization budget conclusions.

The stars' ages were found by comparing pulsars' right ascension, declination, and distance with the ages of nearby star clusters \cite{kharchenko_astrophysical_2005}. The thresholds were set to be within 10 degrees for the right ascension and declination and 20 pc for the distance. Only pulsars with documented masses were used in this process, and the radius was found using the mass-radius relation from Belvedere et al. 2012 \cite{belvedere_neutron_2012}.

Globular clusters themselves form from a singular massive molecular cloud in a rather short time period. A neutron star within a cluster is a result of the evolution and inevitable core collapse supernova of a massive star within the cluster \cite{kuranov_neutron_2006, ivanova_formation_2008}. This does indicate that the age of the neutron stars within a cluster is slightly younger; however, the lifetime of the progenitor is short due to its extreme mass. There may be exceptions to this. In particular, star clusters may contain multiple star populations, or a neutron star may be gravitationally captured by an unrelated star cluster.

Although the age of stars in binary systems may usually be reliably computed, when present in star clusters, they may experience a nonuniform mass transfer that may affect these calculations, as was the case for PSR J0514 4002A \cite{freire_timing_2007}. Further, many MSPs are considered to be recycled pulsars, in which their companion is a white-dwarf that is only born after the MSP accretes mass from a main-sequence companion, see Fig.~\ref{recyled}.

\begin{figure*}[th!]
\centering
\includegraphics[width=\textwidth]{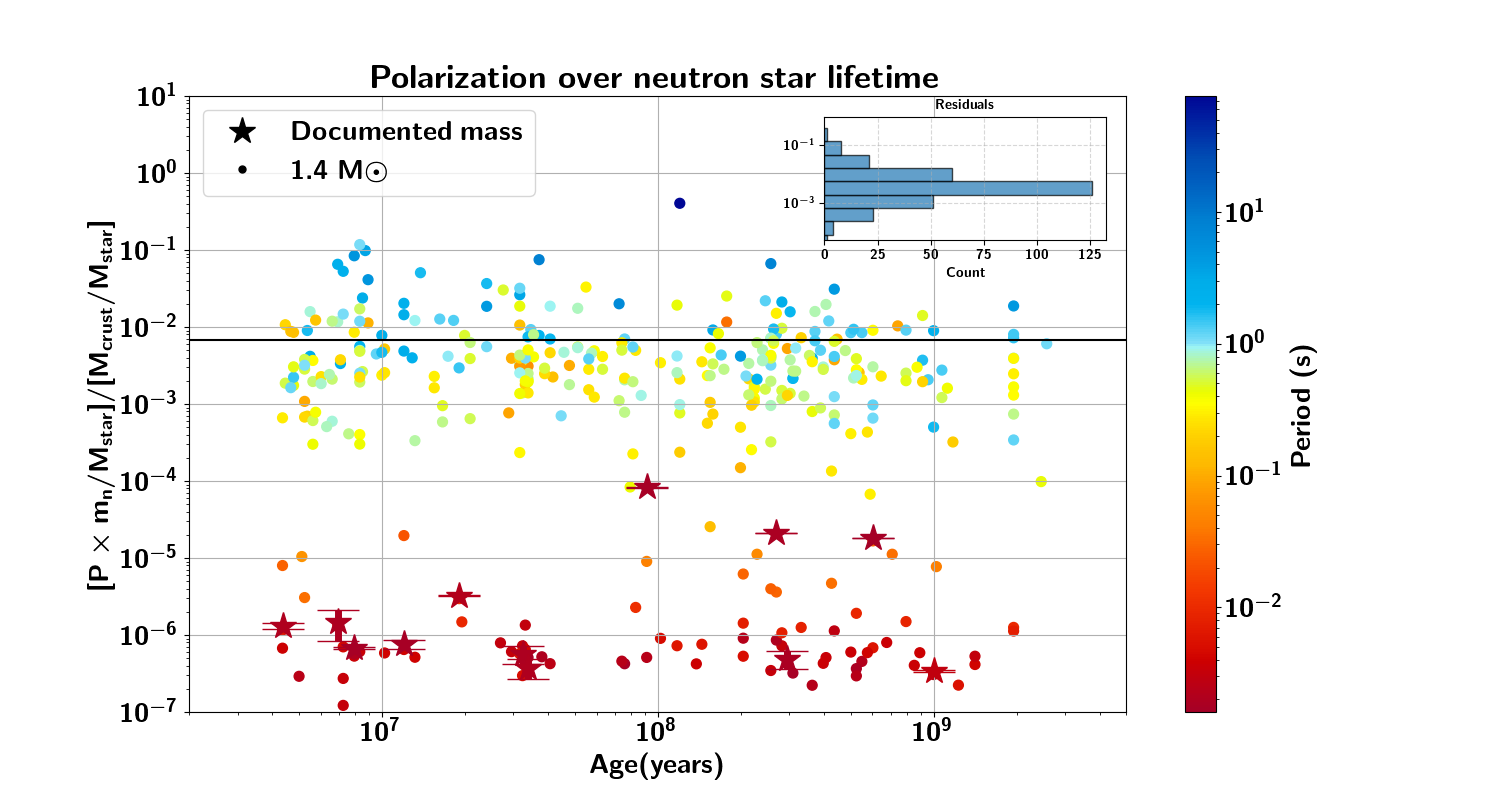}
\caption{Polarization for stars with documented mass, as well as all stars from the ATNF catalog \cite{manchester_australia_2005} with an assumed canonical mass. The age was derived by cross-checking the pulsars' right ascension, declination, and distance. The color scale on the right denotes the relationship between period and polarization.}
\label{fig-pol}
\end{figure*}

As seen in Figure~\ref{fig-pol}, for stars with documented mass values (denoted by star markers), the ratio between the spin polarization mass and crustal mass remains significantly under one, showing it is feasible for these stars to contain their entire magnetic field within their crusts. However, we are limited to a small sample because, as mentioned earlier, most stars with documented mass are millisecond pulsars, and the polarization is proportional to the period of a star.

To look at the behaviors of higher period stars that would approach the ratio limit, we took all stars \cite{manchester_australia_2005} and similarly cross-checked their age and found the polarization. A canonical $1.4~M_{\odot}$ was used for all unknown masses. This provided some stars with larger periods, with the majority of them having a period of around 1 second. Nonetheless, for all of these stars, the ratio of the mass-weighted polarization and crustal mass remained under one, suggesting that all of these stars are capable of containing the entirety of the star's magnetic field in its crust.

There is no apparent relationship between polarization and age, nor between period and age, but this may be due to limitations in age determination. In order to observe if these neutron stars are capable of containing their magnetic field for the entirety of their life, we must model the period over a neutron star's life. To do this in future investigations, one can look at the angle sweep of the position of the linearly polarized radio pulse \cite{guillemot_improving_2023}. This provides the magnetic inclination and allows us to find the individualized spin-down law for each star.

As for the apparent gap between polarization values between $10^{-5}$ and $10^{-3}$, this can be explained by limitations in measurement rather than a consequence of polarization behavior \cite{miller_psr_2019}. This is best visualized with the residuals diagram coupled with Figure~\ref{fig-pol}. The gap in points correlates to periods around 0.01s to 0.1s. Neutron stars in this period range are not often measured because frequencies associated with that period are very common man-made frequencies. During measurement, these frequencies are often filtered out so as not to mistake a frequency on Earth for a neutron star. The Parkes Multibeam Survey, which heavily contributed to ATNF, demonstrated that the optimal sensitivity results from stars ranging from 0.1 s $\leq$ P $\leq$ 2 s \cite{pons_pulsar_2012, manchester_parkes_2001}.

However, one can note a significant cluster in the millisecond range. This can be attributed to multiple measurement advantages presented by millisecond pulsars (MSPs) as well as their relatively long lifetime compared to other neutron stars \cite{bhattacharya_formation_1991}. MSPs are very common in globular clusters and more likely to be observed \cite{camilo_observations_2000}. Also, more recent surveys, unlike the Parkes survey, focused on optimizing MSP measurements. Further, MSPs produce bright gamma rays, and Fermi LAT has been a large contributor to identifying MSPs to unknown gamma ray sources \cite{abdo_detection_2009}. MSPs are approximately 20$\%$ of documented neutron stars; however, they are predicted to be less than 5 $\%$ of all neutron stars.

\paragraph{Uncertainties and limitations.}
The profile in Eq.~\eqref{eq:rho} is a surrogate anchored to $(M,R)$ and microphysical landmarks; it is not a full TOV solution for a single tabulated EoS. As such, it is designed to capture the leading-order geometry relevant for $I$ and the crustal inventory. The error budget reported in Figure~\ref{fig-3} propagates uncertainties from $(M,R)$ and from the density-shape parameters. The global-neutrality jump is treated as a small allowed discontinuity at $R_{\rm core}$ rather than solved from a coupled Poisson–TOV system; this keeps the model
agnostic to specific inner-core compositions while retaining the key interface physics. Finally, $\alpha$ is often not known for many stars. The spin-down inferences use the inclination $\alpha$-dependence explicitly; where $\alpha$ is unknown, a broad prior is assumed, which widens the $B_p$ band but does not alter the finding that $\mathcal{P}_{\rm req}\ll 1$ for the systems considered.

\subsection{Nuclear EDMs from accreted outer-crust nuclei}
The nEDM estimate method developed throughout this work focuses on polarized neutrons as the relevant microscopic dipole. A possible additional contribution could arise from nuclei in the envelope or outer crust. This is worth considering because a recycled millisecond pulsar such as PSR J0437-4715 may not have a pristine cold-catalyzed crust. In the standard recycling picture, the neutron star is spun up through accretion from a binary companion, so at least the outer layers may be better described by accreted or partially accreted matter. The composition of these layers is therefore model dependent, and there is no single unique set of nuclei that can be assigned to the envelope and outer crust.

Table \ref{tab:outer_crust_envelope_nuclei_density_ratio} summarizes several representative composition models. The cold catalyzed sequence gives the standard non-accreted outer-crust baseline, containing neutron-rich nuclei such as Fe, Ni, Se, Ge, Zn, and Ni near neutron drip \cite{baym_ground_1971}. By contrast, the Haensel-Zdunik accreted-crust sequence begins with iron-group ashes and follows the compression-driven electron-capture chain through the outer crust. In the simple A = 56 version, the dominant nuclei evolve approximately as
\[{}^{56}{\rm Fe} \rightarrow{}^{56}{\rm Cr} \rightarrow{}^{56}{\rm Ti} \rightarrow{}^{56}{\rm Ca} \rightarrow{}^{56}{\rm Ar},
\]
before neutron drip \cite{haensel_models_2008}. Heavier rp-process ashes motivate a separate class of accreted-crust models, represented in the table by the A = 106 chain \cite{haensel_non-equilibrium_1990}. Finally, light-element envelope models allow H, He, C, and O to exist in the shallow heat-blanketing layer above the deeper crust \cite{potekhin_internal_1997}.

The density column in Table \ref{tab:outer_crust_envelope_nuclei_density_ratio} is written as $\rho_{\rm max}/\rho_{\rm nuc}$. For the one-component crust sequences, $\rho_{\rm max}$ denotes the maximum density at which a given nuclide dominates before the next electron-capture transition. These densities should therefore be interpreted as model-dependent transition densities, not as directly measured properties of PSR J0437-4715. The hydrogen entry is different: hydrogen belongs to the shallow envelope rather than the bulk crust, so its maximum density depends on the accreted-envelope mass and on nuclear burning. A representative range $\sim 10^{-10}-10^{-8}$ is included to emphasize that, although hydrogen has nonzero spin, any surviving hydrogen is confined to a very shallow layer compared with the bulk outer crust.

For EDM purposes, the important question is not simply whether heavy nuclei are present, but whether the relevant isotopes have nonzero nuclear spin and whether those spins can be polarized. Many of the dominant nuclei in the standard envelope and outer-crust models are even-even. Their protons and neutrons are paired, so their ground states are typically $J^\pi = 0^+$. Such nuclei do not provide a permanent spin-aligned nuclear EDM reservoir. This includes many of the most important species in the table, such as $^4$He, $^{12}$C, $^{16}$O, $^{56}$Fe, $^{56}$Ni, the dominant A=56 accreted-crust nuclei, the representative even-even A=106 rp-process ash chain, and $^{208}$Pb.
\begin{table}[p]
    \centering
    \caption{
    Representative nuclei that may appear in neutron-star envelope and outer-crust
    composition models, with approximate maximum density scales written as
    \(\rho_{\max}/\rho_{\rm nuc}\), where
    \(\rho_{\rm nuc}=2.8\times10^{14}\,{\rm g\,cm^{-3}}\). For one-component
    crust sequences, \(\rho_{\max}\) is the maximum density at which the listed
    nuclide dominates before the next transition. For hydrogen, the quoted range
    is a model-dependent envelope depth scale rather than a crustal transition
    density. Pb and odd-\(A\) Ni/Pb isotopes are included as benchmark heavy or
    impurity species relevant to possible nuclear EDM contributions.
    }
    \label{tab:outer_crust_envelope_nuclei_density_ratio}
    \footnotesize
    \renewcommand{\arraystretch}{1.15}
    \setlength{\tabcolsep}{6pt}
    \begin{tabular}{>{\raggedright\arraybackslash}p{0.22\textwidth}
                    >{\centering\arraybackslash}p{0.17\textwidth}
                    >{\raggedright\arraybackslash}p{0.27\textwidth}
                    >{\raggedright\arraybackslash}p{0.26\textwidth}}
        \toprule
        \textbf{Class} &
        \textbf{Nuclide} &
        \textbf{\(\rho_{\max}/\rho_{\rm nuc}\)} &
        \textbf{Spin / EDM relevance} \\
        \midrule

        \rowcolor{gray!15}
        \multicolumn{4}{l}{\textbf{Light-element envelope}} \\
        Envelope & \({}^{1}{\rm H}\) &
        \(\sim 3.6\times10^{-10}\)--\(3.6\times10^{-8}\) &
        nonzero spin; shallow envelope only \\
        Envelope & \({}^{4}{\rm He}\) &
        model-dependent &
        spin-zero \\
        Envelope & \({}^{12}{\rm C}\) &
        model-dependent &
        spin-zero \\
        Envelope & \({}^{16}{\rm O}\) &
        model-dependent &
        spin-zero \\
        \addlinespace[0.35em]

        \rowcolor{gray!15}
        \multicolumn{4}{l}{\textbf{Cold-catalyzed / non-accreted outer crust}} \\
        BPS sequence & \({}^{56}{\rm Fe}\) &
        \(2.9\times10^{-8}\) &
        spin-zero \\
        BPS sequence & \({}^{62}{\rm Ni}\) &
        \(9.6\times10^{-7}\) &
        spin-zero \\
        BPS sequence & \({}^{64}{\rm Ni}\) &
        \(4.3\times10^{-6}\) &
        spin-zero \\
        BPS sequence & \({}^{84}{\rm Se}\) &
        \(2.9\times10^{-5}\) &
        spin-zero \\
        BPS sequence & \({}^{82}{\rm Ge}\) &
        \(7.9\times10^{-5}\) &
        spin-zero \\
        BPS sequence & \({}^{80}{\rm Zn}\) &
        \(1.7\times10^{-4}\) &
        spin-zero \\
        BPS sequence & \({}^{78}{\rm Ni}\) &
        \(5.7\times10^{-4}\) &
        spin-zero \\
        \addlinespace[0.35em]

        \rowcolor{gray!15}
        \multicolumn{4}{l}{\textbf{Accreted crust: \(A=56\) Haensel--Zdunik sequence}} \\
        \(A=56\) chain & \({}^{56}{\rm Fe}\) &
        \(5.3\times10^{-6}\) &
        spin-zero \\
        \(A=56\) chain & \({}^{56}{\rm Cr}\) &
        \(4.0\times10^{-5}\) &
        spin-zero \\
        \(A=56\) chain & \({}^{56}{\rm Ti}\) &
        \(2.8\times10^{-4}\) &
        spin-zero \\
        \(A=56\) chain & \({}^{56}{\rm Ca}\) &
        \(8.9\times10^{-4}\) &
        spin-zero \\
        \(A=56\) chain & \({}^{56}{\rm Ar}\) &
        \(2.2\times10^{-3}\) &
        spin-zero \\
        \addlinespace[0.35em]

        \rowcolor{gray!15}
        \multicolumn{4}{l}{\textbf{Accreted crust: heavy \(A=106\) rp-process ash sequence}} \\
        \(A=106\) chain & \({}^{106}{\rm Pd}\) &
        \(1.3\times10^{-6}\) &
        spin-zero \\
        \(A=106\) chain & \({}^{106}{\rm Ru}\) &
        \(2.0\times10^{-5}\) &
        spin-zero \\
        \(A=106\) chain & \({}^{106}{\rm Mo}\) &
        \(8.6\times10^{-5}\) &
        spin-zero \\
        \(A=106\) chain & \({}^{106}{\rm Zr}\) &
        \(2.4\times10^{-4}\) &
        spin-zero \\
        \(A=106\) chain & \({}^{106}{\rm Sr}\) &
        \(5.2\times10^{-4}\) &
        spin-zero \\
        \(A=106\) chain & \({}^{106}{\rm Kr}\) &
        \(9.9\times10^{-4}\) &
        spin-zero \\
        \(A=106\) chain & \({}^{106}{\rm Se}\) &
        \(1.7\times10^{-3}\) &
        spin-zero \\
        \(A=106\) chain & \({}^{106}{\rm Ge}\) &
        \(2.8\times10^{-3}\) &
        spin-zero; near neutron drip \\
        \addlinespace[0.35em]

        \rowcolor{gray!15}
        \multicolumn{4}{l}{\textbf{Benchmark heavy nuclei / impurity cases}} \\
        Impurity case & \({}^{61}{\rm Ni}\) &
        not a dominant sequence species &
        nonzero spin \\
        
        Benchmark & \({}^{208}{\rm Pb}\) &
        not a dominant sequence species &
        spin-zero \\
        
        Impurity case & \({}^{207}{\rm Pb}\) &
        not a dominant sequence species &
        nonzero spin \\
        \bottomrule
    \end{tabular}
\end{table}
The nonzero-spin entries in the table should therefore be treated as possible corrections rather than as leading contributions. Hydrogen has nonzero spin, but it is limited to the shallow heat-blanketing envelope and represents only a small fraction of the stellar mass. Odd-A nuclei such as $^{61}$Ni or $^{207}$Pb can also have nonzero spin but they are not dominant species in the simple cold-catalyzed, A=56 accreted, or A=106 one-component sequences. They would have to enter through impurities, multi-component ashes, or a more detailed nuclear-reaction network. Their EDM contribution would therefore scale with their impurity abundance rather than with the total number of nuclei in the crust.

The conclusion is that envelope and outer-crust nuclei do not significantly modify the leading neutron EDM estimate under standard composition assumptions. The dominant nuclei in the commonly used outer-crust and accreted-crust sequences are largely spin-zero. Nonzero-spin species are either confined to a shallow envelope layer or enter as impurities whose effect is suppressed by both abundance and spin polarization. Nuclear EDMs from crustal nuclei are therefore best regarded as a subleading systematic possibility, unless a separate composition model demonstrates a large population of nonzero-spin nuclei together with an efficient mechanism for aligning their spins.

\section{Conclusion}
The majority of the stars provided are considered millisecond or ordinary pulsars; specifically, none of them are magnetars. This suggests that when not looking at magnetars, we may be able to model the magnetic field such that it is crust-exclusive. There are currently many theories on the magnetic field origins and evolution of neutron stars, such as field-induced paramagnetism \cite{peng_physics_2007}, spontaneous ferro-crust or Landau-Stoner ferromagnetism \cite{uma_maheswari_spin_1997}, anisotropic $^3P_2$ superfluid magnetization \cite{mizushima_spin-polarized_2021}, chiral magnetic instability \cite{dehman_magnetar_2025}, spin-polarized ferromagnetic core phase \cite{kutschera_emergence_1999}, and magnetized nuclear-pasta glass \cite{yakovlev_electron_2015}. The models that include a core magnetic field were formed with the behaviors of magnetars in mind. 

These findings do not discredit these models, as their results reflect magnetar behavior better than the crust-exclusive models. However, when investigating neutron properties under extreme densities and magnetic fields, a crust-exclusive model allows for precise calculations, as the structure and contents of the inner crust are known much better than the core. This is significant, especially in nuclear astrophysical calculations, as millisecond and ordinary pulsars have more mass and radius measurements, allowing for precise calculations on the properties of neutrons in the crust of these stars.

%% file: chap3.tex
\chapter{Propagation-informed magnetic-field determination for PSR~J0437$-$4715}
\label{chap:j0437_bfield}

The residual-power analysis requires a magnetic-field normalization that is not obtained from the same spin-down torque law being tested.  The commonly quoted pulsar field,
\begin{equation}
B_{\rm sd}=3.2\times10^{19}\sqrt{P\dot P}~{\rm G},
\label{eq:bsd_standard}
\end{equation}
is a torque-inferred field.  It follows from assigning the observed rotational-energy loss to a specified electromagnetic braking model.  Using this field to normalize the electromagnetic torque and then subtracting that torque from the observed spin-down power would make the residual-power calculation circular.  For this reason, the field scale used below is obtained from radio-polarization propagation rather than from $\dot P$.

The central distinction in this chapter is the following.  The radio data identify a frequency interval over which the orthogonal-polarization-mode (OPM) behavior changes.  They do not directly measure the surface field $B_{\rm surf}$, the polarization-limiting radius $r_{\rm pol}$, the pair multiplicity $\kappa$, or the pair Lorentz factor $\Gamma_\pm$.  Those quantities enter through an explicitly stated propagation model and prior stack.  The result should therefore not be described as a model-free magnetic-field measurement.

\begin{figure}
    \centering
    \includegraphics[width=\linewidth]{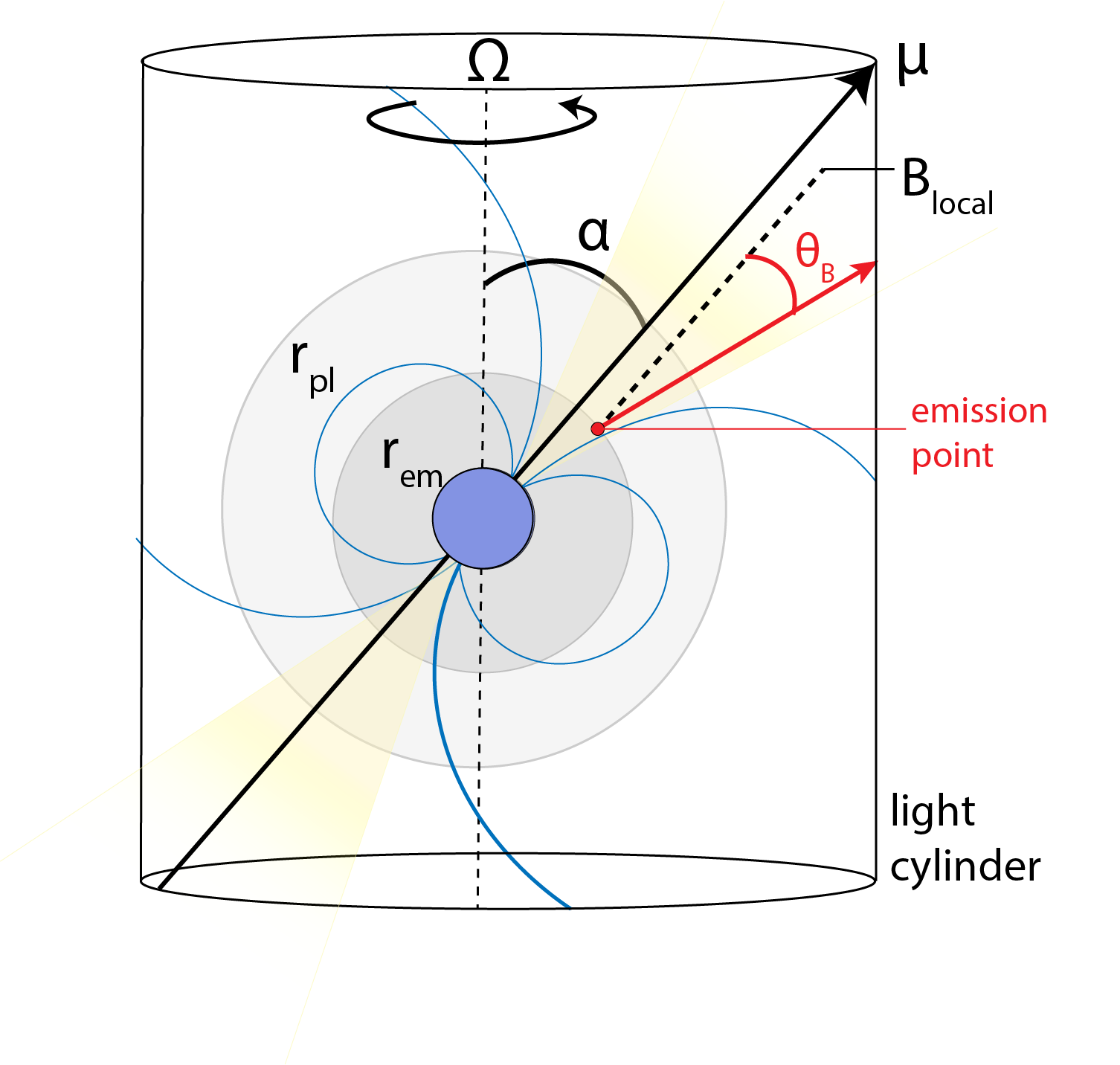}
    \caption{Diagram depicting the geometry used for the propagation-based magnetic-field estimate. The emission radius is shown by the darker shaded circle, and the polarization-limiting radius is shown by the lighter shaded circle. The angle $\theta_B$ denotes the angle between the emitted beam and the local magnetic field.}
    \label{fig:rpl_diag}
\end{figure}

\section{Radio data and directly measured products}
\label{sec:bfield_radio_data}

Pulsar radio emission is polarized.  At each pulse phase \(\phi\), the
measured Stokes parameters \((I,Q,U,V)\) define the linearly polarized
intensity
\[
L(\phi)=\sqrt{Q(\phi)^2+U(\phi)^2}
\]
and the polarization angle
\[
\psi(\phi)=\frac{1}{2}\tan^{-1}\!\left[\frac{U(\phi)}{Q(\phi)}\right],
\]
with the usual \(180^\circ\) ambiguity.  The PA curve \(\psi(\phi)\) is
therefore a directly constructed radio observable, not a magnetic-field model.

Many pulsars show orthogonal polarization modes (OPMs): at a fixed pulse
phase, the radiation preferentially appears on two PA tracks separated by
approximately \(90^\circ\).  These modes are usually interpreted as signatures
of propagation through the magnetospheric pair plasma, where the emitted wave
can be decomposed into natural propagation modes whose relative dominance
changes with phase and observing frequency \cite{petrova_origin_2001,petrova_nature_2006,
oswald_pulsar_2023}.  PSR~J0437$-$4715 is known to show OPM behavior in
single-pulse polarization studies \cite{oslowski_timing_2014}, so the frequency
evolution of its PA curve provides a useful empirical probe of the
magnetospheric propagation environment.

The first radio quantity used in this chapter is the OPM-transition interval,
\(\nu_{\rm OPM}\).  This is not a plasma frequency and not a direct magnetic
field measurement.  It is the observing-frequency range over which the folded
polarization profiles transition from low-frequency mode mixing toward a
stable high-frequency PA branch.  Operationally, a high-frequency PA template,
\(\psi_{\rm ref}(\phi)\), is constructed from the most stable high-frequency
profiles.  Each lower-frequency PA curve is then compared to this reference
branch.  Two simple diagnostics are used:
\[
{\rm PA}_{\rm rms}
=
\text{average PA mismatch from } \psi_{\rm ref}(\phi),
\]
and
\[
{\rm PA}_{\rm corr}
=
\text{alignment with } \psi_{\rm ref}(\phi).
\]
Thus \({\rm PA}_{\rm rms}\) asks how far the PA curve is from the
high-frequency reference, while \({\rm PA}_{\rm corr}\) asks how well it lines
up with that reference. 
The remaining quantities enter through the propagation model.  The radio
emission height, \(r_{\rm em}\), is the radius where the relevant radio
emission is produced.  The polarization-limiting radius, \(r_{\rm pol}\), is
the radius where the polarization stops adiabatically following the plasma and
effectively freezes out.  These are not the same quantity: the wave can be
emitted at \(r_{\rm em}\) and continue to evolve until \(r_{\rm pol}\).  In
this work \(r_{\rm pol}\) is therefore treated as an explicit propagation
prior rather than as a directly measured radio observable.

The pair-plasma quantities describe the medium through which the radio wave
propagates.  The pair multiplicity is
\[
\kappa\equiv\frac{n_\pm}{n_{\rm GJ}},
\]
where \(n_\pm\) is the pair density and
\[
n_{\rm GJ}=\frac{\Omega B}{2\pi e c}
\]
is the Goldreich--Julian reference density \cite{goldreich_pulsar_1969}.
The quantity that enters the final field scaling is \(\kappa\), not the
intermediate density \(n_\pm\).  The effective pair Lorentz factor
\(\Gamma_\pm\) describes the bulk streaming of the pair plasma.  Together,
\(\Gamma_\pm\) and \(\kappa\) control the strength of the plasma birefringence
through the combination \(\Gamma_\pm^3/\kappa\).

The angle \(\theta\) is also a propagation quantity.  It is the local angle
between the photon wave vector and the ambient magnetic field, not the
spin--magnetic obliquity of the pulsar.  The surface magnetic field inferred
below is therefore not obtained directly from the PA curve alone.  Instead,
the radio data determine \(\nu_{\rm OPM}\), and a conditional propagation
model maps
\[
(\nu_{\rm OPM},\, r_{\rm pol},\, \Gamma_\pm,\, \kappa,\, \theta)
\]
to a surface-field posterior for \(B_{\rm surf}\).

The radio data are full-Stokes Ultra-Wideband Low (UWL) observations of PSR~J0437$-$4715 from the CSIRO/Parkes Pulsar Timing Array data release \cite{reardon_neutron_2024}.  The UWL receiver on Murriyang/Parkes provides continuous frequency coverage from approximately $704$ to $4032~{\rm MHz}$ \cite{hobbs_ultra-wide_2020}.  The OPM-transition analysis uses folded, frequency-resolved polarization profiles, while separately reduced single-pulse products are used only to validate the OPM-core branch selection.

The folded-profile diagnostic uses sixteen reduced 64-channel Stokes products,
which together give 970 usable frequency-channel measurements after filtering.  The usable frequency range is $0.719236$--$4.006~{\rm GHz}$.  A high-frequency PA reference template is constructed from the $3.174$--$4.006~{\rm GHz}$ band, using 245 reference channels and 256 valid PA-template phase bins.

The fitted folded-profile transition diagnostics are summarized in Table~\ref{tab:opm_transition_diagnostics}.  The $L/I$ diagnostic identifies the low-frequency onset of the OPM-core transition, while the PA-template diagnostics locate the upper part of the transition near $2.66$--$2.69~{\rm GHz}$.  

\begin{table}[t]
\centering
\caption{Folded-profile diagnostics used to define the adopted OPM-core transition interval.  Values are transition frequencies from logistic fits to frequency-binned polarization diagnostics.  The $L/I$ transition motivates the lower edge of Eq.~\eqref{eq:nu_opm}, while the PA-template diagnostics motivate the upper edge.}
\label{tab:opm_transition_diagnostics}
\begin{tabular}{lccc}
\hline
Diagnostic & $p_{16}$ (GHz) & Median (GHz) & $p_{84}$ (GHz) \\
\hline
$L/I$ & 2.410 & 2.474 & 2.522 \\
PA-template correlation & 2.635 & 2.661 & 2.685 \\
PA rms residual & 2.661 & 2.689 & 2.716 \\
PA coherence & 2.768 & 2.785 & 2.798 \\
\hline
\end{tabular}
\end{table}

The diagnostics do not all measure the same stage of the polarization
evolution. The PA-template correlation and PA rms residual directly test
whether the phase-resolved PA curve has converged toward the stable
high-frequency reference branch. Their transitions at
$2.661^{+0.023}_{-0.027}$ and
$2.689^{+0.027}_{-0.029}~{\rm GHz}$ therefore identify the upper portion
of the OPM-core branch transition.

The PA-coherence statistic instead measures the circular concentration of
the PA values,
\begin{equation}
C_{\rm PA}
=
\frac{
\left|\sum_k w_k\exp(2i\psi_k)\right|
}{
\sum_k w_k
},
\label{eq:pa_coherence}
\end{equation}
where $\psi_k$ is the PA in phase bin $k$, $w_k$ is its polarization
weight, and the factor of two accounts for the $180^\circ$ ambiguity of
linear-polarization position angles. A large value of $C_{\rm PA}$
indicates a narrowly concentrated PA distribution, but does not by itself
establish agreement with the particular high-frequency PA template.

The later PA-coherence break,
\begin{equation}
\nu_{\rm br,PA\,coh}
=
2.785^{+0.013}_{-0.017}~{\rm GHz},
\label{eq:pacoherence_transition}
\end{equation}
is therefore interpreted as the frequency at which the already selected
high-frequency branch becomes fully internally stabilized. It is retained
as a corroborating cross-check, rather than as the upper edge of the
OPM-core transition. Including it in the adopted interval would mix the
initial convergence onto the high-frequency branch with the later
completion of PA narrowing.

The adopted interval is consequently
\begin{equation}
\nu_{\rm OPM}=2.472\text{--}2.680~{\rm GHz},
\label{eq:nu_opm}
\end{equation}
where the lower edge represents the onset of the $L/I$ transition and the
upper edge represents convergence onto the stable high-frequency
PA-template branch.
\section{Emission height versus polarization-limiting radius}
\label{sec:bfield_rpol}

The conversion from $\nu_{\rm OPM}$ to $B_{\rm surf}$ requires a length scale.  The relevant scale in the propagation model is the polarization-limiting radius $r_{\rm pol}$, the radius at which the polarization state effectively decouples from the magnetospheric plasma.  This is distinct from the radio-emission height $r_{\rm em}$.  The emission height is where the radiation is produced; the polarization-limiting radius is where the polarization state freezes out after propagation through the magnetosphere.

For PSR~J0437$-$4715, the source-specific emission-height anchor is the aberration--retardation analysis from Gangadhara--Thomas \cite{gangadhara_millisecond_2006}.  They decompose the 1440 MHz profile into a core plus nested conal components and infer an altitude ladder extending from the core at approximately $0.07R_{\rm LC}$ to the outermost cone at approximately $0.30R_{\rm LC}$.  The relevant middle-cone entries are cone 2 near $0.13R_{\rm LC}$ and cone 3 near $0.23R_{\rm LC}$.  For $P=5.757451941593412~{\rm ms}$,
\begin{equation}
R_{\rm LC}=\frac{cP}{2\pi}=274.7~{\rm km}.
\label{eq:rlc}
\end{equation}
Thus the Gangadhara--Thomas \cite{gangadhara_millisecond_2006} altitude ladder corresponds approximately to
\begin{equation}
0.07R_{\rm LC}\simeq19~{\rm km},\qquad
0.13R_{\rm LC}\simeq36~{\rm km},\qquad
0.23R_{\rm LC}\simeq63~{\rm km},\qquad
0.30R_{\rm LC}\simeq82~{\rm km}.
\end{equation}

The fiducial analysis adopts a local OPM-core freeze-out prior,
\begin{equation}
r_{\rm pol}\sim 0.15R_{\rm LC}-0.22R_{\rm LC},
\label{eq:rpol_prior_uniform}
\end{equation}
or numerically
\begin{equation}
r_{\rm pol}\sim 41.2-60.4~{\rm km}.
\end{equation}
This prior is a deliberately trimmed interior subset of the Gangadhara--Thomas cone-2--cone-3 altitude interval.  The reason for trimming inward is that $r_{\rm pol}$ is not assigned to an emission-component centroid.  Instead, it is treated as a freeze-out scale above the core/inner emission region but below the outer conal region.  This is a source-anchored propagation prior, not a direct measurement of $r_{\rm pol}$.

The Monte Carlo posterior corresponding to this prior is
\begin{equation}
r_{\rm pol}=50.8^{+8.7}_{-8.7}~{\rm km}
\qquad (90\%~{\rm C.L.}),
\label{eq:rpol_result}
\end{equation}
and the associated emission-height posterior is
\begin{equation}
r_{\rm em}=17.9^{+2.7}_{-2.2}~{\rm km}
\qquad (90\%~{\rm C.L.}).
\label{eq:rem_result}
\end{equation}
Equivalently,
\begin{equation}
\frac{r_{\rm pol}}{r_{\rm em}}=2.83^{+0.68}_{-0.58}
\qquad (90\%~{\rm C.L.}).
\label{eq:rpol_over_rem}
\end{equation}

This is the most important modeling assumption in the field inference.  The prior in Eq.~\eqref{eq:rpol_prior_uniform} does not follow from a theorem requiring $r_{\rm pol}<0.30R_{\rm LC}$.  Rather, it defines the fiducial branch in which the OPM-core transition freezes out locally in the same inner/middle magnetospheric region that produces the observed core/cone radio structure.  This branch is motivated by two source-specific facts.  First, PSR~J0437$-$4715 has long been noted to show ordinary-pulsar-like single-pulse behavior despite being a millisecond pulsar \cite{jenet_radio_1998}; \cite{gangadhara_millisecond_2006} explicitly cite this point when motivating a normal core/cone emission-altitude analysis.  Second, their A/R decomposition gives a strong core and clear nested conal components, with the core lower than the conal emission and with different cones occurring at different altitudes.  These facts make a normal-pulsar-style inner/middle cone prior reasonable for a fiducial OPM-core branch, but they do not rule out more distant polarization freeze-out.

The broader interpretation is therefore kept explicit.  General polarization-propagation theory allows the polarization-limiting radius to be a separate propagation scale outside the emission region.  Hence broader alternatives should be treated as systematic branches rather than averaged with the fiducial branch.  Two useful comparison branches are
\begin{align}
\text{full emission-envelope branch:}\quad &0.07R_{\rm LC}<r_{\rm pol}<0.30R_{\rm LC},\\
\text{outer freeze-out branch:}\quad &r_{\rm pol}\gtrsim0.30R_{\rm LC}.
\end{align}
Because the inferred surface field scales as $B_{\rm surf}\propto r_{\rm pol}^3$, these alternatives broaden or shift the inferred field substantially.

\section{Pair-plasma and multiplicity priors}
\label{sec:bfield_plasma}

The propagation model requires a plasma density.  The density is written in terms of the Goldreich--Julian density,
\begin{equation}
n_{\rm GJ}=\frac{\Omega B}{2\pi e c},
\label{eq:ngj}
\end{equation}
and the pair multiplicity $\kappa$,
\begin{equation}
n_\pm=\kappa n_{\rm GJ}.
\label{eq:kappa_def}
\end{equation}
The fiducial multiplicity scale is obtained from a pair-luminosity prior,
\begin{equation}
L_{\rm pair}=10^{-2.0\pm0.25}\dot E_{\rm int}.
\label{eq:lpair_prior}
\end{equation}
The central value is motivated by millisecond-pulsar pair-cascade calculations, which find $L_{\rm pair}\sim10^{-2}L_{\rm sd}$ for MSPs \cite{harding_pulsar_2011}.  The $\pm0.25$ dex width is an adopted modeling uncertainty used in this work, not a directly measured scatter for J0437.

Given the open-field particle flux and pair Lorentz factor, the luminosity scale is related schematically to
\begin{equation}
L_{\rm pair}\sim \Gamma_\pm m_e c^2\,\kappa\dot N_{\rm GJ},
\label{eq:lpair_scaling}
\end{equation}
where $\dot N_{\rm GJ}$ is the Goldreich--Julian particle outflow rate on the selected open-field bundle.  Propagating the OPM-transition interval, emission-height scale, geometry, and pair-luminosity prior gives
\begin{align}
\Gamma_\pm&=149^{+10}_{-9},\label{eq:gamma_result}\\
\kappa&=1.31^{+0.91}_{-0.54}\times10^4.\label{eq:kappa_result}
\end{align}
These are derived plasma posteriors, not direct observables.

\begin{figure*}[t]
\centering
\includegraphics[width=\textwidth]{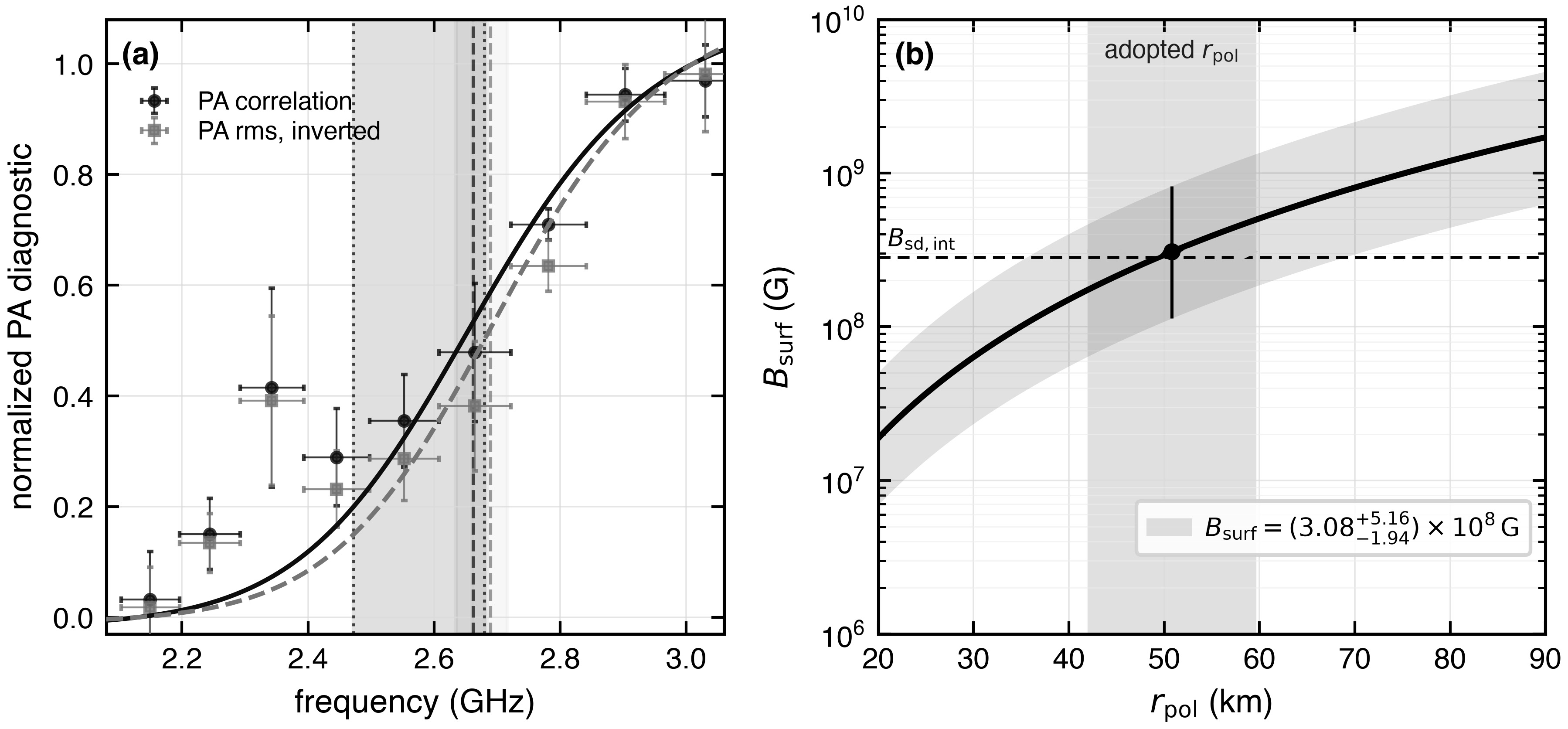}
\caption{\label{fig:bfield_diagnostics}
Magnetic-field diagnostics for PSR~J0437--4715.  (a) Folded-profile polarization diagnostics used to empirically bracket the OPM-core transition.  The PA-template correlation measures agreement with the high-frequency PA template, while the inverted PA-rms curve displays the corresponding residual on the same visual trend.  The shaded band marks the adopted interval $\nu_{\rm OPM}=2.472$--$2.680~{\rm GHz}$.  (b) Propagation-inferred surface field as a function of polarization-limiting radius.  The vertical band shows the adopted $r_{\rm pol}$ prior, the gray band isolates the sampled $\kappa$ range, and the dashed horizontal line shows $B_{\rm sd,int}$ as an external comparison only.}
\end{figure*}

\section{Propagation scaling and surface-field posterior}
\label{sec:bfield_scaling}

The numerical field conversion used in the analysis comes from a local adiabatic-threshold form of the cold-pair propagation scaling.  For a wave of angular frequency \(\omega=2\pi\nu\), the natural-mode wavenumber splitting was implemented as
\begin{equation}
\Delta k \simeq
\frac{\omega_p^2}{2\Gamma_\pm^3 c\omega}\sin^2\theta,
\label{eq:deltak_implemented}
\end{equation}
where \(\Gamma_\pm\) is the effective pair streaming Lorentz factor, \(\theta\) is the local ray--field angle, and
\begin{equation}
\omega_p^2=\frac{4\pi n_\pm e^2}{m_e},
\qquad
n_\pm=\kappa n_{\rm GJ},
\qquad
n_{\rm GJ}=\frac{\Omega B(r)}{2\pi e c}.
\label{eq:plasma_frequency_chain}
\end{equation}
Here \(n_\pm\) is the effective pair density entering the plasma-frequency convention used in the code.  Alternative one-species versus two-species normalizations change only an order-unity coefficient and are included in the propagation-normalization factor defined below.

The polarization-limiting condition was written in terms of an adiabaticity parameter,
\begin{equation}
\Gamma_{\rm ad}=\frac{\Delta k}{|d\psi_B/ds|},
\label{eq:gamma_ad_def}
\end{equation}
where \(\psi_B\) is the projected magnetic-field direction along the ray.  The implementation used
\begin{equation}
|d\psi_B/ds|\simeq \frac{\eta_{\rm rot}}{R_{\rm LC}},
\qquad
B(r)=B_{\rm surf}\left(\frac{R}{r}\right)^3,
\label{eq:geometry_rate_and_dipole}
\end{equation}
with \(R=11.36~{\rm km}\).  Solving \(\Gamma_{\rm ad}=\Gamma_0\) at \(r=r_{\rm pol}\) gives
\begin{equation}
B_{\rm surf}
=
\eta_{\rm prop}
\frac{m_e c\omega}{e}
\frac{\Gamma_\pm^3}{\kappa}
\left(\frac{r_{\rm pol}}{R}\right)^3
\frac{1}{\sin^2\theta},
\qquad
\eta_{\rm prop}\equiv \Gamma_0\eta_{\rm rot}.
\label{eq:bstar_derived_from_gammaad}
\end{equation}
Thus \(B_{\rm surf}\propto \nu_{\rm OPM}\Gamma_\pm^3 r_{\rm pol}^3/(\kappa\sin^2\theta)\).  The angle \(\theta\) is not the magnetic obliquity.  It is the local angle between the photon wave vector and the ambient magnetic field in the plasma-propagation calculation.  This angular dependence is part of the standard wavemode-coupling scaling: \(\theta\) is the wave-vector tilt to the ambient magnetic field, and the coupling radius depends sensitively on it \cite{petrova_nature_2006}.  For the fiducial J0437 branch we adopt \(\theta=15^\circ\), which is also consistent with the dipolar field-line geometry of the adopted local freeze-out range.  In a dipole,
\begin{equation}
\frac{r}{R_{\rm LC}}\simeq \sin^2\vartheta,
\qquad
\rho=\vartheta+\arctan\left(\frac{\tan\vartheta}{2}\right),
\end{equation}
where \(\vartheta\) is the magnetic colatitude of the field line and \(\rho\) is the tangent opening angle used in radio-width geometry \cite{petri_multi-wavelength_2026}.  The local ray--field angle for approximately radial propagation is therefore
\begin{equation}
\theta_{kB}\simeq \rho-\vartheta
=\arctan\left(\frac{\tan\vartheta}{2}\right).
\end{equation}
For \(r_{\rm pol}=0.15\text{--}0.22R_{\rm LC}\), this gives \(\theta_{kB}=11.9^\circ\text{--}14.9^\circ\).  The reference value \(15^\circ\) is therefore a rounded upper-end effective ray--field angle for the same fiducial local OPM-core branch, not an independent measured J0437 parameter.

The factor \(\eta_{\rm prop}\) is not treated as a source prior.  In the fiducial result it is set to unity:
\begin{equation}
\eta_{\rm prop}=1.
\end{equation}
This simply means that freeze-out is defined by \(\Gamma_{\rm ad}=1\) and that the projected-field direction is taken to vary on the light-cylinder scale, \(|d\psi_B/ds|\sim R_{\rm LC}^{-1}\).  A factor-of-order-unity change in this convention would rescale \(B_{\rm surf}\) linearly, but because it has no independent observational source it is kept out of the prior table and treated only as a possible systematic check.

The number \(9.8\times10^8~{\rm G}\) is therefore not an empirical field prior.  It is the reference-value normalization of Eq.~\eqref{eq:bstar_derived_from_gammaad} before the angular-frequency factor is applied:
\begin{equation}
\begin{aligned}
B_0^{\rm(ord)}
&=
\frac{m_e c(2.65\times10^9~{\rm s}^{-1})}{e}
\left(\frac{50^3}{100}\right)
\left(\frac{80~{\rm km}}{11.36~{\rm km}}\right)^3
\frac{1}{\sin^2(15^\circ)} \\
&=9.82\times10^8~{\rm G}.
\end{aligned}
\label{eq:B0_reference_value}
\end{equation}
Using the angular-frequency convention \(\omega=2\pi\nu\) multiplies this reference coefficient by \(2\pi\).  The implemented angular-frequency scaling is therefore
\begin{equation}
\begin{aligned}
B_{\rm surf}
={}&
2\pi(9.8\times10^8~{\rm G})\,
\eta_{\rm prop}
\left(\frac{\nu_{\rm OPM}}{2.65~{\rm GHz}}\right)
\left[\frac{\Gamma_\pm^3/\kappa}{50^3/100}\right]
\\
&\times
\left(\frac{r_{\rm pol}}{80~{\rm km}}\right)^3
\left[
\frac{\sin^2(15^\circ)}{\sin^2\theta}
\right].
\end{aligned}
\label{eq:bstar_scaling}
\end{equation}
No timing-derived spin-down field enters Eqs.~\eqref{eq:bstar_derived_from_gammaad}--\eqref{eq:bstar_scaling}.  The coefficient is fixed by physical constants, the adopted cold-pair propagation convention, the neutron-star radius, and the chosen reference values \((\nu,\Gamma_\pm,\kappa,r_{\rm pol},\theta)=(2.65~{\rm GHz},50,100,80~{\rm km},15^\circ)\).

Monte Carlo propagation of the inputs listed above gives the adopted surface-field posterior
\begin{equation}
B_{\rm surf}=3.08^{+5.14}_{-1.93}\times10^8~{\rm G}
\qquad (90\%~{\rm C.L.}).
\label{eq:bstar_result}
\end{equation}
This posterior is the magnetic-field normalization used in the residual-power analysis.  It is independent of the spin-down torque law in the sense that \(B_{\rm sd}\) is not used to set the field normalization.  It remains conditional on the propagation prior stack in Table~\ref{tab:bfield_prior_stack}.

For comparison only, the intrinsic spin-down field after kinematic correction is
\begin{equation}
B_{{\rm sd,int}}\simeq2.84\times10^8~{\rm G},
\label{eq:bsd_int_comparison}
\end{equation}
using \(P=0.005757451941593412~{\rm s}\) and \(\dot P_{\rm int}\simeq1.37\times10^{-20}~{\rm s~s^{-1}}\).  This timing-derived value is not used as a prior in Eq.~\eqref{eq:bstar_result}.

\section{Adopted priors and model choices}
\label{sec:bfield_prior_table}

Table~\ref{tab:bfield_prior_stack} lists only the inputs that are not directly fitted from the radio data.  The measured radio products are the OPM-transition interval and the branch-validation diagnostics described above.  The quantities below are the assumptions that turn those radio products into a magnetic-field posterior.

\begin{table}[t]
\centering
\small
\renewcommand{\arraystretch}{1.25}
\caption{Non-data priors and model choices entering the propagation-informed
\(B_{\rm surf}\) inference.}
\label{tab:bfield_prior_stack}
\begin{tabularx}{\textwidth}{
@{}
>{\raggedright\arraybackslash}p{0.20\textwidth}
>{\raggedright\arraybackslash}p{0.24\textwidth}
>{\raggedright\arraybackslash}X
@{}}
\toprule
Input & Adopted value & Basis / role \\
\midrule

Polarization-limiting radius &
\(r_{\rm pol}=0.15\text{--}0.22R_{\rm LC}\) &
Fiducial OPM-core freeze-out prior.  Chosen as a trimmed interior subset of
the Gangadhara--Thomas cone-2--cone-3 altitude range, placing freeze-out above
the core/inner emission and below the outer-cone region. \\

Pair luminosity &
\(L_{\rm pair}/\dot E_{\rm int}=10^{-2.0\pm0.25}\) &
Central value motivated by MSP pair-cascade calculations
\cite{harding_pulsar_2011}; the \(\pm0.25\) dex width is an adopted
uncertainty in this work. \\

Pair multiplicity &
\(\kappa \equiv n_\pm/n_{\rm GJ}\) &
Multiplicity entering the field scaling through \(\Gamma_\pm^3/\kappa\).
The reference density is
\(n_{\rm GJ}=\Omega B/(2\pi ec)\) \cite{goldreich_pulsar_1969}. \\

Ray--field angle &
\(\theta=15^\circ\) &
Local ray--field angle, not the spin--magnetic obliquity.  The angular
dependence follows from wavemode-coupling theory
\cite{petrova_nature_2006}; \(15^\circ\) is the rounded dipolar-geometry
value for the fiducial \(r_{\rm pol}\) branch. \\

Propagation scaling &
Eq.~\eqref{eq:bstar_scaling} &
Cold-pair propagation convention mapping
\((\nu_{\rm OPM},r_{\rm pol},\Gamma_\pm,\kappa,\theta)\) to
\(B_{\rm surf}\).  The numerical coefficient is a reference normalization,
not an empirical field prior. \\

\bottomrule
\end{tabularx}
\end{table}

\section{Interpretation}
\label{sec:bfield_interpretation}

The adopted field posterior is independent of the spin-down torque law being tested, but it is not model-free. The UWL radio data identify an OPM-core transition interval.  Under an explicitly stated local OPM-core propagation prior, pair-luminosity prior, and cold-pair plasma scaling, this transition maps to a surface-field posterior centered near $3\times10^8~{\rm G}$.  The posterior is not obtained from $P\dot P$, but it remains conditional on the adopted propagation model. The most important sensitivity is the polarization-limiting radius.  Since
\begin{equation}
B_{\rm surf}\propto r_{\rm pol}^3,
\end{equation}
using the full Gangadhara--Thomas emission envelope or an outer freeze-out branch would broaden or raise the field posterior.  Those alternatives should be reported as systematic checks.  They should not be averaged with the fiducial branch, because they correspond to different physical interpretations of where the observed OPM-core transition freezes out.

The scale of this systematic is large.  The fiducial posterior has median $r_{\rm pol}=50.8~{\rm km}\simeq0.185R_{\rm LC}$.  If the same propagation scaling were evaluated at the outer edge of the Gangadhara--Thomas emission envelope, $r_{\rm pol}=0.30R_{\rm LC}=82.4~{\rm km}$, the field normalization would increase by
\begin{equation}
\left(\frac{0.30}{0.185}\right)^3\simeq4.3.
\end{equation}
Thus the median field would move from $3.08\times10^8~{\rm G}$ to roughly $1.3\times10^9~{\rm G}$ if all other propagation inputs were held fixed.  An explicitly outer-magnetosphere branch, for example $r_{\rm pol}=0.50$--$0.59R_{\rm LC}$, would increase the field by approximately a factor of $20$--$33$, giving a characteristic scale of $6\times10^9$--$1\times10^{10}~{\rm G}$.  However, one should keep in mind the relatively small spin-down luminosity of this pulsar. Magnetic fields on the scale of $10^9-10^{10}~\rm G$ cause the magnetic dipole radiation to exceed the overall spin-down of the star. This is not to claim that this is evidence toward our emission model, as propagation and magnetosphere models are still evolving. This is why the fiducial result should be described as conditional on the inner/mid-zone OPM-core freeze-out prior, while the $r_{\rm pol}>0.30R_{\rm LC}$ case should be shown as a separate systematic branch rather than folded into the same posterior.

%% file: chap4.tex
\chapter{NICER-Constrained Surface-Field Geometry of PSR J0437--4715}
\label{chap:j0437_geometry}

The canonical pulsar picture often begins with a centered, tilted magnetic dipole. This idealization is useful for relating the rotation axis, magnetic axis, and observer line of sight, and it underlies many standard estimates of pulsar magnetic fields and spin-down torques. However, the surface field of a neutron star need not be a centered dipole. The near-surface magnetic topology can contain offset-dipole and higher-multipole structure even when the large-scale magnetosphere is approximately dipolar at the light cylinder. This distinction is especially important for millisecond pulsars, where thermal X-ray pulse-profile modeling can resolve the locations and angular extents of heated surface regions.

For rotation-powered millisecond pulsars, localized thermal X-ray emission is generally interpreted as surface heating by magnetospheric return currents. In this picture, the emitting regions trace the stellar footprints of open magnetic-field-line bundles rather than arbitrary thermal patches. The relationship between hot-spot morphology and magnetic topology is not one-to-one, because the current distribution depends on magnetospheric electrodynamics, pair creation, and relativistic ray tracing. Nevertheless, the positions and sizes of the thermal emitting regions provide valuable geometric information about the near-surface field. This has motivated the use of X-ray hot-spot maps as observational probes of non-dipolar magnetic structure \cite{bogdanov_nearest_2012,bilous_nicer_2019,miller_psr_2019,lockhart_x-ray_2019}.

PSR J0437--4715 is particularly useful for this purpose. It is the nearest and brightest known rotation-powered millisecond X-ray pulsar and has precise external constraints from radio timing. The recent NICER analysis of PSR J0437--4715 inferred the stellar mass, radius, and hot-region geometry from pulse-profile modeling, conditional on informative timing priors for the mass, distance, and binary inclination \cite{choudhury_nicer_2024,reardon_neutron_2024}. The resulting hot-region geometry is non-antipodal, which disfavors a pure centered dipole as a complete description of the near-surface magnetic field. In this chapter, the published NICER hot-region posterior is used as a geometric constraint on low-order surface-field families. The goal is not to refit the raw NICER event data, but to propagate the posterior-level information about hot-region locations and extents into a quantitative estimate of the surface multipolar content.

This geometry analysis serves two roles in the broader thesis. First, it tests whether simple low-order field families can reproduce the NICER hot-region configuration. Second, it supplies a geometry-weighted quadrupolar participation factor, \(f_{\ell=2}\), used later in the spin-down energy-budget constraint on an effective neutron magnetic quadrupole moment. The analysis therefore bridges the X-ray pulse-profile result and the torque calculation: the NICER posterior determines which surface geometries are admissible, while the spherical-harmonic decomposition determines how much centered quadrupolar surface-field power those geometries contain.
\begin{figure}
    \centering
    \includegraphics[width=.85\linewidth]{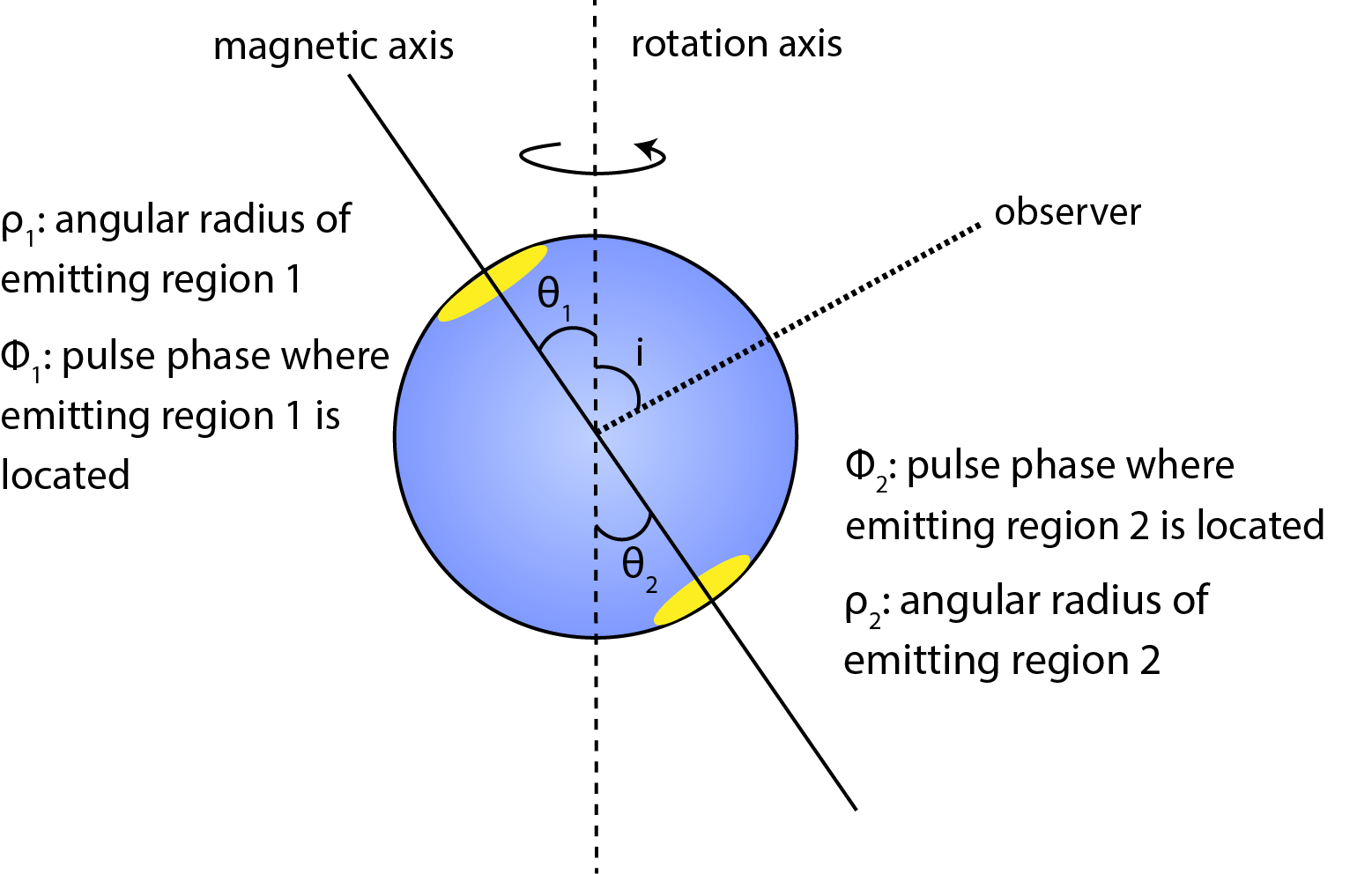}
    \caption{Diagram depicting the geometric parameters being fit. $\theta_{1,2}$ represent the colatitudes of emitting regions 1 and 2, respectively. $\phi_{1,2}$ represent the phases within one cycle at which emitting regions 1 and 2 are located. $\rho_{1,2}$ represent the angular radii of emitting regions 1 and 2.}
    \label{fig:chap4_geometry_parameters}
\end{figure}
\section{NICER hot-region information used in this work}
\label{sec:nicer_data_used}

The NICER input used here is the published hot-region posterior for PSR J0437--4715 \cite{choudhury_nicer_2024}, not the raw NICER event likelihood. Each posterior draw specifies the locations and angular extents of two dominant thermal emitting regions. In the analysis files, the relevant posterior columns are
\begin{equation}
\{\theta_1,\phi_1,\rho_1,\theta_2,\phi_2,\rho_2\},
\end{equation}
where \(\theta_i\) is the spot colatitude, \(\phi_i\) is the rotational phase or longitude coordinate, and \(\rho_i\) is the angular spot extent. The stored phase coordinates are converted to degrees through
\begin{equation}
\phi_i^{\rm deg}=360^\circ \phi_i^{\rm cycle}.
\end{equation}
Thus each posterior draw supplies two finite hot-region targets on the stellar surface.
\begin{figure*}
\centering
\includegraphics[width=\textwidth,clip]{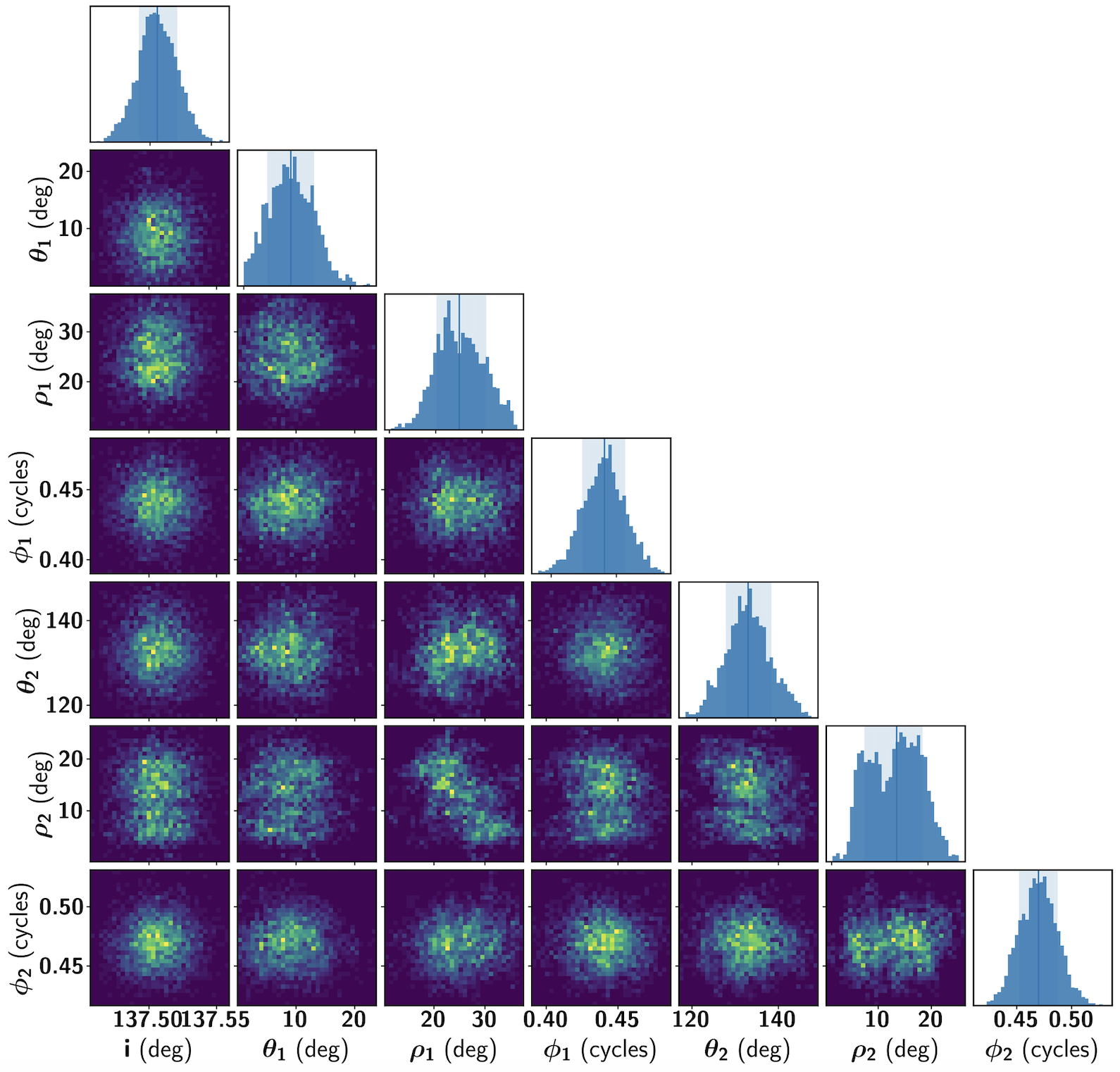}
\caption{Posterior distributions for the hotspot-geometry model used in the field-family comparison for PSR J0437--4715. The plotted parameters are the observer inclination $i$, the colatitudes $(\theta_1,\theta_2)$, phases $(\phi_1,\phi_2)$, and angular sizes $(\rho_1,\rho_2)$ of the two hot emitting regions, together with the wrapped radio-component separation $\Delta\phi_{\rm sep}$ and the nuisance radio phase offset $\phi_{\rm off}$. One-dimensional marginalized histograms are shown on the diagonal, with the shaded region marking the central 68\% credible interval and the vertical line marking the posterior median. Off-diagonal panels show the corresponding two-dimensional marginalized posterior densities.}
\label{fig:chap4_geometry_corner} 
\end{figure*}
The NICER quantities are used only as geometric constraints. For each posterior draw, a trial magnetic-field family is optimized so that its two strongest high-\(|B_r|\) surface regions coincide with the NICER primary and secondary hot-region centers. The angular spot extents enter through finite-area overlap terms, so the comparison is not based solely on two point locations. The raw NICER photon likelihood is not evaluated in this chapter. Instead, the NICER posterior is treated as an already-marginalized representation of the allowed surface-emission geometry.

This distinction is important. The NICER analysis marginalizes over many quantities that are not refit here, including the stellar compactness, observer inclination, hot-region morphology, temperatures, background, and nuisance parameters. The present analysis takes the resulting hot-region posterior as input and asks a narrower question: can a low-order surface magnetic field place its high-\(|B_r|\) regions at the NICER-inferred emitting regions while also remaining consistent with the independently inferred obliquity information?

\section{Physical interpretation of the hot regions}
\label{sec:hot_region_interpretation}

The interpretation of the NICER hot regions as magnetic-topology tracers is motivated by polar-cap heating. In rotation-powered pulsars, the global magnetosphere is supplied by charges extracted from the stellar surface and by pair creation in the magnetosphere. A fraction of the accelerated particles returns to the star and deposits energy near the footprints of open magnetic field lines. These return currents heat localized regions of the surface, producing thermal X-ray emission. Detailed calculations show that even a dipolar magnetosphere can produce nonuniform polar-cap heating, and that multipolar or offset fields can substantially distort the polar-cap shape and relative hotspot locations \cite{goldreich_pulsar_1969, arons_pair_1981,harding_pulsar_2011,gralla_inclined_2017,lockhart_x-ray_2019}.

The inference made in this chapter is therefore deliberately geometric. It does not assume that the NICER spot boundary is exactly the open-field-line boundary, nor that the local temperature is uniform within the polar cap. Instead, the hot-region centers and angular extents are used as observational tracers of the regions where the near-surface magnetic geometry directs return-current heating. This is a weaker and more robust assumption than attempting to infer a unique magnetospheric current distribution from the NICER posterior alone.

This approach is also consistent with the broader NICER literature. For PSR J0030+0451, pulse-profile modeling found hot regions that are strongly non-antipodal, motivating interpretations involving global-scale non-dipolar magnetic structure \cite{bilous_nicer_2019,miller_psr_2019,riley_nicer_2019}. Similar considerations apply to PSR J0437--4715: the observed hot-region geometry is difficult to reconcile with a pure centered dipole, but it can be reproduced by modest departures from centered dipolar symmetry. The purpose of the present analysis is to quantify those departures in a controlled low-order expansion.

\section{Surface-field families}
\label{sec:field_families}

Three low-order magnetic-field families are considered:
\begin{enumerate}
    \item a centered dipole,
    \item an offset dipole, and
    \item a centered dipole plus a general quadrupole.
\end{enumerate}
The centered dipole is the minimal reference model. Its two magnetic extrema are antipodal on the stellar surface, so it cannot reproduce two hot regions that are significantly non-antipodal. It is therefore used mainly as a null model.

The offset dipole allows the magnetic dipole to be displaced from the stellar center. A displaced dipole is not a pure centered \(\ell=1\) field when expanded about the stellar center. Instead, it contains higher centered multipoles, including a quadrupolar component. This family is physically useful because it provides a compact description of a globally dipole-like field whose surface polar caps are shifted and distorted.

The dipole-plus-general-quadrupole family is the explicit quadrupolar model. It allows the centered dipole and centered quadrupole components to vary independently, rather than forcing the quadrupolar content to arise from a translation of the dipole. This distinction is important for the later torque calculation: the offset dipole gives a reference estimate of the centered quadrupolar content implied by displacement, while the dipole-plus-quadrupole family gives an explicit low-order quadrupolar surface geometry.

Higher-order families were checked separately but are not used in the final model comparison. The reason is not that higher-order structure is physically impossible. Rather, once enough degrees of freedom are added, the field can fit localized hot-region geometry without yielding a uniquely interpretable quadrupolar measurement. Since the later MQM calculation requires a controlled estimate of the centered \(\ell=2\) surface-field power, the final comparison is restricted to the lowest-order families that reproduce the NICER geometry.

\section{Fitting procedure}
\label{sec:geom_fitting_procedure}

For each family, the radial surface field is evaluated on the stellar surface,
\begin{equation}
B_r = B_r(\theta,\phi),
\end{equation}
and the fitted magnetic locations are identified with the dominant high-\(|B_r|\) regions. The optimization searches over the allowed parameters of each field family and minimizes a geometric score that combines three pieces of information:
\begin{enumerate}
    \item the angular separation between the NICER primary hot-region center and the corresponding magnetic high-\(|B_r|\) region,
    \item the angular separation between the NICER secondary hot-region center and the corresponding magnetic high-\(|B_r|\) region, and
    \item the finite-area overlap between the NICER hot-region weights and the model high-field regions.
\end{enumerate}
The radio/multiwavelength obliquity prior is imposed during the fit, including the supplementary-angle branch, because the X-ray hot-region geometry alone does not uniquely determine the orientation of the magnetic axis relative to the spin axis \cite{petri_multi-wavelength_2026}.

Both assignments of magnetic extrema to the primary and secondary NICER hot regions are tested. This is necessary because the thermal geometry does not determine the absolute magnetic polarity. The assignment giving the lower geometric score is retained. The procedure therefore tests whether a field family can reproduce the observed hot-region arrangement without imposing an arbitrary polarity convention.

For each posterior draw and each field family, the optimization returns a best-fitting surface map, the primary and secondary angular residuals,
\begin{equation}
d_{\rm P},\qquad d_{\rm S},
\end{equation}
and the corresponding spherical-harmonic power fractions. A draw is counted as geometrically successful if
\begin{equation}
d_{\rm P}\leq 10^\circ,\qquad d_{\rm S}\leq 5^\circ .
\label{eq:geom_acceptance_cuts}
\end{equation}
The primary tolerance is broader because its posterior residuals are less tightly localized than those of the secondary.

For the representative maps shown in Fig.~\ref{fig:geometry_heatmap}, we also quote a compact map-level goodness-of-fit probability. This quantity is not a raw NICER likelihood p-value. It is an approximate geometric p-value computed from the median hotspot-center residuals,
\begin{equation}
\chi^2_{\rm geom}
=
\left(\frac{d_{\rm P}}{10^\circ}\right)^2+
\left(\frac{d_{\rm S}}{5^\circ}\right)^2,
\label{eq:chi2_geom_chapter}
\end{equation}
with
\begin{equation}
p_{\rm geom}=P(\chi^2_4\geq \chi^2_{\rm geom}).
\label{eq:pgeom_chapter}
\end{equation}
The four degrees of freedom correspond to the two angular coordinates of each of the two hot-region centers.
\section{NICER-constrained surface-field maps}
\label{sec:surface_field_maps}

\begin{figure*}[t]
\centering
\includegraphics[width=\linewidth]{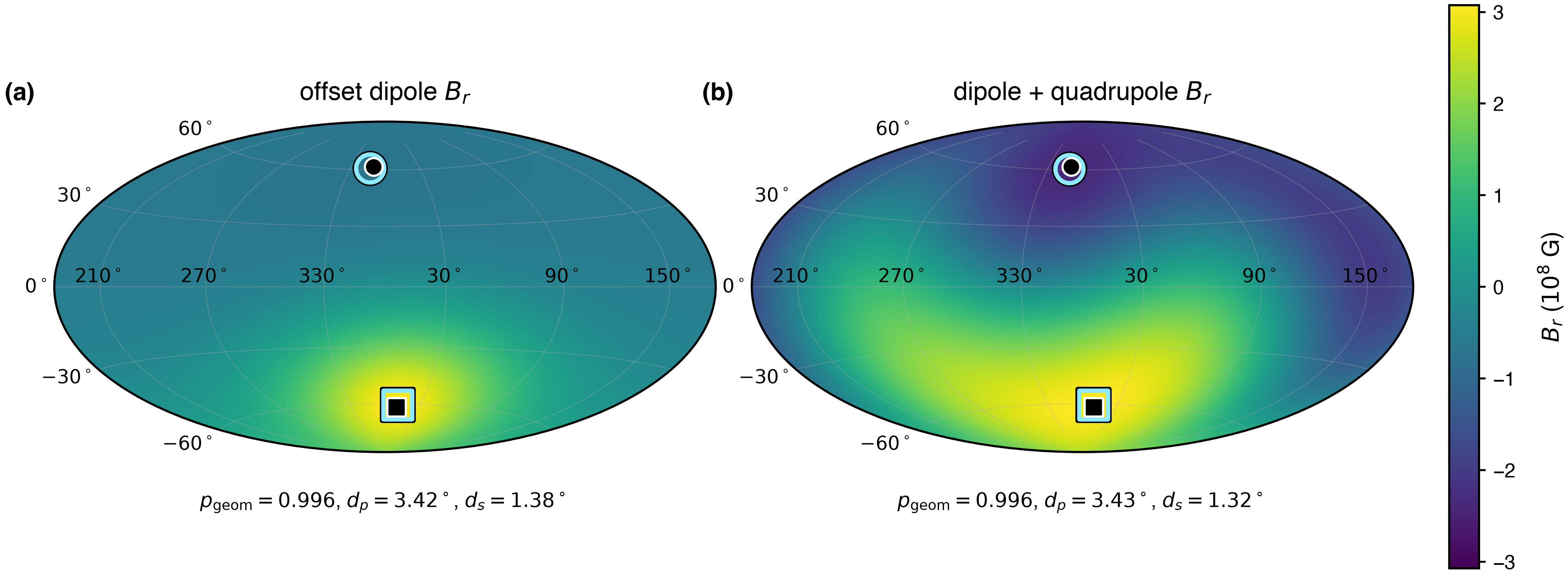}
\caption{\label{fig:geometry_heatmap}
Low-order surface radial-field geometries compatible with the published NICER hot-region posterior.  (a) Offset-dipole reference solution.  (b) Dipole-plus-general-quadrupole solution.  Maps are shown in an oblique Hammer projection and scaled to $B_{\rm surf}=3.08\times10^8~{\rm G}$.  Filled black markers show NICER hot-region centers, open cyan markers show the fitted high-$|B_r|$ locations, and dashed circles show the median residuals.  The maps illustrate compatible phenomenological geometries rather than a unique reconstruction of the surface field.}
\end{figure*}

Figure~\ref{fig:geometry_heatmap} shows the two successful low-order field families. The maps are plotted in a projection centered on the NICER hot-region geometry. This projection is used only for visualization: it rotates the display frame so that both hot regions are visible away from the map boundary, but it does not change the fitted surface coordinates or the spherical-harmonic decomposition.

The color scale gives the radial surface field \(B_r\) in units of \(10^8~{\rm G}\). Each normalized field geometry is scaled to the median timing-independent surface-field normalization,
\begin{equation}
B_{\rm surf} = 3.08\times 10^8~{\rm G}.
\end{equation}
The absolute scale of the map is set by the independent radio-polarization field estimate, while the relative surface pattern is determined by the NICER-constrained geometry fit.

The offset-dipole and dipole-plus-general-quadrupole families both pass the posterior validation test. The median residuals are nearly identical. The offset dipole gives
\begin{equation}
(d_{\rm P},d_{\rm S})=(3.42^\circ,1.38^\circ),
\end{equation}
while the dipole-plus-general-quadrupole family gives
\begin{equation}
(d_{\rm P},d_{\rm S})=(3.43^\circ,1.32^\circ).
\end{equation}
Both representative maps therefore have
\begin{equation}
p_{\rm geom}=0.996.
\end{equation}
The centered dipole fails this test, missing the NICER primary and secondary regions by tens of degrees in the posterior-median fit. This confirms that the NICER hot-region geometry requires either a displacement of the dipole or explicit higher multipolar structure near the stellar surface.

\section{Spherical-harmonic decomposition}
\label{sec:surface_harmonic_decomposition}

To connect the geometric fit to the later torque calculation, each fitted radial surface field is decomposed into real spherical harmonics,
\begin{equation}
B_r(\theta,\phi)=
\sum_{\ell=0}^{\ell_{\rm max}}
\sum_{m=-\ell}^{\ell}
a_{\ell m}Y_{\ell m}(\theta,\phi).
\label{eq:br_harmonic_chapter}
\end{equation}
The power in degree \(\ell\) is
\begin{equation}
P_\ell=
\sum_{m=-\ell}^{\ell}a_{\ell m}^2,
\label{eq:pl_chapter}
\end{equation}
and the fractional non-monopolar power is
\begin{equation}
f_\ell=
\frac{P_\ell}{\sum_{\ell'\geq1}P_{\ell'}}.
\label{eq:fl_chapter}
\end{equation}
The monopole term is excluded from the denominator because a physical magnetic field has no net magnetic monopole flux; any residual \(\ell=0\) contribution is a numerical or fitting artifact. The resulting \(f_\ell\) values quantify the geometric distribution of surface-field power among centered multipoles and are independent of the absolute magnetic-field scale.

This definition separates two questions. The NICER posterior determines which surface patterns are geometrically admissible. The independent radio-polarization analysis determines the absolute surface-field normalization. The harmonic decomposition then determines what fraction of the admissible surface-field geometry is carried by the centered quadrupole. If a map is scaled to a characteristic surface field \(B_{\rm surf}\), the corresponding RMS field associated with degree \(\ell\) scales approximately as
\begin{equation}
B_{\ell,{\rm rms}}\simeq B_{\rm surf}\sqrt{f_\ell}.
\label{eq:bell_rms_chapter}
\end{equation}
In the MQM calculation, the quadrupolar fraction \(f_{\ell=2}\) is used as a geometry-weighted participation factor.

\section{Quadrupolar surface-field fraction}
\label{sec:f2_results}

The offset-dipole family gives
\begin{equation}
f_{\ell=2}^{\rm off.~dip.}
=
0.232^{+0.046}_{-0.039},
\label{eq:f2_offset_chapter}
\end{equation}
while the dipole-plus-general-quadrupole family gives
\begin{equation}
f_{\ell=2}^{\rm dip.+quad.}
=
0.183^{+0.113}_{-0.041}.
\label{eq:f2_dipquad_chapter}
\end{equation}
These values are obtained from the NICER posterior ensemble. Varying the harmonic cutoff over
\begin{equation}
\ell_{\rm max}=6,8,10,12
\end{equation}
leaves the quoted \(f_{\ell=2}\) values unchanged to the displayed precision, indicating that the inferred quadrupolar fraction is not an artifact of the decomposition cutoff.

We therefore adopt
\begin{equation}
f_{\ell=2}\simeq 0.18\text{--}0.23
\label{eq:f2_adopted_chapter}
\end{equation}
as the median family bracket for the centered quadrupolar surface-field component. The quoted uncertainties in Eqs.~\eqref{eq:f2_offset_chapter} and \eqref{eq:f2_dipquad_chapter} define the broader posterior-plus-family envelope.

The interpretation of \(f_{\ell=2}\) differs slightly between the two successful families. In the dipole-plus-quadrupole family, the \(\ell=2\) contribution is an explicit centered quadrupole. In the offset-dipole family, the \(\ell=2\) contribution is the quadrupolar content induced when a displaced dipole is expanded about the stellar center. Both quantities are therefore centered surface-harmonic power fractions, but they arise from different parameterizations of the surface field. Their agreement at the level of \(f_{\ell=2}\sim0.2\) indicates that the NICER geometry requires a modest but non-negligible centered quadrupolar component, independent of whether it is represented as an explicit quadrupole or as the centered expansion of a displaced dipole.

\section{Uncertainties and limitations}
\label{sec:geometry_limitations}

Several limitations should be explicit. First, the NICER hot-region posterior is used as a geometric constraint, not as a full likelihood. This is appropriate for the present goal, but it means that the analysis does not propagate all correlations in the original event-level inference. Second, the mapping between return-current heating and the surface magnetic field is not unique. The emitting regions are treated as tracers of the near-surface topology, but detailed current closure, pair creation, atmospheric beaming, and temperature gradients can alter the relationship between the open-field-line footprint and the observed thermal map \cite{gralla_pulsar_2016, gralla_inclined_2017,lockhart_x-ray_2019}. Third, the field families are intentionally low-order. Higher multipoles could reproduce the NICER geometry, but they would weaken the interpretability of the \(\ell=2\) component used in the MQM calculation.

The adopted residual tolerances also have a diagnostic character. The cuts \(d_{\rm P}\le10^\circ\) and \(d_{\rm S}\le5^\circ\) are chosen to reflect the relative localization of the two hot-region centers in the posterior ensemble and in the low-order field-family fits. They should not be interpreted as a substitute for the original NICER likelihood.

Finally, the plotted field scale depends on the independently inferred surface normalization \(B_{\rm surf}\). The geometry and harmonic fractions are dimensionless and do not depend on this absolute scale. The map in Fig.~\ref{fig:geometry_heatmap} is scaled to \(B_{\rm surf}=3.08\times10^8~{\rm G}\) for visualization, while the torque calculation later propagates the full uncertainty in \(B_{\rm surf}\), \(f_{\ell=2}\), and the polarized neutron reservoir.

%% file: chap5.tex
\chapter{Constraining the nEDM upper bound using PSR J0437-4715 spin-down}
Chapters 3 and 4 provide a propagation-informed surface-field normalization and identify two low-order magnetic-field families compatible with the NICER hot-region geometry of PSR J0437-4715: an offset dipole and a dipole plus a general quadrupole. The field normalization is not obtained by equating the measured spin-down to the electromagnetic torque. In this chapter, these results are combined with the source-specific stellar structure model from Chapter 2 to construct the present-day torque budget and refine the effective polarized-neutron reservoir. Any positive power remaining after the modeled standard losses are subtracted is then assigned, as an unscreened benchmark, to a possible collective electric-dipole-radiation channel and converted into a conditional neutron-EDM bound.

\section{Torque evolution}
\subsection{Magnetic dipole radiation}
Electromagnetic spin-down is the dominant modeled standard-loss channel for PSR J0437-4715. The central question is therefore not only the field strength, but also which field component and magnetospheric state determine the torque acting on the star. 

When studying magnetic dipole radiation, the starting point is usually the vacuum retarded dipole model \cite{deutsch_electromagnetic_1955, satherley_pedagogical_2022}. In this picture, the star rotates in empty space and loses rotational energy through electromagnetic radiation from an oblique dipole, and it is expressed as
\begin{equation}
    L_{\rm vac}=\frac{2}{3}\frac{\mu^2 \Omega^4}{c^3}\sin^2 \alpha.
\end{equation}

This model is attractive because it is analytic and leads directly to the familiar textbook field estimate $B\propto \sqrt{P\dot P}$, but it has strict physical assumptions: it neglects plasma, magnetospheric currents, and pair creation. Its most important qualitative feature is that the torque vanishes for an exactly aligned rotator, because the loss depends entirely on the transverse time-varying dipole moment.

Modern pulsar theory instead favors a plasma-filled magnetosphere. In the ideal force-free limit, the star is embedded in enough plasma that the large-scale magnetosphere nearly satisfies $\mathbf{E\cdot B}=0$, while currents and open magnetic flux carry energy and angular momentum away from the star. For oblique rotators, Spitkovsky found \cite{spitkovsky_time-dependent_2006}
\begin{equation}
    L_{\rm ff} \approx \frac{\mu^2 \Omega^4}{c^3}(1+\sin^2\alpha).
\end{equation}

Unlike the vacuum case, the aligned rotator still spins down in this limit because the torque is carried by the global current system and the Poynting flux through the open field region. For small $\alpha$, this makes the force-free torque much larger than the vacuum torque for the same $(\mu, \alpha)$.

The tension between these two endpoint models motivated the development of resistive or dissipative magnetosphere models. Li, Spitkovsky, and Tchekhovskoy introduced finite-conductivity magnetospheres that interpolate smoothly between the vacuum and force-free limits \cite{li_resistive_2012}. In these models, the spin-down luminosity, the open-field-line potential drop, and the amount of magnetic flux all vary continuously as the plasma conductivity is changed. Physically, this means that the vacuum and force-free laws should be viewed as limiting cases rather than as the only two realistic choices. A real pulsar may occupy an intermediate magnetospheric state in which some accelerating electric field survives while the large-scale structure remains force-free.

A second major refinement is that the field geometry near the star need not be a centered dipole, as we have shown. A true multipolar field contains higher spherical-harmonic components such as quadrupole and octupole terms, while an off-centered dipole displaces the dipole moment away from the stellar center. Pétri showed that an off-centered dipole can be expanded into a sum of centered multipoles, so even a geometrically simple displacement naturally induces higher-order structure \cite{petri_multipolar_2015, petri_theory_2016}. Multipolar fields can strongly reshape the surface field, cap morphology, and local curvature without necessarily dominating the far-zone field that controls the global spin-down.

This distinction between near-surface structure and far-zone torque content is crucial. Higher multipoles decay more rapidly with radius than the dipole, so they can dominate hotspot geometry and local surface field strength while contributing much less to the field near the light cylinder. Pétri's exact multipole solutions show that the Poynting flux and braking behavior depend on multipole order and on the compactness ratio $R/r_{\rm L}$, where $r_{\rm L} =c/\Omega$, but they also make clear that one should not identify a strong local surface field with the global dipole moment that enters the standard spin-down laws. In practical terms, the torque law should be fed an effective large-scale dipole moment, not the strongest cap-scale surface field inferred from hotspot or polarization data. Multipolar contributions, however, can affect local magnetic fields, and this is explored later in this chapter.

In plasma-filled magnetospheres, multipolar structure matters not only through direct electromagnetic radiation but also through the polar-cap and current geometry. Gralla, Lupsasca, and Philippov showed that force-free polar-cap location and current distributions depend sensitively on the underlying magnetic topology, and that mixed dipole-quadrupole fields can produce qualitatively new cap shapes such as annular polar caps \cite{gralla_inclined_2017, lockhart_x-ray_2019}. This matters because the magnetosphere couples to the star through these caps: changes in cap geometry alter the return current, surface heating, and likely the pair-production environment. Thus multipoles can change the torque indirectly even when the far-zone field remains mostly dipolar.

Offset dipole structure may also change the pair supply that helps determine whether the magnetosphere sits closer to the vacuum or force-free limit. Harding and Muslimov found that modest polar-cap offsets can substantially increase pair multiplicity by enhancing the accelerating electric field on part of the cap. This provides a natural physical bridge between field geometry and torque state: a star with distorted or offset caps may sustain a more plasma-rich, more force-free-like magnetosphere than a centered-dipole estimate may suggest \cite{harding_pulsar_2011}. In that sense, geometry and magnetospheric state are not independent problems. 

For PSR J0437-4715, these general considerations are especially important because the star is unusually well constrained observationally. Precision timing gives a tightly measured binary inclination and a strong timing framework for the intrinsic spin-down budget, while recent NICER modeling finds non-antipodal hot regions inconsistent with a pure centered dipole \cite{reardon_neutron_2024, choudhury_nicer_2024}. The most recent multi-wavelength modeling goes further and finds that a slightly off-centered dipole augmented by small-scale structure can reproduce the hotspot morphology, radio polarization, and $\gamma$-ray behavior, with a magnetic obliquity $\alpha \approx 42 \pm 5^\circ$ and a line-of-sight angle $\zeta\approx 136 \pm 5^\circ$ \cite{petri_multi-wavelength_2026}. This makes PSR J0437-4715 a particularly strong case where the surface topology is clearly more complex than the canonical centered-dipole picture.

To determine the proper torque model, the problem is separated into three layers: geometry, global torque field, and magnetospheric state. The geometry layer is what we constrained with the hotspot modeling, radio component separation, and polarization arguments. The global torque field is the effective large-scale dipole moment $\mu_{\rm dip}$ and obliquity $\alpha$. The magnetospheric state is where the star behaves more like a vacuum, force-free, or intermediate resistive system. 

\subsubsection{Dipolar field component entering the torque}

Chapter 3 provides a propagation-informed characteristic surface-field scale,
\begin{equation}
B_{\rm surf}
=
3.08^{+5.14}_{-1.93}\times10^{8}~{\rm G},
\label{eq:bstar_torque_input}
\end{equation}
where the quoted range is the central 90\% posterior interval. This field is not assumed to be entirely dipolar. Chapter 4 instead decomposes each accepted surface-field realization into centered spherical harmonics and supplies the fraction $f_{\ell=1}$ of the surface-field power carried by the dipole. Because $f_\ell$ is a power fraction, the corresponding dipolar field amplitude is
\begin{equation}
B_{\rm dip}=B_{\rm surf}\sqrt{f_{\ell=1}}.
\label{eq:bdip_from_bstar}
\end{equation}

The magnetic dipole moment entering the torque is then defined by
\begin{equation}
\mu_{\rm dip}=\frac{1}{2}B_{\rm dip}R_\star^3,
\label{eq:mudip_from_bdip}
\end{equation}
and the force-free electromagnetic luminosity is
\begin{equation}
L_{\rm dip}
=
\frac{\mu_{\rm dip}^{2}\Omega^4}{c^3}
\left(1+\sin^2\alpha\right).
\label{eq:ldip_forcefree_final}
\end{equation}
Equations~\eqref{eq:bdip_from_bstar}--\eqref{eq:ldip_forcefree_final} are evaluated sample by sample while propagating the posteriors for $B_{\rm surf}$, $f_{\ell=1}$, $R_\star$, and $\alpha$. The two NICER-compatible field families are carried through separately. Their resulting $L_{\rm dip}$ posteriors, reported in Table~\ref{tab:torque_budget}, agree closely; no additional family-dependent correction factor is applied after the harmonic projection.

This construction separates the propagation-informed surface normalization from the large-scale dipolar amplitude that enters the torque and ensures that the field amplitude, stellar radius, magnetic moment, and luminosity are propagated within one consistent posterior calculation.

\subsection{Gravitational waves and other torque mechanisms}

In addition to the electromagnetic torque, the present-day energy-loss
budget includes gravitational-wave emission from a permanent equatorial
deformation and from a possible saturated $r$-mode. The total
gravitational-wave luminosity is written as
\begin{equation}
L_{\rm GW}=L_{\epsilon}+L_r.
\label{eq:Lgw_total}
\end{equation}

For a star with equatorial ellipticity $\epsilon$, the luminosity from a
steadily rotating mass quadrupole is
\begin{equation}
L_{\epsilon}
=
\frac{32G}{5c^5}
I^2\epsilon^2\Omega^6,
\label{eq:Lgw_ellipticity}
\end{equation}
where $I$ is the stellar moment of inertia and
$\Omega=2\pi/P$ is the angular spin frequency. The corresponding torque
magnitude is
\begin{equation}
N_{\epsilon}
=
\frac{L_{\epsilon}}{\Omega}
=
\frac{32G}{5c^5}
I^2\epsilon^2\Omega^5.
\label{eq:Ngw_ellipticity}
\end{equation}
The Monte Carlo calculation adopts
\begin{equation}
\epsilon=(1.0\pm0.5)\times10^{-9}.
\label{eq:epsilon_prior}
\end{equation}

The $l=m=2$ $r$-mode contribution is evaluated using the
gravitational-radiation growth-timescale approximation implemented in the
torque calculation. The magnitude of the gravitational-radiation
timescale is
\begin{equation}
\left|\tau_{\rm GR}\right|
=
3.26~{\rm s}
\left(\frac{P}{1~{\rm ms}}\right)^6
\left(\frac{M}{1.4\,M_{\odot}}\right)^{-1}
\left(\frac{R_\star}{10~{\rm km}}\right)^{-4}.
\label{eq:rmode_tau}
\end{equation}
The associated torque is
\begin{equation}
N_r
=
\frac{
3\alpha_r^2\widetilde J
M R_\star^2\Omega
}{
\left|\tau_{\rm GR}\right|
},
\label{eq:rmode_torque}
\end{equation}
and the corresponding luminosity is
\begin{equation}
L_r
=
N_r\Omega
=
\frac{
3\alpha_r^2\widetilde J
M R_\star^2\Omega^2
}{
\left|\tau_{\rm GR}\right|
}.
\label{eq:rmode_luminosity}
\end{equation}
Here $\alpha_r$ is the dimensionless $r$-mode amplitude and
\begin{equation}
\widetilde J
=
\frac{1}{M R_\star^4}
\int_0^{R_\star}\rho(r)r^6\,dr
\label{eq:Jtilde_definition}
\end{equation}
is a dimensionless stellar-structure constant. The numerical
implementation adopts
\begin{equation}
\widetilde J=1.635\times10^{-2},
\qquad
\alpha_r=(1.0\pm0.5)\times10^{-10}.
\label{eq:rmode_inputs}
\end{equation}
Both gravitational-wave contributions are propagated in the Monte Carlo
calculation and subtracted from the intrinsic spin-down power. The
residual used for the nonstandard radiation-channel inference is therefore
\begin{equation}
\Delta
=
\dot E_{\rm int}
-
L_{\rm dip}
-
L_{\epsilon}
-
L_r.
\label{eq:residual_with_gw}
\end{equation}
For the adopted priors, the $r$-mode contribution is negligible, while
the ellipticity contribution remains subdominant to the electromagnetic
loss.

\subsubsection{Other potential torque mechanisms}
The present-day budget is restricted to channels that directly remove stellar rotational energy during the post-recycling radio-pulsar phase. Binary gravitational-wave emission is omitted because it removes orbital, rather than stellar spin, energy and angular momentum \cite{peters_gravitational_1963, peters_gravitational_1964}. Accretion and propeller torques are also omitted because PSR J0437-4715 is modeled in its detached radio-pulsar state, with no active accretion disk.

\begin{table*}[t]
\centering
\caption{\label{tab:torque_budget}
Present-day luminosity budget for PSR~J0437--4715. The residual is $\Delta=\dot E_{\rm int}-L_{\rm dip}-L_{\rm GW}$; only the positive branch subsequently assigned to a nonstandard radiation channel is reported. The intervals are the sampled central 68\% ranges. $L_{\rm quad}$ is retained as a comparison luminosity and is not subtracted. The gravitational-wave priors are $\epsilon=(1.0\pm0.5)\times10^{-9}$ and $\alpha_r=(1.0\pm0.5)\times10^{-10}$.}
\setlength{\tabcolsep}{18pt}
\renewcommand{\arraystretch}{1.15}
\begin{tabular}{lll}
\hline\hline
Category & Quantity & $L$ $(\mathrm{erg~s^{-1}})$ \\
\hline
Spin-down
& $\dot E_{\rm int}$
& $(8.74^{+0.86}_{-0.85})\times10^{33}$ \\
\hline
Standard losses
& offset-dipole $L_{\rm dip}$
& $(2.18^{+5.87}_{-1.60})\times10^{33}$ \\
& dipole+quadrupole $L_{\rm dip}$
& $(2.28^{+6.15}_{-1.67})\times10^{33}$ \\
& GW ellipticity
& $(2.88^{+3.64}_{-2.07})\times10^{32}$ \\
& GW $r$-mode
& $(2.93^{+4.61}_{-2.15})\times10^{25}$ \\
\hline
Residual, $\Delta>0$
& offset-dipole branch
& $(6.54^{+1.51}_{-2.81})\times10^{33}$ \\
& dipole+quadrupole branch
& $(6.49^{+1.53}_{-2.84})\times10^{33}$ \\
\hline
Quadrupolar comparison
& offset-dipole $L_{\rm quad}$
& $(6.96^{+19.19}_{-5.11})\times10^{32}$ \\
& dipole+quadrupole $L_{\rm quad}$
& $(5.84^{+16.36}_{-4.32})\times10^{32}$ \\
\hline\hline
\end{tabular}
\end{table*}
\section{Refining neutron polarization}
In Chapter 2, we introduced the polarization, $\mathcal{P}$, of a neutron star that reflected the fraction of neutrons within the crust required such that their individual magnetic dipoles would sum to that of the overall star. For the purpose of investigating the potential of crust-confined field models, this was sufficient. However, using that fraction to calculate the nEDM warrants some further microphysical constraints. In particular, the neutrons relevant to the present problem are not just the total neutrons in the full crust, but the subset associated with the free-neutron component of the inner crust, weighted by the local magnetic field environment \cite{chamel_physics_2008, lattimer_nuclear_2012}.

Below the neutron drip density, matter is organized into nuclei embedded in an electron background. Above neutron drip, however, a fraction of the neutrons is no longer confined to nuclei and instead forms a neutron fluid coexisting with nuclear clusters. As described in Chapter 2, this defines the inner crust. The free-neutron population, therefore, is the dominant neutron population that can be modeled as an extended, bulk reservoir for polarization, whereas bound neutrons are mostly locked into nuclei whose internal spin structure strongly cancels any net alignment. This is not to say that free neutrons are trivially polarizable; rather, the free-neutron component is the relevant effective reservoir for a bulk polarization model.

A more physical counting begins from the local free-neutron number density,
\begin{equation}
n_{n,\mathrm{free}}(r) \;=\; x_{\mathrm{free}}(r)\,\frac{\rho(r)}{m_n},
\label{eq:nfree_def}
\end{equation}
where $\rho(r)$ is the mass density, $m_n$ is the neutron mass, and
\(x_{\mathrm{free}}(r)\) is the free-neutron fraction. To distinguish the
original global bookkeeping parameter $\mathcal{P} $ from the final
shell-resolved polarization weighting, we denote the latter by
$f_{\rm pol}$. The polarized-neutron count is then obtained by integrating
this reservoir over the inner crust with an appropriate local polarization
weighting,
\begin{equation}
N_{\mathrm{pol}}=\int_{R_{\mathrm{core}}}^{R_{\mathrm{drip}}}\!\int_{\Omega} n_{n,\mathrm{free}}(r)
f_{\rm pol}\!\big[B(r),\rho(r),T(r)\big]\, r^2\,d\Omega\,dr.
\label{eq:npol_full}
\end{equation}
Here, $f_{\rm pol}$ represents the local response factor that determines what
fraction of the available free-neutron reservoir contributes effectively under
the local field and matter conditions. In this form, the estimate makes
explicit that $N_{\rm pol}$ depends on an overlap: one must know both where
the free neutrons reside, and how the magnetic field is distributed through
those same layers.
\begin{figure}
    \centering
\includegraphics[width=.70\linewidth]{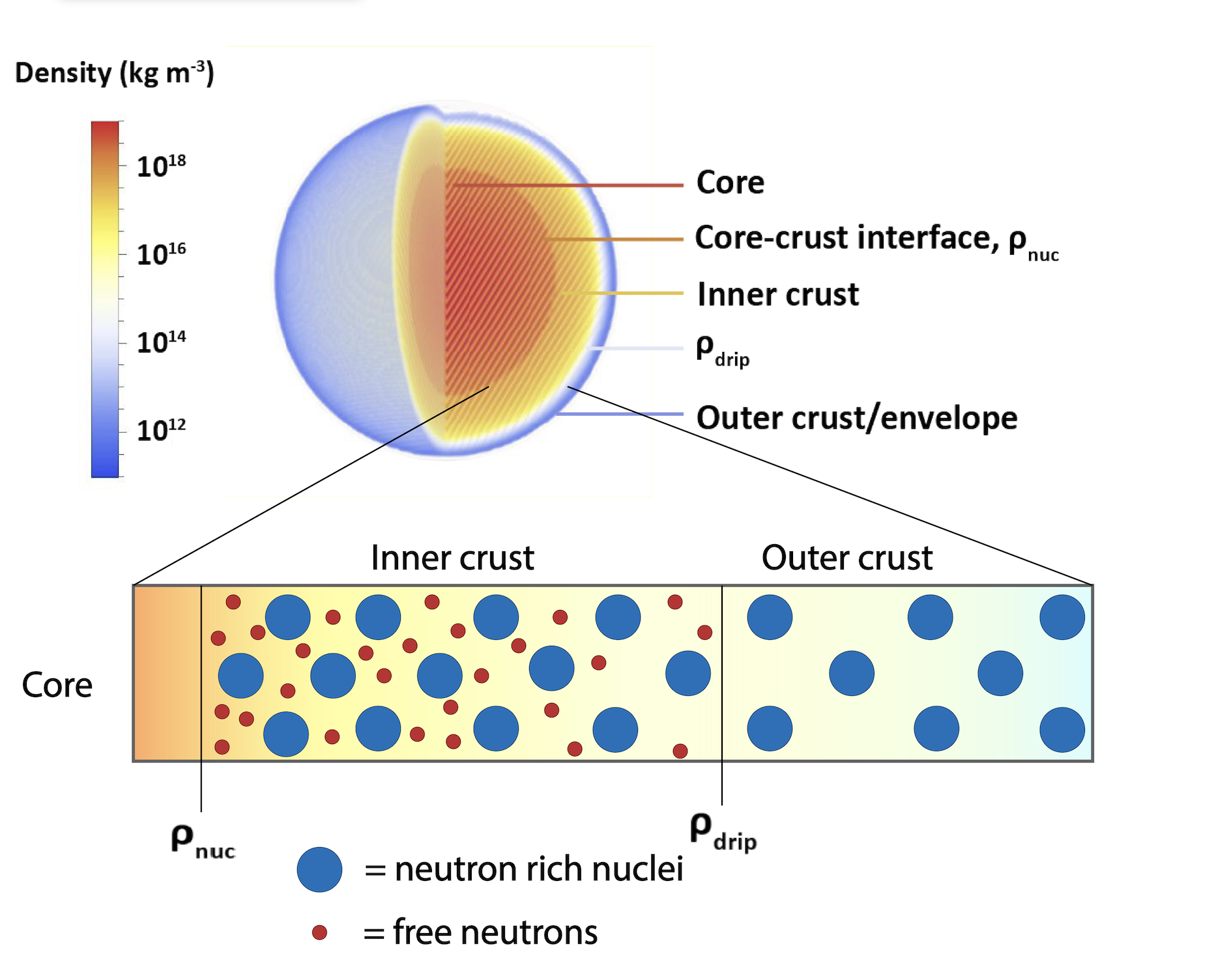}
    \caption{
Schematic structure of the neutron-star crust and core-crust interface. The upper panel shows the radial density profile, marking the core, inner crust, neutron-drip transition \(\rho_{\rm drip}\), and outer crust/envelope. The lower panel zooms in on the crust, where neutron-rich nuclei occupy the outer crust and free neutrons appear beyond \(\rho_{\rm drip}\). The crust joins the dense liquid core near nuclear saturation density, \(\rho_{\rm nuc}\).
}
\label{fig:neutron_star_crust_structure}
\end{figure}
For PSR J0437-4715, the adopted inner-crust profile gives a total free-neutron
reservoir
\begin{equation}
N_{n,\mathrm{free}}^{\mathrm{(ic)}} \;=\; 2.6989\times10^{55}.
\end{equation}
After propagating the uncertainties in the crustal structure and magnetic
geometry through the refined inner-crust Monte Carlo, the effective
participating reservoir for the fiducial offset-dipole case was found to be
\begin{equation}
N_{\rm eff,part} = 1.74^{+0.69}_{-0.67}\times10^{55},
\end{equation}
while the corresponding adjusted polarization fraction was
\begin{equation}
f_{\rm pol}^{\rm adj} = 1.26^{+0.50}_{-0.39}\times10^{-4}.
\end{equation}
This yields the final polarized-neutron count used in the subsequent nEDM
propagation,
\begin{equation}
N_{\rm pol} = 2.14^{+1.26}_{-0.84}\times10^{51}.
\label{eq:npol_final}
\end{equation}

The field now relevant for the polarized subset is a weighted mean,
\begin{equation}
\langle B\rangle_{\rm pol}= \frac{\int_{V_{\rm ic}} B(\mathbf{r}) n_{n,\mathrm{free}}(\mathbf{r}) f_{\rm pol}\left[B(\mathbf{r}),T(\mathbf{r})\right]\,
dV}{\int_{V_{\rm ic}}n_{n,\mathrm{free}}(\mathbf{r})\,f_{\rm pol}\!\left[B(\mathbf{r}),T(\mathbf{r})\right]\,
dV }.
\label{eq:Bpol_weighted}
\end{equation}
This quantity is not the large-scale dipole field inferred from the spin-down
torque, but the effective local field sampled by the polarized inner-crust
reservoir.

To emphasize the conceptual distinction from the torque calculation:
\begin{enumerate}
    \item \(\mu_{\rm dip}\) determines the large-scale electromagnetic luminosity and torque;
    \item \(B(\mathbf{r})\) determines the local polarization environment inside the crust;
    \item \(N_{\rm pol}\) determines how a bulk electric dipole moment would map onto the inferred neutron EDM.
\end{enumerate}
\paragraph{$N_{\rm pol}$ multipolar comparison}
For comparison, the dipole+quadrupole realization gives
$N_{\rm pol}=1.88^{+1.09}_{-0.73}\times10^{51}$, showing that the final
polarized reservoir remains of order $10^{51}$ across the non-dipolar comparison geometries explored here.
\section{Residual-power Monte Carlo and nEDM estimator}
Table~\ref{tab:torque_budget} summarizes the present-day luminosity budget. The force-free dipolar luminosity is the dominant modeled standard loss, while the gravitational-wave contributions are subdominant for the adopted priors.

The signed residual power available to an additional channel is
\begin{equation}
\Delta
\equiv
\dot E_{\rm int}-L_{\rm dip}-L_{\rm GW},
\label{eq:delta_ch5}
\end{equation}
where
\begin{equation}
\dot E_{\rm int}
=
-I\Omega\dot\Omega
=
4\pi^2I\frac{\dot P_{\rm int}}{P^3}.
\label{eq:edot_intrinsic_ch5}
\end{equation}
The Monte Carlo propagates the uncertainties in $I$, $\dot P_{\rm int}$, $P$, $B_{\rm surf}$, $f_{\ell=1}$, $R_\star$, $\alpha$, and the gravitational-wave inputs. Samples with $\Delta<0$ are not interpreted as negative exotic luminosity; they identify combinations for which the modeled standard losses exceed the measured spin-down budget.

As an unscreened benchmark, each sample with $\Delta>0$ is assigned entirely to coherent electric-dipole radiation,
\begin{equation}
L_{\rm ED}
=
\frac{2}{3c^3}p_0^2\Omega^4\sin^2\alpha_{\rm ED},
\label{eq:led}
\end{equation}
where $p_0$ is the collective stellar electric dipole participating in the channel. The benchmark assumes that this dipole is tied to the same magnetic/polarization axis used in the torque model, so that $\alpha_{\rm ED}=\alpha$. It is therefore inclined relative to the rotation axis; a dipole exactly aligned with the rotation axis would be stationary in the inertial frame and would not radiate through Eq.~\eqref{eq:led}. Identifying
\begin{equation}
p_0=N_{\rm pol}d_n,
\label{eq:p0_npol_dn}
\end{equation}
the signed estimator propagated in the Monte Carlo is
\begin{equation}
Q_{\rm signed}
\equiv
\frac{3c^3\Delta}
{2N_{\rm pol}^2\Omega^4\sin^2\alpha_{\rm ED}}.
\label{eq:Qsigned_ch5}
\end{equation}
For $\Delta>0$, Eq.~\eqref{eq:Qsigned_ch5} is the inferred value of $d_n^2$. For $\Delta<0$, it remains a signed diagnostic of how far a sampled standard-loss model lies beyond the available spin-down power. For compactness, $Q$ below denotes $Q_{\rm signed}$.

\begin{figure}
    \centering
    \includegraphics[width=1\linewidth]{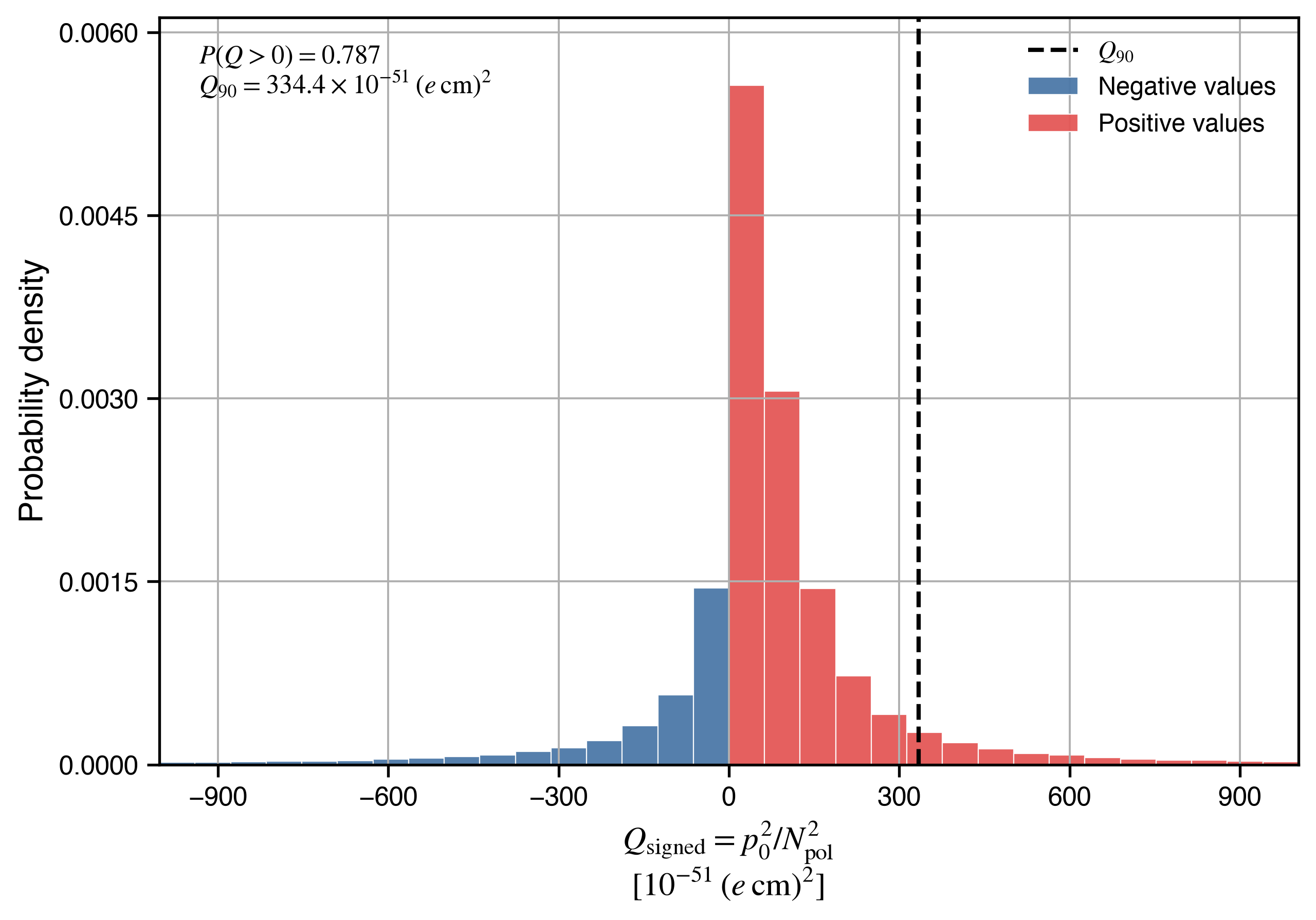}
    \caption{Histogram of the signed Monte Carlo estimator $Q=3c^3\Delta/[2N_{\rm pol}^2\Omega^4\sin^2\alpha_{\rm ED}]$. Negative and positive realizations are retained together to display the full propagated uncertainty relative to the physical boundary at $Q=0$. On the positive branch, $Q=d_n^2$. The dashed vertical line marks the empirical 90th percentile of the positive branch, $Q_{90}=3.344\times10^{-49}\,(e\,{\rm cm})^2$, corresponding to $d_n=5.783\times10^{-25}~e\,{\rm cm}$. The fraction of Monte Carlo realizations with positive residual power is $P(Q>0)=0.787$.}
    \label{fig:hist}
\end{figure}

\subsection{Positive-branch selection and reported upper limit}
The fraction of Monte Carlo draws with positive residual power is
\begin{equation}
P(Q>0)=0.787.
\label{eq:PQpos_ch5}
\end{equation}
The remaining 21.3\% of samples have $\Delta<0$ and are retained in the signed distribution as a diagnostic of tension between sampled standard-loss models and the measured spin-down budget. The reported benchmark is defined by the empirical 90th percentile of the positive branch,
\begin{equation}
P\!\left(Q\leq Q_{90}\mid Q>0\right)=0.90,
\label{eq:Q90_conditional_ch5}
\end{equation}
which gives
\begin{equation}
Q_{90}=3.344\times10^{-49}\,(e\,{\rm cm})^2.
\label{eq:Q90_value_ch5}
\end{equation}
Thus,
\begin{equation}
|d_n|_{\rm EDR}
<
\sqrt{Q_{90}}
=
5.783\times10^{-25}~e\,{\rm cm}.
\label{eq:dn_edr_ch5}
\end{equation}
Equation~\eqref{eq:dn_edr_ch5} is a conditional 90th-percentile upper-limit-style benchmark for the positive-residual branch, not a conventional 90\% confidence interval for the full signed model.

\subsection{Conclusion}
This chapter combines the source-specific moment of inertia, the propagation-informed surface-field posterior, the NICER-constrained harmonic decomposition, and the refined polarized-neutron reservoir in a single present-day energy budget. The electromagnetic torque is evaluated consistently from $B_{\rm surf}\sqrt{f_{\ell=1}}$, rather than from a fixed magnetic moment or from a spin-down-derived field. The two NICER-compatible geometry families yield nearly identical dipolar luminosities and residual-power distributions.

The full signed $Q$ distribution is retained because it displays both the physical positive-residual branch and the subset of sampled models for which the standard losses exceed the available spin-down power. Conditional on $Q>0$, the empirical 90th percentile gives the unscreened benchmark in Eq.~\eqref{eq:dn_edr_ch5}.

Most importantly, the conversion assumes that the collective electric dipole radiates without suppression by the conducting crust or the charge-filled magnetosphere. Chapter 6 therefore examines crustal and magnetospheric screening and develops a screening-aware magnetic-quadrupole interpretation of the same residual-power framework. The direct-EDR result obtained here is consequently an effective unscreened benchmark rather than a screening-independent constraint on the intrinsic neutron EDM.

%% file: chap6.tex
\chapter{Crustal and magnetospheric screening}

The neutron-star electric-dipole-radiation calculation presented in this
work should be interpreted as an effective unscreened estimate.  In the
idealized model, a fraction of the polarized inner-crust neutron reservoir
contributes coherently to a macroscopic electric dipole moment. However, this does not, by itself, include the response of the dense conducting crust or the charge-filled pulsar magnetosphere. For this reason, the result quoted in this work should be
understood as the bound obtained in the limit of no additional screening.
\section{Crustal screening}
There are two physically distinct screening effects.  The first is crustal screening: the response of the charged particles in the neutron-star crust to a microscopic or macroscopic electric dipole polarization. The crust is not an electrically empty medium.  It contains a
lattice of neutron-rich nuclei immersed in a highly degenerate electron gas, with free neutrons appearing in the inner crust above neutron drip
\cite{chamel_physics_2008, pethick_superfluid_2010}.  The electrons are mobile on microscopic length scales and therefore respond to electrostatic fields. In ordinary neutral systems this response is closely related to Schiff screening: a static electric dipole moment embedded in a neutral, nonrelativistic, electrostatic system is screened by rearrangement of the surrounding charges, leaving only finite-size and relativistic corrections such as the Schiff moment \cite{webster_schiffs_1963,liu_atomic_2007}. The neutron-star crust is obviously not an atom, but the same physical warning applies: a permanent electric dipole moment placed in a conducting or polarizable charged medium does not necessarily produce the full unscreened external electric field that a vacuum calculation would imply.

\subsubsection{Phenomenological screening factors}
In the present model this uncertainty can be represented by a phenomenological crustal screening factor \(S_{\rm crust}\). If the vacuum calculation requires an electric dipole moment \(D_{\rm}\), but the dense crust allows only a fraction of that moment to participate coherently in the radiating field, then the observable effective moment may be written as
\begin{equation}
    D_{\rm rad}
    =
    \frac{D_{}}{S_{\rm crust}} ,
    \qquad
    S_{\rm crust}\geq 1 .
\end{equation}
Equivalently, the inferred microscopic neutron EDM scales as
\begin{equation}
    d_n(S_{\rm crust})
    =
    S_{\rm crust}\,
    d_n^{(0)} ,
\end{equation}
where \(d_n^{(0)}\) is the value inferred under the unscreened assumption. Thus crustal screening weakens the microscopic interpretation of the constraint.  The value quoted in this thesis corresponds to \(S_{\rm crust}=1\), and should therefore be read as an effective local
upper bound rather than as a fully self-consistent calculation of electric fields inside a conducting neutron-star crust.
\subsubsection{Thomas-Fermi screening and coherence of the polarized neutron reservoir}

A more microscopic way to phrase crustal screening is in terms of the Thomas-Fermi screening length of the degenerate electron gas. The inner crust contains neutron-rich nuclei, dripped neutrons, and a highly degenerate background of electrons.  Electrostatic perturbations in such a medium are not transmitted over arbitrary distances. Instead, mobile electrons rearrange over a characteristic Thomas-Fermi length,
\begin{equation}
    \lambda_{\rm TF}
    =
    k_{\rm TF}^{-1},
\end{equation}
where, in linear response,
\begin{equation}
    k_{\rm TF}^{2}
    =
    4\pi e^{2}\frac{\partial n_{e}}{\partial \mu_{e}} .
    \label{eq:tf_general}
\end{equation}
Here \(n_e\) is the electron number density and \(\mu_e\) is the electron chemical potential.  This expression makes clear that screening is controlled by the compressibility of the electron gas: a highly degenerate electron fluid can rearrange charge efficiently, thereby reducing long-range electrostatic fields. Thomas-Fermi screening is commonly used in neutron star crust calculations to describe how the degenerate electron background screens Coulomb interactions between ions \cite{chamel_physics_2008,chugunov_thermal_2007}.

This matters because the electric-dipole-radiation estimate assumes that the polarized neutron population contributes coherently to one macroscopic effective dipole moment,
\begin{equation}
    D_{\rm eff}^{(0)}
    =
    N_{\rm pol} d_n .
\end{equation}
This is the maximally coherent limit. If the electron response screens the electric field over a length scale much smaller than the global stellar scale, then different regions of the polarized crust may no longer add as a single phase-coherent electric dipole.  In that case, the relevant question is not only how many neutrons are polarized, but how many of them contribute coherently to the far-zone radiating multipole.

A useful worst-case way to parameterize this uncertainty is to replace the fully coherent sum by a random-walk sum over incoherent screened domains. If all \(N_{\rm pol}\) microscopic dipoles add with the same sign and phase, the effective dipole scales as
\begin{equation}
    D_{\rm coh}
    \sim
    N_{\rm pol} d_n .
\end{equation}
If, however, Thomas-Fermi screening breaks the crust into many domains whose local contributions have uncorrelated phases or signs, then the net moment scales only as the root-mean-square fluctuation,
\begin{equation}
    D_{\rm rms}
    \sim
    \sqrt{N_{\rm pol}}\, d_n .
    \label{eq:sqrt_npol_limit}
\end{equation}
This does not mean that the physical number of polarized neutrons becomes \(\sqrt{N_{\rm pol}}\). Rather, it means that the coherently radiating number of polarized neutrons may be reduced from \(N_{\rm pol}\) to an effective root-mean-square value. This expression should be interpreted only as an extreme loss-of-coherence diagnostic, not as a reported bound, because the actual coherence length and domain structure have not been modeled. In the most pessimistic screening-and-incoherence limit, this can be represented schematically as
\begin{equation}
    N_{\rm pol}
    \longrightarrow
    N_{\rm pol}^{\rm eff}
    \sim \sqrt{N_{\rm pol}} .
\end{equation}
With this replacement, the inferred microscopic neutron EDM becomes
\begin{equation}
    d_n^{\rm incoh}
    \sim
    \frac{N_{\rm pol}}{\sqrt{N_{\rm pol}}}
    d_n^{(0)}
    =
    \sqrt{N_{\rm pol}}\,d_n^{(0)} ,
\end{equation}
where \(d_n^{(0)}\) is the unscreened, fully coherent value. This is an extreme limit, not the default assumption. It is included to show how sensitive the microscopic interpretation of the constraint is to the coherence assumption. The result quoted earlier corresponds to the opposite limit, namely that the polarized crustal reservoir sources a single effective dipole moment.
\subsection{Global neutrality model consequences}
A related caveat concerns the assumption of strict local charge neutrality. In the simplest treatment, the crust is taken to be locally neutral, so that the proton and electron charge densities cancel point by point and no large internal electric field is introduced.  In models that impose only global charge neutrality, however, the total stellar charge still vanishes, but small local charge separations can exist, especially near the core-crust interface. Such a charge separation would carry an electrostatic energy cost and would have to be balanced against the nuclear, gravitational, and composition-dependent energies of the star.

This matters for the present calculation because the effective polarized neutron number is not purely a geometric count. If an interface electric field is present, the energetically favored orientation of neutron electric dipoles could depend on position within the star. In principle, this could make the effective polarized population slightly different in the two magnetic hemispheres: one side of the star could contain a larger energy-favored aligned population, while the opposite side could contain a smaller one.  This should not be interpreted as a prediction of an enhancement or suppression.  Rather, it is a reminder that the mapping between the microscopic neutron EDM and the macroscopic radiating dipole depends on the assumed charge-neutrality model.

For this reason, the calculation in this work does not attempt to assign a numerical correction for global-neutrality effects. The fiducial result assumes local neutrality and treats the polarized neutron population as a single effective quantity.  A self-consistent global-neutrality treatment would require solving for the proton, electron, and neutron distributions together with the electric field at the core-crust interface, and could modify the effective value of the polarized neutron population used in the dipole estimate.
\section{Magnetospheric screening}
The second effect is magnetospheric screening. A pulsar is not surrounded by vacuum. In the standard Goldreich-Julian picture, rotation of a magnetized neutron star induces electric fields strong enough to pull charges from the stellar surface and fill the magnetosphere with plasma \cite{goldreich_pulsar_1969}. The characteristic charge density required for corotation is
\begin{equation}
    \rho_{\rm GJ}
    \simeq
    -\frac{\boldsymbol{\Omega}\cdot\mathbf{B}}{2\pi c},
\end{equation}
up to geometry-dependent corrections. In a charge-filled magnetosphere, plasma currents rearrange to screen electric fields parallel to the magnetic field, regulate particle acceleration, and set the global current structure. This is why modern pulsar spin-down models differ from the vacuum magnetic dipole formula.

For the electric-dipole-radiation constraint, magnetospheric screening means that the outgoing electromagnetic field associated with a crustal electric dipole may be modified, reduced, or reprocessed by the surrounding pair plasma before it contributes to the far-zone luminosity.  This effect can be encoded by a second phenomenological factor \(S_{\rm mag}\), so that
\begin{equation}
    L_{\rm EDR}^{\rm obs}
    =
    \frac{1}{S_{\rm mag}^2}
    L_{\rm EDR}^{\rm vac}.
\end{equation}
Because the inferred dipole moment scales as the square root of the available luminosity, the neutron EDM inferred from a given residual torque
then scales as
\begin{equation}
    d_n(S_{\rm crust},S_{\rm mag})
    =
    S_{\rm crust} S_{\rm mag}
    d_n^{(0)} .
    \label{eq:screened_dn_scaling}
\end{equation}
Equation~\ref{eq:screened_dn_scaling} makes explicit why the bound reported in this work is conservative only within the chosen unscreened model. If either the dense crust or the magnetosphere substantially suppresses the radiating electric dipole field, then the same observed residual torque would correspond to a larger underlying microscopic neutron EDM.

This distinction is also useful when comparing electric and magnetic multipoles. Electrostatic moments are especially sensitive to charge screening. Schiff's theorem shows that a nuclear EDM is screened in a neutral electrostatic system, with observable effects arising only through
corrections such as finite nuclear size and the Schiff moment \cite{webster_schiffs_1963,liu_atomic_2007}. Electric octupole moments are likewise modified by screening corrections, while magnetic quadrupole moments are not screened in the same way because they couple magnetically rather than through a static electrostatic potential \cite{lackenby_time_2018}. Therefore, screening arguments that apply to electric dipole or electric octupole fields should not be automatically applied to magnetic quadrupole
contributions.

In summary, crustal screening refers to the microscopic response of dense charged matter inside the neutron-star crust, while magnetospheric screening refers to the plasma response of the exterior pulsar magnetosphere. The calculation in this thesis adopts the \(S_{\rm crust}=S_{\rm mag}=1\) limit. The resulting constraint is therefore best described as an effective
unscreened, local-neutrality neutron-EDM bound. A fully self-consistent calculation of \(S_{\rm crust}\) and \(S_{\rm mag}\) would require modeling both the electrodynamic response of the inner crust and the propagation of the induced multipolar field through a charge-filled, current-carrying pulsar magnetosphere.

\section{Neutron EDM constraint from the magnetic quadrupole channel}
\label{sec:mqm_to_nedm}

The screening discussion above motivates treating the electric-dipole
estimate with care.  A static electric dipole embedded in conducting or
polarizable matter can be strongly modified by charge rearrangement in
the crust and by plasma response in the magnetosphere.  The magnetic
quadrupole channel provides a complementary way to connect the same
neutron-star spin-down budget to a CP-odd neutron observable.  In this
case the constrained microscopic quantity is not the neutron EDM itself,
but an effective single-neutron magnetic quadrupole moment \(M_n^0\).  The
result can then be translated, under an assumed CP-violating source, into
a bound on the QCD angle \(\bar{\theta}\) and into a corresponding
pure-\(\bar{\theta}\) comparison value for \(d_n\).

The physical reason to consider this channel is that magnetic quadrupole
moments are not screened by the same electrostatic Schiff-screening
argument that suppresses static electric dipole moments in neutral
systems.  Nuclear electric dipole moments couple to electric fields and
are screened by the surrounding charge distribution, while magnetic
quadrupole moments couple magnetically and can generate observable
\(P,T\)-violating effects \cite{lackenby_time_2018}.  The electric
octupole channel is also not used here as an unscreened replacement,
because electric octupole fields receive screening corrections analogous
to Schiff-moment corrections \cite{flambaum_extension_2012}.  The magnetic
quadrupole channel is therefore the natural non-dipolar CP-odd
electromagnetic moment to test with the spin-down energy budget.

\subsection{Spin-down power assigned to the MQM channel}
\label{subsec:quad_power_to_m0n}

The residual power available to a nonstandard radiative channel is just as described in Chapter 5:
\begin{equation}
    \Delta
    =
    \dot{E}_{\rm int}
    -
    L_{\rm dip}
    -
    L_{\rm GW}.
    \label{eq:delta_mqm_chap}
\end{equation}
The difference is that now we attribute the residual power to a neutron MQM. We use the same positive-residual branch as in Chapter~5, for which
$P(\Delta>0)=0.787$, and assign
\begin{equation}
P_{\rm MQM}=\Delta
\end{equation}
for each realization satisfying $\Delta>0$.

The quadrupolar electromagnetic luminosity inferred directly from the
surface-field decomposition is retained only as a comparison luminosity.
The primary bound uses the residual-power construction in
Eq.~\eqref{eq:delta_mqm_chap}, because it asks how much present-day
spin-down power remains after the modeled standard losses are subtracted.
This is the more conservative upper-limit interpretation.

\subsection{Conversion from residual power to \(M_n^0\)}
\label{subsec:m0n_conversion_chap}

The MQM-like luminosity is parameterized as radiation from an effective
oscillating magnetic quadrupole,
\begin{equation}
    P_{\rm MQM}
    =
    \frac{K_{\rm quad}\omega_{\rm rad}^{6}}{c^5}
    \left(f_{\ell=2}N_{\rm pol}M_n^0\right)^2 .
    \label{eq:PMQM_chap}
\end{equation}
Here \(M_n^0\) is the effective single-neutron magnetic quadrupole moment,
\(N_{\rm pol}\) is the effective aligned inner-crust neutron reservoir,
and \(f_{\ell=2}\) is the centered quadrupolar fraction obtained from the
surface-field decomposition.  For the radiation convention used in this work, \(K_{\rm quad}=1/(60\pi)\).  The factor \(f_{\ell=2}\) should be
read as a geometry-weighted participation prescription, not as a
microscopic derivation of quadrupolar ordering in dense matter.

The radiation formula requires a characteristic angular frequency for the
effective quadrupolar electromagnetic mode. This frequency is not directly
measured as an MQM spectral line, nor is it assumed to describe the rigid
rotation of a static body-fixed quadrupole. We therefore introduce an
effective mode frequency,
\begin{equation}
    \omega_{\rm rad}=\omega_{\rm eff}=2\pi\nu_{\rm eff}.
    \label{eq:omega_eff_mqm}
\end{equation}
For the fiducial source-specific calculation, this frequency is identified
with the OPM-core transition scale measured in Chapter~3,
\begin{equation}
    \nu_{\rm eff}\equiv\nu_{\rm OPM}
    =
    2.472\text{--}2.680~{\rm GHz}.
    \label{eq:nueff_opm}
\end{equation}
The motivation is that $\nu_{\rm OPM}$ is the observed magnetoionic scale
at which the outgoing polarization state approaches the stable
high-frequency branch and at which the propagation-based surface-field
normalization is fixed. It therefore provides a source-specific
characteristic frequency for the effective magnetospheric mode used in
the MQM-like luminosity model.

This identification is a phenomenological mode prescription rather than
a microscopic derivation of coherent CP-odd neutron motion at GHz
frequencies. The resulting constraint should therefore be interpreted as
conditional on $\nu_{\rm eff}=\nu_{\rm OPM}$.

Setting \(P_{\rm MQM}=\Delta\) on the positive residual branch gives
\begin{equation}
    |M_n^0|
    =
    \frac{1}{f_{\ell=2}N_{\rm pol}}
    \left[
        \frac{\Delta c^5}
        {K_{\rm quad}\omega_{\rm rad}^{6}}
    \right]^{1/2}.
    \label{eq:M0n_resid_chap}
\end{equation}
The corresponding decomposition-only comparison replaces \(\Delta\) by
the quadrupolar luminosity inferred from the same surface-field
decomposition.  It is useful as an internal consistency check, but it is
not adopted as the primary result.

The effective MQM is translated to hadronic CP violation using \cite{lackenby_time_2018}
\begin{equation}
    M_n^0
    =
    2.5\times10^{-29}\,\bar{\theta}\ e\,{\rm cm}^{2},
    \label{eq:M_to_theta}
\end{equation}
and the pure-\(\bar{\theta}\) neutron-EDM comparison is obtained from \cite{chupp_electric_2019}
\begin{equation}
    d_n
    =
    1.48\times10^{-16}\,\bar{\theta}\ e\,{\rm cm}.
    \label{eq:theta_to_dn}
\end{equation}
The EDM value reported below is therefore conditional: it is the value
that would correspond to the same \(\bar{\theta}\) source, not a direct
measurement of the neutron EDM.

\subsection{Numerical results}
\label{subsec:m0n_quad_results}

The main result is the residual-power upper limit on the effective
single-neutron MQM,
\begin{equation}
    |M_n^0|_{90}^{\rm resid}
    =
    7.47\times10^{-38}\ e\,{\rm cm}^{2}.
    \label{eq:M0n90_resid_final}
\end{equation}
Using Eq.~\eqref{eq:M_to_theta}, this gives
\begin{equation}
    |\bar{\theta}|_{90}^{\rm resid}
    =
    2.99\times10^{-9},
    \label{eq:theta90_resid_chap}
\end{equation}
and the corresponding pure-\(\bar{\theta}\) neutron-EDM comparison value is
\begin{equation}
    |d_n|_{90}^{\rm resid}
    =
    4.42\times10^{-25}\ e\,{\rm cm}.
    \label{eq:dn90_resid_final}
\end{equation}

The decomposition-based quadrupolar-power calculation gives the comparison
limit
\begin{equation}
    |M_n^0|_{90}^{\rm quad}
    =
    4.29\times10^{-38}\ e\,{\rm cm}^{2},
    \label{eq:M0n90_quad_comparison_chap}
\end{equation}
which corresponds to
\begin{equation}
    |\bar{\theta}|_{90}^{\rm quad}
    =
    1.72\times10^{-9},
\end{equation}
and
\begin{equation}
    |d_n|_{90}^{\rm quad}
    =
    2.54\times10^{-25}\ e\,{\rm cm}.
    \label{eq:dn90_quad_comparison_chap}
\end{equation}
Although this comparison bound is numerically stronger, the residual-power
result is adopted as the primary constraint because it is the direct
energy-budget upper limit on the maximum additional luminosity allowed by
the present-day spin-down.

\subsection{Uncertainty budget}
\label{subsec:m0n_quad_uncertainty}

The dominant uncertainty is the aligned neutron reservoir.  This follows
directly from Eq.~\eqref{eq:M0n_resid_chap}, since
\begin{equation}
    M_n^0 \propto N_{\rm pol}^{-1}.
\end{equation}
The next-largest uncertainty is the propagation-inferred magnetic-field
normalization.  It affects the standard force-free torque subtraction
through \(L_{\rm dip}\propto B_{\rm dip}^{2}\), and therefore changes both
the size of the residual and the probability that a Monte Carlo sample has
\(\Delta>0\).  The field-decomposition uncertainty is smaller because the
successful low-order geometry families give similar dipolar and
quadrupolar fractions.  The gravitational-wave priors and the effective
mode frequency are subdominant at the present precision.

\begin{table}[h!]
\centering
\caption{Dominant numerical uncertainty contributions for the
residual-power \(M_n^0\) result.  Entries are absolute contributions to
the final \(M_n^0\) uncertainty in units of \(e\,{\rm cm}^{2}\).}
\label{tab:m0n_residual_uncertainty_absolute}
\begin{tabular}{lc}
\hline
Source & Contribution \((e\,{\rm cm}^{2})\) \\
\hline
\(N_{\rm pol}\) & \(3.88\times10^{-38}\) \\
\(B_{\rm surf}\) & \(1.28\times10^{-38}\) \\
Field decomposition & \(5.73\times10^{-39}\) \\
\(\dot{E}_{\rm int}\) & \(4.58\times10^{-39}\) \\
\(R\) & \(8.33\times10^{-39}\) \\
\(\alpha\) & \(3.72\times10^{-40}\) \\
\(L_{\rm GW,\epsilon}\) & \(1.59\times10^{-39}\) \\
\(L_{\rm GW,r}\) & \(2.00\times10^{-46}\) \\
\(\omega_{\rm rad}\) & \(1.41\times10^{-40}\) \\
\hline
\end{tabular}
\end{table}

\subsection{Interpretation}
\label{subsec:m0n_quad_interpretation}

The result in Eq.~\eqref{eq:dn90_resid_final} is weaker than the direct
laboratory neutron-EDM bound.  Its significance is therefore
methodological rather than competitive: it demonstrates that a
source-specific neutron-star energy budget can constrain an nEDM-related
CP-odd neutron moment using a macroscopic astrophysical observable.  The
method combines radio-polarization propagation, NICER surface geometry,
timing, and neutron-star structure while avoiding the circular use of
\(B_{\rm sd}\) to normalize the magnetic field.

Several qualifications are important.  The quantity directly constrained
is the effective magnetic quadrupole moment \(M_n^0\), not \(d_n\).  The
reported \(d_n\) value is obtained only after assuming that both the MQM
and the EDM arise from the same pure-\(\bar{\theta}\) source.  The
calculation also assumes that the surface quadrupolar fraction can be used
as a geometry-weighted participation factor for the aligned neutron
reservoir and that the relevant electromagnetic mode is characterized by
the OPM-core transition scale.  These assumptions make the result an
effective, source-specific upper limit rather than a complete microscopic
calculation of CP-odd ordering in dense matter.

Future improvements would mainly require a better calculation of
\(N_{\rm pol}\), a predictive magnetospheric transfer model for
\(B_{\rm surf}\), and a microscopic mapping between the observed
surface-field decomposition and coherent neutron participation.  Larger
theory uncertainties in the \(M_n^0/\bar{\theta}\) or
\(d_n/\bar{\theta}\) conversion factors should be propagated separately as
model systematics.

%% file: chap7.tex
\chapter{Conclusion}

This thesis investigated whether the spin-down of a neutron star
can be used to constrain the neutron electric dipole moment. The
central idea was to treat PSR~J0437--4715 as a source-specific
physical system rather than relying only on its measured period and
period derivative. By combining stellar-structure modeling,
wideband radio polarization, NICER hot-region constraints, and a
present-day torque budget, the analysis estimated the rotational
power available to a possible CP-odd radiative channel and the
effective neutron population that could participate in that
channel.

The original method developed in Chapters 2--5 was based on direct
electric-dipole radiation. A permanent neutron EDM would define an
electric direction aligned with the neutron spin. If a sufficiently
large population of polarized neutrons contributed coherently, the
star could then possess an effective bulk electric dipole whose
radiation would draw from the rotational-energy budget. Under this
model, the observed intrinsic spin-down power was compared with the
modeled conventional losses, and any remaining positive residual was
assigned to an electric-dipole-radiation channel.

The first requirement for this calculation was a source-specific
model of the star. The measured mass and radius of
PSR~J0437--4715 were used to construct a reduced-form density
profile resolving the core, inner crust, and outer crust. This model
provided the moment of inertia entering the spin-down luminosity and
the inner-crust free-neutron inventory entering the collective
dipole estimate. Weighting that inventory by the adopted magnetic
polarization response gave
\begin{equation}
N_{\rm pol}
=
2.14^{+1.26}_{-0.84}\times10^{51}.
\end{equation}
This quantity represents the effective aligned-neutron reservoir
used to map a macroscopic CP-odd moment to a neutron-level
coefficient.

A second requirement was a magnetic-field normalization that did
not come from the same spin-down torque being modeled. Wideband
radio-polarization observations were used to identify a transition
from low-frequency orthogonal-mode mixing toward a stable
high-frequency position-angle structure. The folded-profile
diagnostics bracketed this transition at
\begin{equation}
\nu_{\rm OPM}
=
2.472\text{--}2.680~{\rm GHz}.
\end{equation}
Under the adopted propagation-height, pair-plasma, and viewing
priors, the transition was mapped to the surface-field posterior
\begin{equation}
B_{\rm surf}
=
3.08^{+5.14}_{-1.93}\times10^8~{\rm G}.
\end{equation}
Because this field was not normalized by assigning the observed
spin-down to a magnetic torque, it could be used in the subsequent
energy budget without making the calculation circular.

The surface-field geometry was constrained independently using the
published NICER hot-region posterior. Interpreting the emitting
regions as return-current-heated polar caps links their
non-antipodal locations to the near-surface magnetic topology. A
centered dipole could not reproduce the observed configuration,
whereas both an offset dipole and a dipole plus a general
quadrupole were compatible with the hot-region posterior. A
centered spherical-harmonic decomposition of the accepted maps
gave
\begin{align}
f_{\ell=2}^{\rm off.dip.}
&=
0.232^{+0.046}_{-0.039},\\
f_{\ell=2}^{\rm dip.+quad.}
&=
0.183^{+0.113}_{-0.041}.
\end{align}
The corresponding dipolar components were used in the conventional
electromagnetic torque, while the quadrupolar components later
provided the geometry factor for the MQM interpretation.

Using the source-specific moment of inertia, propagation-informed
field, and NICER-compatible geometries, the intrinsic spin-down
luminosity was compared with the electromagnetic and
gravitational-wave losses. The residual was defined as
\begin{equation}
\Delta
=
\dot E_{\rm int}-L_{\rm dip}-L_{\rm GW}.
\end{equation}
In the original EDR calculation, the physically allowed residual
was assigned to coherent radiation from a bulk electric dipole
$p_0=N_{\rm pol}d_n$. The full signed Monte Carlo distribution was
retained to show realizations on both sides of the physical
boundary. Restricting the reported upper limit to the positive
branch gave
\begin{equation}
P(Q>0)=0.787
\end{equation}
and
\begin{equation}
\boxed{
|d_n|_{\rm EDR}
<
5.783\times10^{-25}~e\,{\rm cm}
}
\end{equation}
at the 90\% level.

This value is the direct result of the Chapter 5 model, but its
physical interpretation depends on an important assumption: that
the collective electric dipole radiates without additional
suppression by the surrounding charged matter. Chapter 6 therefore
examined the response of the dense conducting crust and the
charge-filled pulsar magnetosphere. Mobile electrons in the crust
can rearrange in response to an electrostatic field, while the
magnetospheric pair plasma can modify or reprocess the outgoing
electric field. The Chapter 5 result should consequently be
understood as an effective unscreened EDR bound rather than a
screening-independent constraint on the intrinsic neutron EDM.

The screening analysis motivated a second application of the same
spin-down framework. A magnetic quadrupole moment is a
$P$- and $T$-odd electromagnetic moment and is therefore sensitive
to CP-violating neutron structure, but it is not suppressed by the
same electrostatic screening argument that complicates a static
electric dipole. The final Chapter 6 calculation therefore assigned
the positive residual spin-down power to an effective MQM-like
radiative channel.

The effective collective moment was modeled as
\begin{equation}
\mathcal M_{\rm bulk}
=
f_{\ell=2}N_{\rm pol}M_n^0,
\end{equation}
where $M_n^0$ is the effective CP-odd neutron MQM coefficient and
$f_{\ell=2}$ weights the aligned-neutron reservoir by the
quadrupolar surface-field geometry. The corresponding radiation
model was
\begin{equation}
P_{\rm MQM}
=
\frac{K_{\rm quad}\omega_{\rm rad}^{6}}{c^5}
\left(
f_{\ell=2}N_{\rm pol}M_n^0
\right)^2,
\end{equation}
with $K_{\rm quad}=1/(60\pi)$ and
$\omega_{\rm rad}=2\pi\nu_{\rm OPM}$. On the positive-residual
branch,
\begin{equation}
P(\Delta>0)=0.787.
\end{equation}
Assigning the available residual power conservatively to the
MQM-like channel gave
\begin{equation}
\boxed{
|M_n^0|
<
7.47\times10^{-38}~e\,{\rm cm^2}
}
\end{equation}
at the 90\% level.

Under the additional hypothesis that the effective neutron MQM is
generated entirely by the QCD $\bar\theta$ term, this becomes
\begin{equation}
|\bar\theta|
<
2.99\times10^{-9}.
\end{equation}
Using the separate pure-$\bar\theta$ relation for the neutron EDM
then gives
\begin{equation}
|d_n|_{\bar\theta}
<
4.42\times10^{-25}~e\,{\rm cm}.
\end{equation}

The two pathways therefore answer related but distinct questions.
The Chapter 5 calculation determines what neutron EDM would be
allowed if the residual spin-down power were emitted coherently as
unscreened electric-dipole radiation. Chapter 6 shows why that
interpretation is sensitive to the electrodynamic response of the
crust and magnetosphere and develops a magnetic-quadrupole
alternative that avoids the same electrostatic-screening problem.
The MQM analysis directly constrains $M_n^0$; its limits on
$\bar\theta$ and $d_n$ require additional hadronic assumptions.

Neither result is competitive with the present laboratory nEDM
limit. The significance of the work is instead methodological. It
demonstrates how neutron-star structure, radio propagation,
X-ray-inferred surface geometry, and precision timing can be
combined to translate a macroscopic rotational-energy budget into
constraints on microscopic CP-odd neutron properties. It also
identifies the assumptions that must be improved before such an
astrophysical constraint can become more robust: the
polarization-limiting-radius and pair-plasma priors, the
aligned-neutron reservoir, the coherence of the collective source,
the mapping between surface quadrupolar geometry and neutron
participation, and the coupling of that source to the outgoing
magnetospheric mode.

The work should therefore be viewed as a development and physical
assessment of two related neutron-star methods. The direct EDR
calculation establishes the original spin-down-to-nEDM framework
and exposes its sensitivity to screening. The MQM calculation
retains the same energy-budget strategy while providing a
screening-aware route to an effective CP-odd neutron moment.
Together, the two analyses establish a foundation for future
source-specific tests of neutron CP violation using astrophysical
systems.
\section{Opportunities for improvement}

\subsection{Effective magnetic torque}
 With the nEDM scaling as $d_n \propto \sqrt{\Delta}$, any reduction in the uncertainty of the residual power has a square-root payoff in the final limit. For example, a reduction of $\Delta$ by a factor of 10 would improve the bound by a factor $\approx 3.2$. Further, many elements within $\Delta$ have potential for improvement. The independent magnetic field estimate could be improved with improved single-pulse polarimetry, stronger constraints on the polarization-limiting radius, more explicit treatment of radio geometry, and a more precise mapping between the measured polarized variability and the field component that eventually enters the torque. The large fraction of unphysical Monte Carlo draws suggests that the fiducial force-free normalization may over-assign the available spin-down power for this source, or that the effective torque lies between the idealized force-free and vacuum prescriptions. 

 A more complete treatment of the magnetic-field surface geometry could also tighten the nEDM upper bound. The hotspot comparison in this work favors an offset-dipole-like morphology and does not statistically require higher-order multipoles. This is significant because the largest uncertainty is unlikely to arise from an octupolar correction, but how the preferred large-scale asymmetric field maps onto the open-field line region and torque-producing magnetic moment. Therefore, while a more thorough investigation of field geometry may improve the nEDM constraint, it is likely by only a few percent unless the geometric constraints offer an improvement in the torque normalization itself. 
\subsection{ Structure constraints}
There is also strong potential for improvements within the effective polarized-neutron count $N_{\rm pol}$. In this work, the relevant neutron population is the free-neutron component of the inner crust weighted by the local polarization response. This is physically better motivated than using the entirety of the star, but it introduces microphysical uncertainty through the free-neutron fraction, the field distribution through the inner crust, and the response factor $f_{\rm pol}$. Any increase in the reliably participating polarized reservoir translates linearly into a tighter nEDM bound. Thus, theoretical improvements in the inner crust microphysics could cause significant improvements within the nEDM upper bound.

Structural constraints can also be applied to the overall $\dot E$ calculation via the moment of inertia. However, the rather tight constraints on the mass and radius values do not provide much room for improvement. Even a generous $10\%$ improvement in the moment of inertia would only lead to a $\sim 5 \%$ improvement on the nEDM bound. 

\subsection{Potential benefits of studying alternative stars}

One of the most significant uncertainties lies in the choice of star. PSR J0437-4715 is ideal as it is nearby, bright, and unusually well characterized. However, that does not indicate that it is the best star for an astrophysical constraint. Stars at varying ages have varying dominant energy loss channels, magnetic fields, structure, and cooling and heating mechanisms. PSR J0437-4715's age offers the advantage of assumptions: many processes have slowed or halted on the Gyr scale, allowing us to ignore significant cooling or magnetic field evolution. Further, the low magnetic field of this star allowed for a crust-exclusive magnetic field model to be a rather safe assumption.

Other younger stars with stronger magnetic fields offer the benefit of potentially multiple orders of magnitude of improvement on the $N_{\rm pol}$ and thus nEDM constraint. However, many current works suggest that with higher magnetic field stars, which are not old enough for one to assume a core magnetic field contribution has dissipated, the assumption of a crust-exclusive field model is not justifiable. Generally, the uncertainties of the effective torque model elements would vastly increase with other stars. However, with  $d_n \propto \sqrt{\Delta}/N_{\rm pol}$, there may be a situation in which the improvement in the effective neutron reservoir may exceed the increased uncertainty in $\Delta$. This would not be the first circumstance in which a star with worse observational constraints offers a better microphysical constraint. In Appendix \ref{app:axion}, we attempted to expand our work with PSR J0437-4715 to constrain the axion mass similarly to Buschmann et al. (2022) \cite{buschmann_upper_2022}. However, they specifically used stars with ages around a Myr, meaning they had not yet entered the recycled millisecond-pulsar phase, but the photon and neutrino luminosities were comparable. Further, while they did not have the vast range of observational constraints that PSR J0437-4715 has, the luminosity and age constraints were sufficient. The benefit of the cooling environment of this star exceeded the uncertainties that came with a lack of structure constraints.

Further, with the work done in Chapter 6 there is a motivation to look into stars whose magnetic field geometry cannot be explained by a simple centered dipole. A star with large quadrupolar components would increase the value of effective polarized neutrons. The magnetic field origins of neutron stars are not well understood. However, the connection between a core-supported quadrupole and the externally observable surface field is not direct. The core is expected to be highly conducting and may contain superconducting protons, so magnetic flux can evolve on very long 
timescales and may be coupled to the crust only through the crust-core boundary.  For this reason, a quadrupolar field inferred from surface hot-spot structure or  near-surface emission geometry should be interpreted as evidence for a non-dipolar  near-surface magnetic geometry, but not necessarily as unambiguous evidence that the core magnetic field itself is quadrupole-dominated. This may be a reason to expand investigations to low-field magnetars, or other stars with larger magnetic fields than MSPs.

This work opens an astrophysical route to constraining the neutron EDM using macroscopic observables of neutron stars. While the present numerical bound is not competitive with the laboratory searches, the method is physically distinct and thus valuable. It probes a different environment, a different systematic regime, and an extremely different neutron ensemble. For those reasons alone, further development of this framework is well justified. If the dominant uncertainties here can be reduced, neutron stars may become a meaningful complementary route for testing CP-violating physics beyond the Standard Model.

%% file: appa.tex
\chapter{Constraining the axion mass using pulsar energy loss}
\label{app:axion}
The axion is one of the best-motivated candidates for physics beyond the Standard Model. It was originally introduced as a dynamical solution to the strong-CP problem, and in many realizations it can also contribute to the cosmic dark-matter abundance \cite{peccei_cp_1977,peccei_constraints_1977,weinberg_new_1978,wilczek_problem_1978}. Neutron stars provide a particularly useful probe of axions because they do not rely on the assumption that axions make up the dark matter. Instead, the dense, highly degenerate matter in the stellar core can thermally produce axions through ordinary many-body processes. Once produced, axions interact so weakly that they escape the star and carry energy away, in close analogy with neutrinos. In this sense, axions act as an additional cooling channel in the thermal evolution of a neutron star \cite{buschmann_upper_2022,yanagi_thermal_2020}.

For isolated neutron stars at ages of order $10^5$--$10^6~{\rm yr}$, axion cooling can be studied as a perturbation of the standard passive-cooling problem \cite{buschmann_upper_2022}. However, for an old recycled millisecond pulsar such as PSR~J0437$-$4715, passive cooling alone is not sufficient: the observed ultraviolet surface temperature requires a persistent internal heat source \cite{durant_spectrum_2012,gonzalez_internal_2010}. In this regime, axions do not simply make the star cooler in isolation. Rather, they modify the late-time balance between internal heating and the neutrino, photon, and axion luminosities. The problem is therefore intrinsically one of heated neutron-star evolution, not passive cooling.

\section{Standard neutron-star cooling and global thermal balance}

The standard thermal evolution of a neutron star is governed by the competition between internal heat capacity, neutrino emission from the core, and photon emission from the surface \cite{yakovlev_neutron_2004,potekhin_neutron_2015}. In the isothermal-interior approximation, the redshifted global thermal-balance equation may be written as
\begin{equation}
    C \frac{dT_b^\infty}{dt}
    =
    L_H^\infty
    -
    L_\nu^\infty
    -
    L_a^\infty
    -
    L_\gamma^\infty ,
\end{equation}
where $C$ is the total heat capacity, $T_b^\infty$ is the redshifted internal temperature, $L_H^\infty$ is the total internal heating power, $L_\nu^\infty$ is the neutrino luminosity, $L_a^\infty$ is the axion luminosity, and $L_\gamma^\infty$ is the redshifted photon luminosity.

The total neutrino luminosity can be written as
\begin{equation}
    L_{\nu}^{\infty}
    =
    \int Q_{\nu}(r,T,\rho,\ldots)\,e^{2\Phi(r)}\,dV ,
\end{equation}
where $Q_{\nu}$ is the local neutrino emissivity, $dV$ is the proper volume element, and $\Phi(r)$ is the metric potential that accounts for gravitational redshift. For nucleonic matter, the dominant slow neutrino mechanisms are typically the modified Urca process and nucleon--nucleon bremsstrahlung \cite{yakovlev_neutron_2004}. These processes operate in ordinary $npe$ matter when the kinematic threshold for direct beta decay is not satisfied. Their emissivities scale strongly with temperature and usually dominate the cooling of low-mass neutron stars in which the faster direct-Urca channel is forbidden. If the proton fraction is high enough to satisfy momentum conservation for beta decay, then the direct Urca process becomes allowed and can increase the neutrino luminosity by orders of magnitude \cite{yakovlev_neutron_2004}.

Superfluidity and superconductivity modify neutrino cooling in two important ways. First, pairing suppresses many standard neutrino reactions because the phase space for quasiparticle excitations is reduced once an energy gap opens. Second, the formation and breaking of Cooper pairs itself becomes a neutrino-emitting process, usually called pair breaking and formation (PBF) \cite{yakovlev_neutron_2004,ho_discovery_2011}. As a result, the choice of superfluid gap model has a large influence on the predicted cooling history.

Photon cooling is the thermal radiation that escapes from the stellar surface and is the dominant cooling channel at late times \cite{yakovlev_neutron_2004,potekhin_neutron_2015}. Unlike neutrino emission, photon cooling depends crucially on how heat is transported from the interior to the surface. The main observable is the redshifted effective temperature $T_s^\infty$, which is related to the internal temperature through the heat-blanketing envelope. The envelope acts as a thermal insulator, so two stars with the same interior temperature may have different surface temperatures depending on the structure of the outer layers. Light-element envelopes generally conduct heat more efficiently than heavy-element envelopes \cite{yakovlev_neutron_2004,potekhin_neutron_2015}. The redshifted photon luminosity is
\begin{equation}
    L_\gamma^\infty
    =
    4\pi R_\infty^2 \sigma \left(T_s^\infty\right)^4 .
\end{equation}

At sufficiently late times, the neutrino luminosity drops rapidly because of its steep temperature dependence, and the surface photon luminosity becomes the dominant energy-loss channel. This marks the photon-cooling stage. The cooling curve then becomes especially sensitive to the crust, the blanketing envelope, and any persistent internal heating terms. For stars like PSR~J0437$-$4715, the late-time surface temperature is therefore determined by a balance between residual internal heating and surface photon losses \cite{gonzalez_internal_2010, durant_spectrum_2012}.

\section{Axion cooling in neutron stars}

In the present work, axions are included as an additional cooling channel. The axion--nucleon interaction is written in derivative form as
\begin{equation}
    \mathcal{L}_{aN}
    =
    \sum_{N=p,n}
    \frac{C_N}{2f_a}
    \bar{\psi}_N \gamma^\mu \gamma_5 \psi_N \,\partial_\mu a ,
\end{equation}
where $f_a$ is the axion decay constant, $C_p$ and $C_n$ are model-dependent proton and neutron couplings, and $a$ is the axion field \cite{cortona_qcd_2016,buschmann_upper_2022}. It is convenient to define the effective dimensionless couplings
\begin{equation}
    g_{aNN} = \frac{C_N m_N}{f_a},
\end{equation}
which in the code are represented by the proton and neutron couplings $g_{app}$ and $g_{ann}$.

For a QCD axion, the mass is related to the decay constant by
\begin{equation}
    m_a \simeq 5.7\,\mu{\rm eV}
    \left(\frac{10^{12}\,{\rm GeV}}{f_a}\right),
\end{equation}
with the precise coefficient determined by low-energy QCD inputs \cite{cortona_qcd_2016}. In DFSZ models, the couplings depend on $\tan\beta$ through
\begin{equation}
    C_n = -0.160 + 0.414 \sin^2\beta,
    \qquad
    C_p = -0.182 - 0.435 \sin^2\beta ,
\end{equation}
so that each point in $(m_a,\tan\beta)$ space maps onto a definite pair $(g_{ann},g_{app})$ \cite{buschmann_upper_2022}.

The total axion luminosity is obtained by integrating the local axion emissivity over the star,
\begin{equation}
    L_a^\infty
    =
    \int Q_a(r)\,e^{2\Phi(r)}\,dV ,
\end{equation}
where $Q_a$ is the local axion emissivity, $\Phi(r)$ is the metric potential, and $dV$ is the proper volume element. In dense nucleonic matter, the dominant axion-production channels are nucleon--nucleon bremsstrahlung and, when the nucleons are paired, Cooper pair breaking and formation \cite{buschmann_upper_2022}. Thus,
\begin{equation}
    L_a^\infty
    =
    L_{a,\mathrm{brems}}^\infty
    +
    L_{a,\mathrm{PBF}}^\infty .
\end{equation}
The bremsstrahlung contribution is present throughout the nucleonic core, while the PBF contribution becomes important when the temperature passes through the superfluid critical temperature. In practice, axions therefore act as an additional sink of internal thermal energy, analogous to neutrinos but with a different coupling structure and parameter dependence.

For nearby isolated neutron stars of age $\sim{\rm Myr}$, Buschmann et al.\ treat axions as an additional loss term in a passive-cooling analysis with no internal heating term \cite{buschmann_upper_2022}. In contrast, for PSR~J0437$-$4715 the relevant question is whether a given axion luminosity is compatible with the amount of heating required to sustain the observed late-time surface temperature \cite{durant_spectrum_2012,gonzalez_internal_2010}. In this sense, the constraint is obtained from the competition between heating and cooling, not from passive cooling alone.

\section{Rotochemical heating}

The dominant heating mechanism considered in this work is rotochemical heating. The basic idea, introduced by Reisenegger and developed further by Fernández \& Reisenegger, is that as a rotating neutron star spins down, it gradually loses centrifugal support and contracts \cite{reisenegger_deviations_1995,fernandez_rotochemical_2005,reisenegger_rotochemical_2006}. This slow compression changes the beta-equilibrium composition corresponding to the local density, but the weak interactions do not restore equilibrium instantaneously. As a result, the matter is driven out of beta equilibrium, storing chemical energy that is later released through nonequilibrium weak reactions.

For npe matter, the chemical imbalance is
\begin{equation}
    \eta_{npe}\equiv \mu_n-\mu_p-\mu_e ,
\end{equation}
and it is often convenient to define the dimensionless imbalance
\begin{equation}
    \xi \equiv \frac{\eta_{npe}}{k_B T^\infty} .
\end{equation}
When $\eta_{npe}\neq 0$, beta reactions proceed at different forward and reverse rates, producing both neutrino emission and heat \cite{fernandez_rotochemical_2005}. In the non-superfluid modified-Urca case, the redshifted nonequilibrium neutrino luminosity and the net reaction rate can be written as
\begin{equation}
    L_\nu^\infty
    =
    \tilde{L}\,F_M(\xi)\,(T^\infty)^8 ,
\end{equation}
\begin{equation}
    \Delta \tilde{\Gamma}
    =
    \frac{\tilde{L}}{k_B}\,H_M(\xi)\,(T^\infty)^7 ,
\end{equation}
where $\tilde{L}$ is the microphysical normalization of the modified-Urca channel, and the dimensionless nonequilibrium functions are
\begin{equation}
    F_M(\xi)
    =
    1
    +
    \frac{22020}{11513\pi^2}\xi^2
    +
    \frac{5670}{11513\pi^4}\xi^4
    +
    \frac{420}{11513\pi^6}\xi^6
    +
    \frac{9}{11513\pi^8}\xi^8 ,
\end{equation}
and
\begin{equation}
    H_M(\xi)
    =
    \frac{14680}{11513\pi^2}\xi
    +
    \frac{7560}{11513\pi^4}\xi^3
    +
    \frac{840}{11513\pi^6}\xi^5
    +
    \frac{24}{11513\pi^8}\xi^7 ,
\end{equation}
following the standard non-superfluid rotochemical-heating formalism \cite{fernandez_rotochemical_2005}. The corresponding heating power is
\begin{equation}
    L_H^\infty
    =
    \eta_{npe}\,\Delta \tilde{\Gamma}
    =
    \tilde{L}\,H_M(\xi)\,(T^\infty)^8 .
\end{equation}

The imbalance itself evolves because spin-down continuously drives the star away from equilibrium while weak reactions restore it. In the simplest npe approximation, this may be written schematically as
\begin{equation}
    \frac{d\eta_{npe}^\infty}{dt}
    =
    2W_{npe}\,\Omega \dot{\Omega}
    -
    Z_{npe}\,\Delta \tilde{\Gamma} ,
\end{equation}
where $W_{npe}$ and $Z_{npe}$ are structure-dependent coefficients and $\Omega$ is the spin frequency \cite{fernandez_rotochemical_2005,reisenegger_rotochemical_2006}. The first term represents spin-down-driven departure from equilibrium, while the second term represents weak-interaction relaxation back toward equilibrium. At late times, the system can approach a quasi-steady state in which the compression-induced departure from equilibrium is balanced by the restoring reactions, so that the heating power, neutrino losses, and surface photon luminosity all vary only slowly with time \cite{fernandez_rotochemical_2005}.

Superfluidity modifies the rotochemical evolution in two important ways. First, pairing suppresses the ordinary beta reactions and reduces the heat capacity. Second, it introduces threshold effects that delay equilibration, allowing the chemical imbalance to grow larger before reactions become efficient \cite{petrovich_rotochemical_2010, gonzalez-jimenez_rotochemical_2015}. As a result, rotochemical heating generally predicts higher late-time temperatures than the non-superfluid case, which is why it is especially relevant for PSR~J0437$-$4715.
\begin{figure*}
\centering
\includegraphics[width=\textwidth,clip]{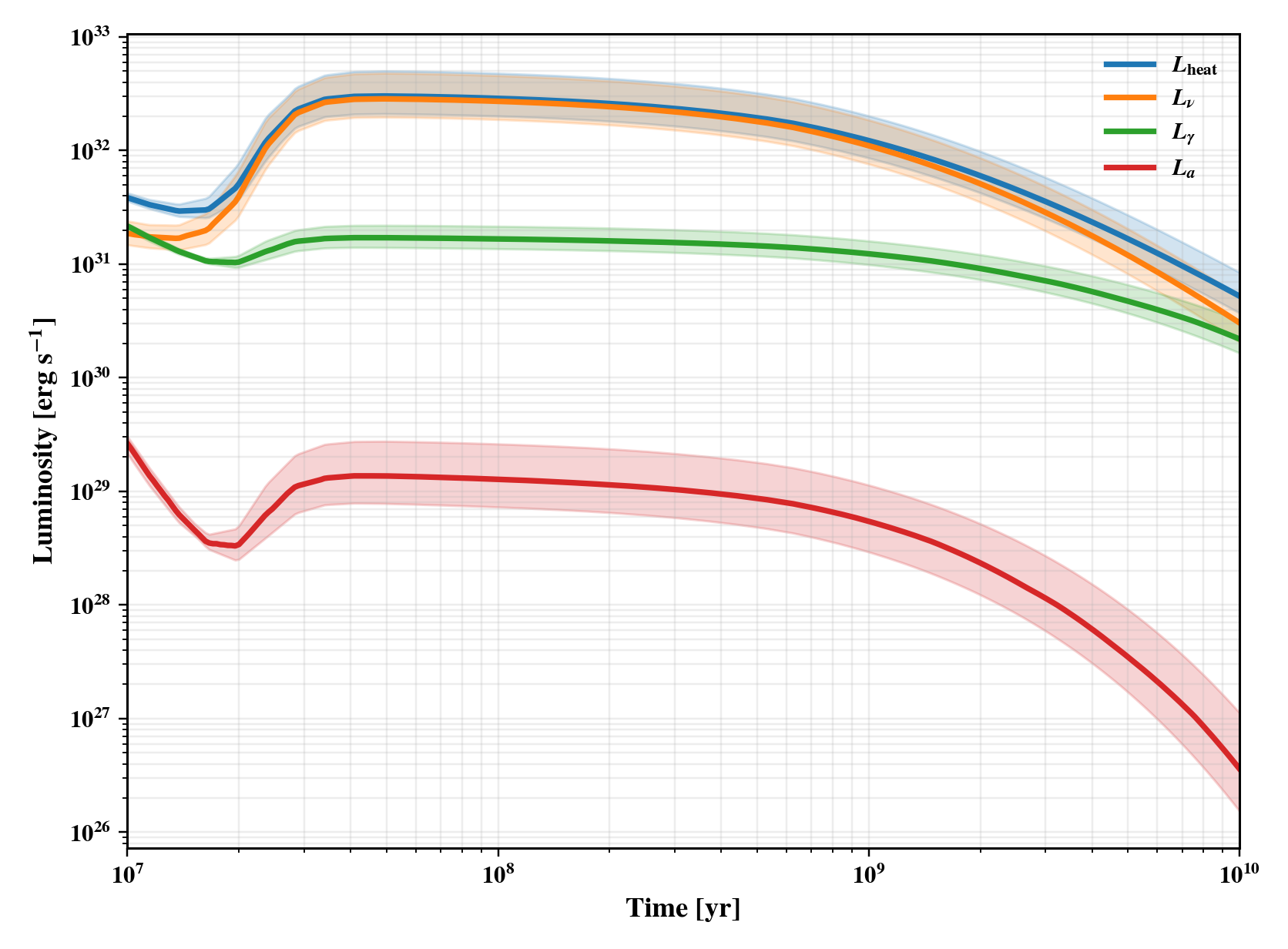}
\caption{Thermal-evolution luminosity bands for PSR~J0437$-$4715 in the heated-evolution framework, showing the relative contributions of the dominant cooling and heating channels over time. The bands summarize the spread obtained from varying the thermal and microphysical inputs in the \textsc{nscool} calculations, including the stellar structure, superfluid gap model, envelope prescription, and heating parameters. The comparison illustrates the epoch at which passive cooling gives way to a late-time regime regulated by the competition between internal heating and photon, neutrino, and axion losses.}
\label{fig:luminosity_bands_all}
\end{figure*}
\section{Vortex heating}

Vortex creep heating refers to the heat produced when a neutron superfluid adjusts to stellar spin-down and its quantized vortices move through pinning sites in the inner crust \cite{alpar_vortex_1984,gonzalez_internal_2010}. The heating comes from frictional dissipation: rotational energy stored in the more rapidly rotating superfluid is converted into heat as the vortex distribution evolves. Although it is often less dominant than rotochemical heating in the very late-time regime relevant here, it can still provide a late-time temperature floor and therefore remains an important systematic ingredient in old millisecond pulsars \cite{gonzalez_internal_2010}.

A commonly used parametrization is
\begin{equation}
    L_{\rm vortex}^{\infty}\simeq J\,|\dot{\Omega}_{\infty}| ,
\end{equation}
where $J$ is an effective excess angular-momentum reservoir associated with pinned superfluid components \cite{gonzalez_internal_2010}. In practical thermal-evolution studies, different assumptions about the superfluid gaps and pinning physics translate into different levels of vortex heating.

\section{Implementation into \textsc{nscool}}

The axion sector is inserted into the \textsc{nscool} model through the couplings $g_{ann}$ and $g_{app}$, which determine the local axion emissivity $Q_a(r)$ and hence the redshifted luminosity $L_a^\infty$ \cite{buschmann_upper_2022,page_nscool_2016}. Rotochemical heating is coupled to the thermal evolution through an effective core-averaged solver. At each timestep, the code computes the redshifted core temperature, reconstructs the instantaneous $\Omega$ and $\dot{\Omega}$ from the adopted spin-down history, evolves the chemical imbalance according to the rotochemical evolution equation, and evaluates the associated nonequilibrium neutrino and heating luminosities.

These luminosities are then distributed over the core with a redshift-weighted normalization
\begin{equation}
    N_{\rm core}
    =
    \sum_{j\le i_{\rm core}} e^{2\phi_j}\,\Delta V_j ,
\end{equation}
so that the additional local source terms are
\begin{equation}
    q_{{\rm heat,roto}}
    =
    \frac{L_{H,{\rm roto}}^\infty}{N_{\rm core}},
    \qquad
    q_{\nu,{\rm roto}}
    =
    \frac{L_{\nu,{\rm roto}}^\infty}{N_{\rm core}} .
\end{equation}
Schematically, the local thermal budget entering the evolution equation is therefore
\begin{equation}
    q_{\rm net}
    =
    q_{\rm heat}
    -
    q_\nu
    -
    q_a .
\end{equation}

\begin{figure*}
\centering
\includegraphics[width=\textwidth,clip]{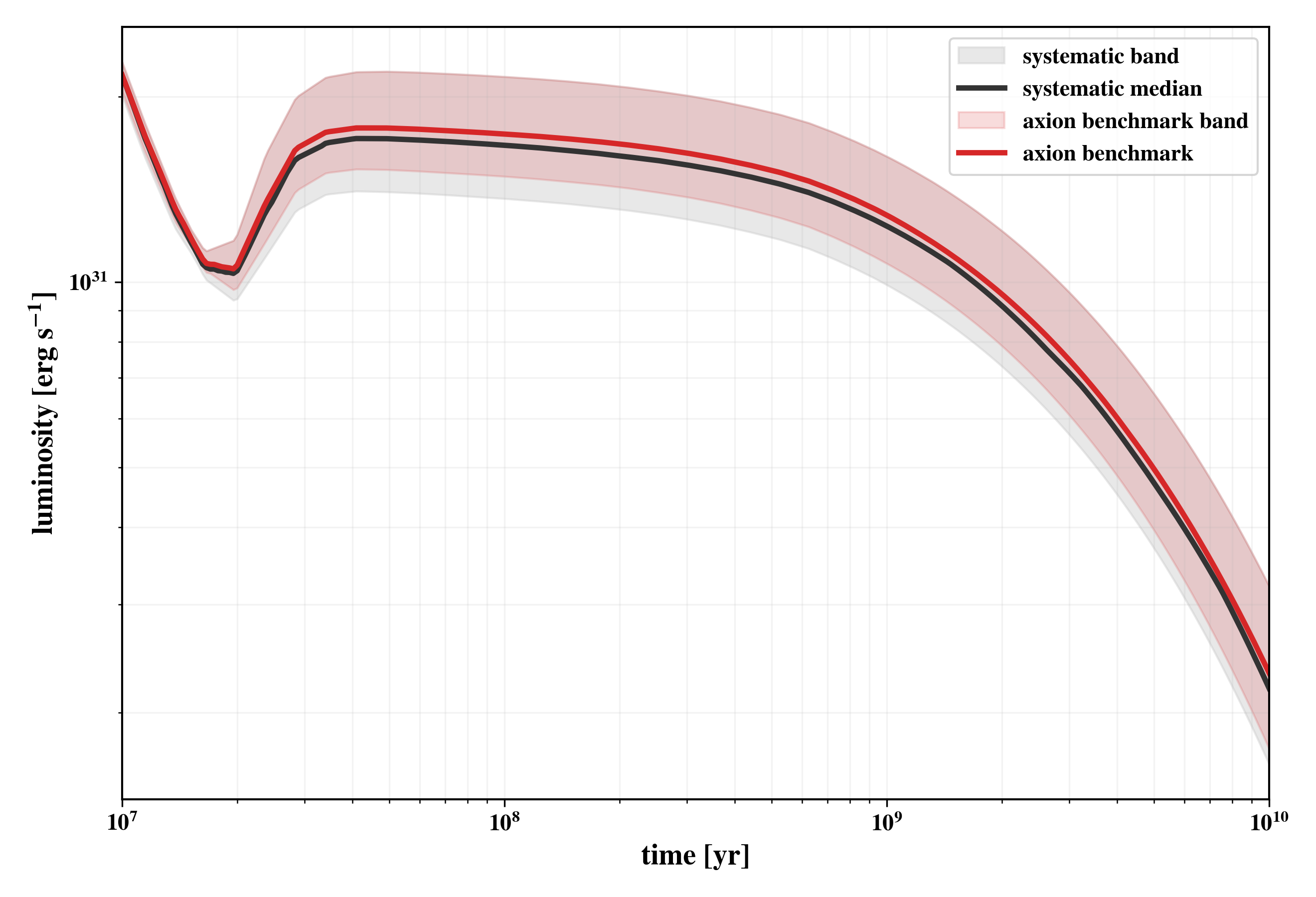}
\caption{Comparison of a benchmark axion model with the full thermal-systematic band for the photon luminosity, $L_{\gamma}$, over $10^{7}$--$10^{10}\,\mathrm{yr}$. The gray shaded region and black median curve show the range produced by uncertainties in the thermal modeling, including the heating prescription and nuisance microphysics, while the red curve and shaded band show the benchmark axion model. Because the benchmark axion prediction remains within, or very close to, the thermal systematic band over the relevant age range, the axion-induced change in the observable photon luminosity is not cleanly separable from the underlying thermal-model uncertainty in the heated PSR~J0437$-$4715 framework. This is the main reason that the present analysis does not yield a strong axion constraint.}
\label{fig:axion_vs_systematic_lgamma}
\end{figure*}
The transition from passive photon-dominated cooling to a heating- and neutrino-regulated regime marks the onset of a nonequilibrium thermal state. Beyond this point, the surface temperature is no longer determined primarily by the residual cooling of the original thermal reservoir, but rather by the balance between spin-down-powered internal heating and core energy losses through neutrinos, photons, and any additional channels such as axions. The timing of this transition depends on the spin-down history, the rotochemical coefficients, the superfluid gap structure, the stellar mass and equation of state, the envelope composition, and the strength of the axion cooling channel.

\subsection{Interpretation and conclusions}

The thermal-evolution results show that PSR~J0437$-$4715 is not well described as a purely passive-cooling neutron star at late times. Instead, by the epoch relevant to the observed ultraviolet temperature, the star has entered a quasi-steady heated regime in which the surface emission is set by the competition between internal heating and the combined neutrino, photon, and axion loss channels. This point is important because it changes the nature of the axion test. In a passive-cooling analysis, an additional axion channel can produce a comparatively clean downward shift in the cooling curve. In the present case, however, the observable temperature is determined by a balance among several poorly constrained components, so the axion contribution must be interpreted relative to that full thermal budget rather than in isolation.

The luminosity-budget plots make this clear. At early times, the evolution follows the familiar cooling sequence in which thermal energy is lost primarily through standard channels. At later times, however, the star approaches a nonequilibrium state in which spin-down-powered heating replenishes the thermal reservoir while photon and neutrino losses regulate the outward energy flow. In this regime, adding axion emission does not simply cool the star in a one-parameter way. Instead, it perturbs an already balanced system whose late-time temperature also depends on the rotochemical-heating efficiency, the adopted superfluid gap model, the envelope relation, the stellar structure, and any additional late-time heating contributions such as vortex dissipation.

For this reason, the relevant comparison is not between an axion curve and a single no-axion baseline, but between the axion prediction and the full thermal systematic band. In the calculations presented here, the benchmark axion model remains within, or very close to, the spread produced by plausible thermal-model variations over the age range of interest. The inferred change in $L_\gamma^\infty$ or $T_{\mathrm{eff}}^\infty$ is therefore not cleanly separable from the uncertainty associated with the nuisance microphysics and heating prescriptions. Put differently, the present thermal uncertainty is large enough that an axion-induced shift can be absorbed by changes in the non-axion model ingredients.

The main conclusion is therefore not that axions have no effect, but that for PSR~J0437$-$4715 the current heated-evolution framework does not provide a competitive or robust axion exclusion in the same sense as passive-cooling studies of younger isolated neutron stars. The limiting factor is not numerical sensitivity alone; it is the physical degeneracy between extra cooling and uncertain late-time heating and transport physics. This explains why the present constraints are substantially weaker than those obtained in systems where the thermal history is closer to passive cooling and the axion signal can be isolated more cleanly.

Nevertheless, the exercise is still informative. It demonstrates explicitly that once persistent internal heating is required by the data, any axion interpretation becomes strongly model dependent. In that sense, PSR~J0437$-$4715 provides a useful counterexample to the passive-cooling paradigm: it shows that old reheated millisecond pulsars are not generically ideal laboratories for precision axion bounds unless the heating mechanism, envelope composition, and superfluid microphysics can be constrained much more tightly than is currently possible. Future improvements would therefore have to come primarily from reducing those thermal systematics, rather than simply from running a denser scan in axion parameter space.

%% file: biblio.tex
%% This defines the bibliography file (main.bib) and the bibliography style.
%% If you want to create a bibliography file by hand, change the contents of
%% this file to a `thebibliography' environment.  For more information 
%% see section 4.3 of the LaTeX manual.
\begin{singlespace}
% \bibliography{main}
%\bibliographystyle{plain}
\printbibliography
\end{singlespace}